\documentclass{aa}  

\usepackage{graphicx}
\usepackage{tabularx}
\usepackage{array}
\usepackage{txfonts}
\usepackage{orcidlink}
\usepackage{gensymb}
\usepackage{natbib}
\bibpunct{(}{)}{;}{a}{}{,} 
\usepackage{hyperref}
\hypersetup{
  colorlinks   = true, 
  urlcolor     = blue, 
  linkcolor    = red, 
  citecolor    = blue 
}

\usepackage{upgreek}

\usepackage{xcolor}
\usepackage{multirow}

\makeatletter
\renewcommand*\aa@pageof{, page \thepage{} of \pageref*{LastPage}}
\makeatother

\begin{document}

   \title{A `MeerKAT-meets-LOFAR' study of the complex multi-component (mini-)halo in the extreme sloshing cluster Abell~2142}

   \author{C.~J.~Riseley\,\orcidlink{0000-0002-3369-1085}\,\inst{1,2},
        A.~Bonafede\,\orcidlink{0000-0002-5068-4581}\,\inst{1,2},
        L.~Bruno\,\orcidlink{0000-0003-1217-6807}\,\inst{1,2},
        A.~Botteon\,\orcidlink{0000-0002-9325-1567}\,\inst{2},
        M.~Rossetti\,\orcidlink{0000-0002-9775-732X}\,\inst{3},
        N.~Biava\,\orcidlink{0000-0001-7947-6447}\,\inst{4,2},
        E.~Bonnassieux\,\orcidlink{0000-0003-2312-3508}\,\inst{5,6,2}, 
        F.~Loi\,\orcidlink{0000-0002-8627-6627}\,\inst{7},
        T.~Vernstrom\,\orcidlink{0000-0001-7093-3875}\,\inst{8,9},
        M.~Balboni\,\orcidlink{0009-0001-3048-0020}\,\inst{4,10}
    }
    \authorrunning{C. J. Riseley et al.}
    \titlerunning{MeerKAT-meets-LOFAR: Abell~2142}
   \institute{
        Dipartimento di Fisica e Astronomia, Università degli Studi di Bologna, via P. Gobetti 93/2, 40129 Bologna, Italy\\
        \email{christopher.riseley@unibo.it}
        \and 
        INAF -- Istituto di Radioastronomia, via P. Gobetti 101, 40129 Bologna, Italy
        \and 
        INAF -- IASF Milano, Via A. Corti 12, 20133 Milano, Italy
        \and 
        Th\"{u}ringer Landessternwarte, Sternwarte 5, 07778 Tautenburg, Germany
        \and 
        Julius-Maximilians-Universit\"{a}t W\"{u}rzburg, Fakult\"{a}t f\"{u}r Physik und Astronomie, Institut f\"{u}r Theoretische Physik und Astrophysik, Lehrstuhl f\"{u}r Astronomie, Emil-Fischer-Str. 31, D-97074 W\"{u}rzburg, Germany
        \and 
        Observatoire de Paris, 5 place Jules Janssen, 92195 Meudon, France
        \and 
        INAF -- Osservatorio Astronomico di Cagliari, via della Scienza 3, Selargius, Italy
        \and 
        ICRAR, The University of Western Australia, 35 Stirling Hw, 6009 Crawley, Australia
        \and 
        CSIRO Space \& Astronomy, PO Box 1130, Bentley, WA 6102, Australia
        \and 
        DiSAT, Universit\`a degli Studi dell’Insubria, via Valleggio 11, I-22100 Como, Italy
        }

   \date{Received 23rd February 2024; accepted 29th February 2024; in original form 13th December 2023}

 
  \abstract
   {Clusters of galaxies are known to be turbulent environments, whether they are merging systems where turbulence is injected via the conversion of gravitational potential energy into the intracluster medium (ICM), or whether they are relaxed systems in which small-scale core sloshing is occurring within the potential well. In many such systems, diffuse radio sources associated with the ICM are found: radio haloes and mini-haloes.}
   {Abell~2142 is a rich cluster undergoing an extreme episode of core sloshing, which has given rise to four cold fronts and a complex multi-component radio halo. Recent work revealed that there are three primary components to the halo, which spans a distance of up to around 2.4\,Mpc. The underlying physics of particle acceleration on these scales is poorly-explored, and requires high-quality multi-frequency data with which to perform precision spectral investigation. We aim to perform such an investigation.}
   {We use new deep MeerKAT L-band (1283\,MHz) observations in conjunction with LOFAR HBA (143\,MHz) data as well as X-ray data from \textit{XMM-Newton} and \textit{Chandra} to study the spectrum of the halo and the connection between the thermal and non-thermal components of the ICM.}
   {We confirm the presence of the third halo component, detecting it for the first time at 1283\,MHz and confirming its ultra-steep spectrum nature, as we recover an integrated spectrum of $\alpha_{\rm H3, \, total} = -1.68 \pm 0.10$. All halo components follow power-law spectra with increasingly steep spectra moving toward the cluster outskirts. We profile the halo along three directions, finding evidence of asymmetry and spectral steepening along an axis perpendicular to the main axis of the cluster. Our thermal/non-thermal investigation shows sub-linear correlations that are steeper at 1283\,MHz than 143\,MHz, and we find evidence of different thermal/non-thermal connections in different components of the halo. In particular, we find both a moderate anti-correlation (H1, the core) and positive correlation (H2, the ridge) between radio spectral index and X-ray temperature.}
   {Our results are broadly consistent with an interpretation of turbulent (re-)acceleration following an historic minor cluster merger scenario in which we must invoke some inhomogeneities. However, the anti-correlation between radio spectral index and X-ray temperature in the cluster core is challenging to explain; the presence of three cold fronts and a generally lower temperature may provide the foundations of an explanation, but detailed bespoke modelling is required to study this further.}


   \maketitle

\section{Introduction}
In the hierarchical model of structure formation, galaxy clusters form from over-dense seeds at high redshift, growing over cosmic time into the largest gravitationally-bound structures in the Universe. Clusters are dominated by dark matter, which makes up some $\sim80\%$ of the mass budget, whereas galaxies comprise only around five per cent of the mass. The remaining $\sim 15\%$ of the mass budget is comprised of the intra-cluster medium (ICM), a hot and rarefied plasma ($T \sim10^7 - 10^8$~K, $n_{\rm{e}} \sim 10^{-3}$~cm$^{-3}$) which emits at X-ray wavelengths via the Bremsstrahlung mechanism and scatters Cosmic Microwave Background (CMB) photons via the Sunyaev-Zel'dovich (SZ) effect.

While the dividing line between galaxy groups and clusters is somewhat arbitrary, clusters typically have a mass $M_{500} \gtrsim 10^{14} \, {\rm M}_{\odot}$ \citep[where $M_{500}$ is the mass contained within a radius of $r_{500}$, and $r_{500}$ is the radius at which the mean cluster overdensity is $500\times$ the critical density of matter in the Universe at the cluster redshift; e.g.][]{Planck2016_SZcat} with the most massive clusters having upwards of $10^{15} \, {\rm M}_{\odot}$. Located at the nodes and intersections of the Cosmic Web, clusters grow through a variety of processes from continuous accretion of matter, to minor mergers and absorption of sub-groups, through to the most dynamic and spectacular events in the Universe since the Big Bang: major mergers with other clusters \citep[e.g.][]{Kravtsov_Borgani_2012}.

During major merger events, tremendous amounts of gravitational potential energy are deposited into the ICM through processes like shocks and turbulent phenomena which drive heating, the mixing of gas, and the acceleration of particles (principally electrons) to the relativistic regime, generating cosmic ray electrons \citep[CRe; e.g.][]{Markevitch_Vikhlinin_2007}. Given the ionised nature of the ICM, these environments are permeated by large-scale magnetic fields, and where CRe are in the presence of magnetic fields, synchrotron emission is generated, which can be detected in the form of radio waves \citep[e.g.][]{Govoni_Feretti_2004,vanWeeren2019}.

As such, many merging clusters are known to host diffuse radio emission on some of the largest scales in the Universe, up to and over 1\,Mpc in physical extent. Such emission is not associated with individual cluster-member galaxies but rather the ICM itself, reflecting the ongoing processes at work. Broadly-speaking, the sources hosted by merging clusters can be classified into two main groups: radio relics, which trace merger shocks, and radio haloes, which are commonly accepted to trace ICM turbulence in the wake of a merger. There exists a third class of diffuse radio emission: `mini-haloes'. These are typically more compact than haloes, with linear sizes typically $\lesssim 300$\,kpc, and historically have been detected primarily in relaxed cool-core clusters \citep[e.g.][]{Giacintucci2014b}. For an observational review of the field, see \cite{vanWeeren2019}; whereas for exploration of the theory of particle acceleration in cluster environments, see for example \cite{Brunetti_Jones_2014} and \cite{Ruszkowski_Pfrommer_2023}.

To-date, around one hundred each of radio relics and haloes have been detected, and around 40 mini-haloes, although these numbers are increasing as a result of next-generation instrumentation like the LOw-Frequency Array \citep[LOFAR;][]{vanHaarlem2013}, the Murchison Widefield Array \citep[MWA;][]{Tingay2013,Wayth2018}, the Australian Square Kilometre Array Pathfinder \citep[ASKAP;][]{Johnston2007,DeBoer2009} and MeerKAT (\citealt{Jonas2016}, though see also \citealt{Camilo2018,Mauch2020}). See for example surveys with LOFAR \citep[e.g.][]{Botteon2022_LOTSS_Planck,Hoang2022_LOTSS_NonPlanck}, the MWA \citep{Duchesne2021_MWA2_ASKAP,Duchesne2021_MWA_EoR0}, ASKAP \citep[e.g.][]{Duchesne2021_MWA2_ASKAP,Duchesne2024_EMU-ES-Clusters} and MeerKAT \citep{Knowles2022}.

Not only are these new instruments detecting ever more relics, haloes and mini-haloes, but diffuse radio emission is being detected on even larger scales, with `mega'-haloes spanning scales of several Mpc \citep{Cuciti2022_megahaloes}, inter-cluster bridges currently detected on scales of $\sim 3$ Mpc \citep{Govoni2019_radio_bridge,Botteon2020_A1758_bridge,deJong2022_A399_A401,Pignataro2023_arXiv}, and radio emission filling volumes up to $r_{200}$ \citep{Botteon2022_Abell2255}. Further, the once-strict division between haloes and `mini'-haloes has become blurred, with several `mini'-haloes showing diffuse emission on larger scales than previously reported, up to $\gtrsim 0.5$\,Mpc. These clusters often show multiple components, hinting at difference physical processes at work on different scales \citep[e.g.][Biava et al., submitted]{Savini2018,Savini2019,Biava2021_RXCJ1720,Lusetti2023,Riseley2022_MS1455,Riseley2023_A1413}.

\subsection{Radio haloes}
Radio haloes are centrally-located and irregular diffuse radio sources that broadly trace the thermal emission from the ICM, implying a direct connection between thermal and non-thermal components. The CRe that trace diffuse radio emission in the ICM have energies in the GeV regime, and so have a relatively short lifetime --- typically $\lesssim 100$\,Myr --- far shorter than the diffusion timescale over Mpc distances, which is $\gtrsim 100$\,Gyr for CRe with GeV energies \citep[e.g.][]{Brunetti_Jones_2014,Ruszkowski_Pfrommer_2023}. As such, these CRe must be accelerated to GeV energies in-situ.

It is generally accepted that the particle acceleration is carried out by a form of magnetohydrodynamic (MHD) turbulence following cluster mergers \citep{Brunetti2001,Petrosian2001} which (re-)accelerates electrons to relativistic energies. This scenario naturally forges the link between the presence of radio haloes and ongoing/recent cluster mergers \citep[e.g.][]{Brunetti_Lazarian_2007,Brunetti2009}. Indeed, simulations suggest that while initial merger shocks traverse the cluster on timescales $\lesssim 2$\,Gyr (the cluster crossing timescale), turbulence can persist in cluster environments for far longer, taking up to $\sim 4$\,Gyr before turbulence decays to the typical level found in relaxed clusters, where the turbulent pressure is less than around $5\%$ of the total pressure budget \citep[e.g.][]{Paul_2011_ClusterShocksTurbulence,Vazza2011}.

Turbulence in the ICM is inherently intermittent both spatially and temporally, and is an inefficient particle acceleration mechanism. However, it successfully explains many observed properties of haloes, such as the often similar spatial extent of radio haloes and the X-ray emitting ICM, the steep radio spectrum --- typically $\alpha \sim -1$, where we adopt the convention that flux density $S$ is proportional to observing frequency $\nu$ as $S \propto \nu^{\alpha}$ --- and the existence of ultra-steep spectrum radio haloes: those with a spectral index $\alpha \lesssim -1.5$.

Ultra-steep spectrum haloes are a natural consequence of turbulent re-acceleration \citep[e.g.][and references therein]{Brunetti2008_A521,Cassano2010}. They may represent ancient radio haloes where the majority of the turbulence has already dissipated and is thus depositing only a very low amount of energy into the ICM. Alternatively, they may be generated by inefficient acceleration in the ICM resulting from minor mergers, low-impact-parameter interactions, and/or mergers involving low-mass systems. Indeed, among the known halo population, several exhibit ultra-steep spectra \citep[e.g.][]{Brunetti2008_A521,Dallacasa2009_A521,Macario2010_A697,Wilber2018_Abell1132,Duchesne2021_A141_A3404,Rajpurohit2023_A2256-Halo}.

However, turbulence has many associated parameters and properties that are poorly constrained by available observations. In order to  constrain these properties and shed light on the underlying MHD turbulence mechanism, we need to study large samples of haloes at high sensitivity and high resolution, across a broad frequency range. We also require detailed observations into the thermal properties of the ICM, provided by X-ray telescopes.

\subsection{The ``MeerKAT-meets-LOFAR'' mini-halo census}
In order to rigorously test the particle acceleration mechanisms responsible for the generation of mini-haloes, we are performing a statistical and homogeneously-selected census of all known mini-haloes catalogued by \cite{vanWeeren2019} that lie in the Declination range $-1\degree$ to $+30\degree$. Within this Declination range we have coverage with deep MeerKAT L-band observations (872$-$1712~MHz) under MeerKAT Project ID (PID) SCI-20210212-CR-01 (P.I. Riseley), as well as LOFAR observations sourced from the LOFAR Two-metre Sky Survey \citep[LoTSS;][]{Shimwell2017_LoTSS_PaperI,Shimwell2019_LOTSS_PaperII,Shimwell2022_LOTSS_PaperIII} or targeted observations where LoTSS coverage is unavailable. Further, this multi-frequency radio dataset is complemented by archival X-ray data from \textit{Chandra} and/or \textit{XMM-Newton} which provides the thermal window. The first two clusters to be studied in our census were MS~1455.0+2232 \citep{Riseley2022_MS1455} and Abell~1413 \citep{Riseley2023_A1413}, with a further ten clusters in the sample, beyond the one we will discuss in this paper: Abell~2142.

\subsection{The galaxy cluster Abell 2142}
Abell~2142 is a nearby (redshift $z = 0.0894$) massive cluster of $M_{500} = (8.8 \pm 0.2) \times 10^{14}$\,M$_{\odot}$ \citep{Planck2016_SZcat}. It is the dominant member of the Abell~2142 galaxy supercluster, which extends for approximately $50 \, h^{-1}$\,Mpc in a roughly straight filament \citep[see][]{Einasto2015_Abell2142}. The cluster itself shows textbook signatures of interaction, such as the presence of cold fronts \citep[e.g.][]{Markevitch2000_Abell2142}.

Optical observations reveal that the environment around the cluster is extremely rich and dynamic. For example, there are over $950$ cluster-member galaxies within $\sim 3$\,Mpc \citep{Owers2011_A2142_RXJ1720,Liu2018_A2142} which show clear evidence of substructure and ongoing accretion and minor merger activity with the main body of Abell~2142 \citep{Owers2011_A2142_RXJ1720,Einasto2015_Abell2142,Einasto2018_A2142}. Similarly, weak-lensing analysis indicates substructure in the mass distribution of Abell~2142 \citep{Okabe2008_weaklensing}. The vast majority of this substructure is aligned along the same axis as the greater supercluster, following a north-west/south-east axis.

At X-ray wavelengths, Abell~2142 shows an extended morphology typical of a disturbed ICM, elongated along a north-west/south-east axis. It is the first cluster in which cold fronts were detected with \textit{Chandra} at X-ray wavelengths \citep{Markevitch2000_Abell2142}; it has subsequently been studied extensively using both \textit{Chandra} and \textit{XMM-Newton} \citep[e.g.][]{Tittley_Henriksen_2005,Markevitch_Vikhlinin_2007,Johnson2011_Thesis,Rossetti2013_Abell2142,Wang_Markevitch_2018}. The cluster appears to be undergoing an extreme form of core sloshing, hosting four cold fronts: three are within $\sim340$\,kpc of the cluster centre, whereas the fourth is at an extremely large distance of around 1\,Mpc to the south-east \citep{Rossetti2013_Abell2142}. It is among the most distant (with respect to the cluster centre) cold fronts yet detected, third behind outer cold fronts in Perseus \citep{Walker2022_Perseus} and Abell~3558 \citep{Mirakhor2023_A3558} which each lie around 1.2\,Mpc from the respective cluster centres.

Further, the cluster appears to have a marginally elevated specific entropy in the centre that marks it as an intermediate between ``cool-core'' and non-cool-core clusters: instead it belongs to a relatively rare class of ``warm-core'' clusters (\citealt{Wang_Markevitch_2018}; along with Abell~1413, see \citealt{Giacintucci2017,Botteon2018_ChandraProfiles}). The interpretation of the available X-ray data is that Abell~2142 underwent an intermediate-mass-ratio merger event in the past, which has disrupted --- but not destroyed --- an existing cool core and triggered large-scale sloshing within the ICM, forming the observed cold fronts. The nature of the cold front far to the south-east is not completely clear, but it is hypothesised to be the result of either the natural long-term evolution of the small-scale sloshing structures in the cluster core, or an intermediate merger event that was more violent than typical off-axis mergers that trigger sloshing in cool-core clusters \citep[see discussion by][]{Rossetti2013_Abell2142}.

\begin{figure*}
\begin{center}
 \includegraphics[width=0.8\textwidth]{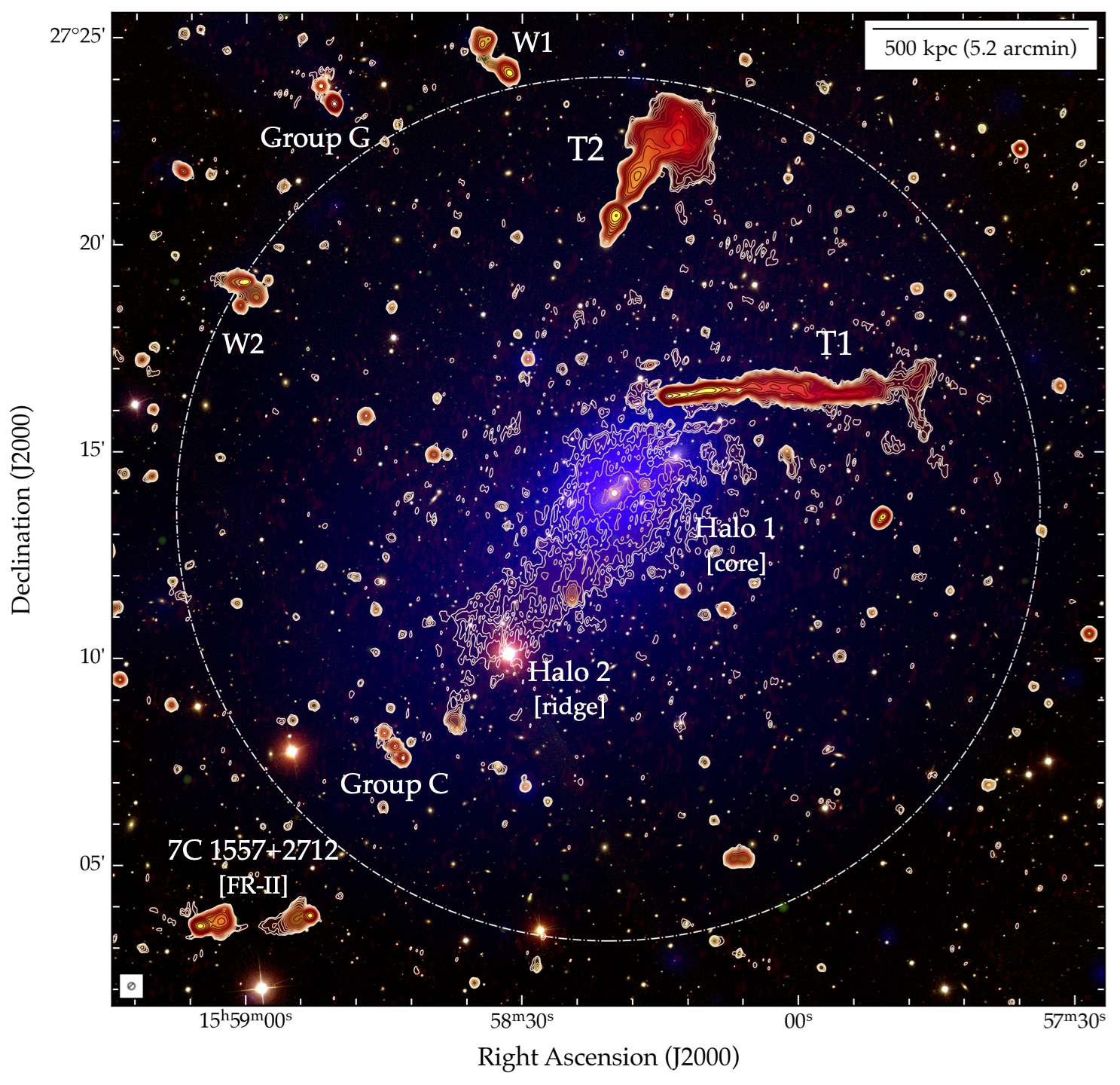}
\caption{Colour-composite image of Abell~2142, overlaying radio colour on optical RGB. Optical RGB image comprises $i$-, $r$- and $g$-bands from the SDSS. Yellow-through-red colours trace the radio surface brightness measured by LOFAR at 143\,MHz and MeerKAT at 1283\,MHz, both shown at 10\,arcsec resolution (indicated by the hatched circle in the lower-left corner), with redder colours denoting increasingly steep-spectrum synchrotron emission. Contours trace the LOFAR surface brightness starting at $4\sigma$ and scaling by a factor of $\sqrt{2}$. Blue colours trace the thermal ICM plasma measured by \textit{XMM-Newton}, adaptively-smoothed to enhance low surface brightness emission. Known features are identified following the established nomenclature \protect\citep{Venturi2017_A2142,Bruno2023_A2142}. The dot-dashed circle indicates a radius of 1\,Mpc centred on the cluster centre.}
\label{fig:A2142_Composite}
\end{center}
\end{figure*}

In the radio regime, Abell~2142 has been studied over the course of several decades \citep{Harris1977_clusters,Giovannini1999_NVSS_clusters,Giovannini_Feretti_2000,Farnsworth2013_GBT,Venturi2017_A2142,Bruno2023_A2142}. Aside from the presence of two rather spectacular tailed radio galaxies that are clearly embedded in the ICM, observations reveal the presence of a diffuse, highly-extended and asymmetric radio halo spanning up to $\sim 2$\,Mpc \citep{Farnsworth2013_GBT}. Sensitive higher-resolution observations of the halo revealed the presence of multiple components, which have been suggested to constitute either a mini-halo-plus-halo system \citep{Venturi2017_A2142} or a multi-component radio halo \citep{Bruno2023_A2142}. We show a colour-composite map of Abell~2142, constructed using optical, radio, and X-ray observations, in Figure~\ref{fig:A2142_Composite}. The optical colourmap was constructed using $i$-, $r$- and $g$-bands from the Sloan Digital Sky Survey (SDSS) Data Release 12 \citep[SDSS DR12;][]{Alam2015_SDSS-DR12}.

The multiple components within the radio halo show different properties: the central component (`Halo 1') is more rounded and is co-located with the brighter part of the ICM, whereas the second component (`Halo 2') is more ridge-like and extends to the south-east following the direction of the large-scale cold front \citep{Venturi2017_A2142,Bruno2023_A2142}. Finally, the third component (`Halo 3') identified at low frequencies by \cite{Bruno2023_A2142} appears to encompass the existing structures and fill a volume out to $\sim 1.2$\,Mpc from the cluster centre. 

All components of the halo show steep radio spectra $\alpha < -1$ (derived using data covering 50\,MHz to 405\,MHz), with Halo 3 showing an ultra-steep radio spectrum of $\alpha \lesssim -1.5$ (detected only at 50\,MHz and 143\,MHz). \cite{Bruno2023_A2142} propose two potential scenarios for the generation of H3: either the natural evolution of the original merger some $\sim 4$\,Gyr ago, which formed the main components of the halo, which has evolved into the present state; alternatively, the continuous accretion of small galaxy groups onto the main structure may provide sufficient turbulence injected over the cluster volume to generate the observed large-scale halo. The likely very low efficiency of such processes would provide a natural explanation for the low surface brightness and ultra-steep spectrum observed, however further theoretical exploration of such a scenario is required.

\section{Observations \& Data Reduction}\label{sec:observations}

\subsection{Radio: MeerKAT}
Abell~2142 was observed with MeerKAT using the L-band receivers covering the frequency range 872$-$1712~MHz under Project ID (PID) SCI-20210212-CR-01 (P.I. Riseley). Due to the high Declination of Abell~2142, it was necessary to perform two observing runs to achieve the full 5.5\,hours on-source time required to achieve our target sensitivity. These observations were performed on 2021~Oct.~10 (capture block ID 1633862771) and 2021~Nov.~12 (capture block ID 1636710370). During each observing run, the primary calibrator PKS~J1939$-$6342 was used to set the flux scale and derive bandpass solutions, and the secondary calibrator source J1609$+$2641 was used to track the time-varying gains. Polarisation calibration solutions were provided by two scans of the gold-standard polarisation calibrator J1331$+$3030 (3C\,286) separated by a large parallactic angle. The total on-source time was 5.5\,hours.

All calibration was performed in the same manner as previous papers from our census \citep{Riseley2022_MS1455,Riseley2023_A1413}. We used the Containerized Automated Radio Astronomy Calibration (\texttt{CARACal}) pipeline\footnote{\url{https://github.com/caracal-pipeline/caracal}} \citep{Jozsa2020,Jozsa2021} which employs calibration tasks from the Common Astronomy Software Application (\texttt{CASA}) package via the Stimela framework \citep{Makhathini2018}. We refer the reader to our previous papers for discussion of initial calibration steps.

Following initial calibration, we performed direction-independent (DI) and direction-dependent (DD) calibration using the \texttt{DDFacet} imager with the sub-space deconvolution algorithm \citep[\texttt{SSD};][]{Tasse2018}, and the \texttt{killMS} \citep{Tasse2014,Smirnov2015} solver to perform the calibration. Throughout, we employed the quality-based weighting scheme from \cite{Bonnassieux2018} to weight our calibration solutions, which expedited our self-calibration convergence. We performed a total of five rounds DI self-calibration (three applying phase-only solutions, two applying amplitude and phase solutions) before transitioning to the DD regime. In our DD calibration, we tessellated the sky into 16 regions and performed out two rounds of DD self-calibration and imaging, applying both amplitude and phase solutions on the fly during imaging. 

Finally, we generated an extracted dataset covering a small region around our target by subtracting our best DD-calibrated sky model of all sources outside the region of interest, namely Abell~2142. The primary motivation for this is to allow for efficient post-processing of our data, as MeerKAT has a large primary beam full-width at half-maximum (FWHM), around 67~arcmin at 1.28~GHz \citep[][]{Mauch2020}. Due to the relatively low redshift of Abell~2142 and the several resolved radio galaxies associated with the cluster, the extraction region of interest constituted a radius of $0.35\degree$ centred on the cluster. 

Finally, to improve the quality of the calibration in the region of our target, we performed four rounds of DI amplitude and phase self-calibration and imaging on the extracted dataset. All postprocessing imaging was performed with \texttt{WSclean} \texttt{v3.0.1} \citep{Offringa2014,Offringa2017}\footnote{WSclean is available at \url{https://gitlab.com/aroffringa/wsclean}} using the in-built \texttt{wgridder} algorithm \citep{Arras2021_wgridder,Ye2021_wgridder} and self-calibration performed on the extracted dataset using the general-purpose Default Preprocessing Pipeline \citep[\texttt{DP3};][]{vanDiepen2018_DP3} software with the \texttt{GainCal} mode.

\subsection{Radio: LOFAR HBA}
LOFAR observations of Abell~2142 were sourced from the LOFAR Two-metre Sky Survey \citep[LoTSS;][]{Shimwell2017_LoTSS_PaperI,Shimwell2019_LOTSS_PaperII,Shimwell2022_LOTSS_PaperIII}. We use the same LOFAR observations as recently presented by \cite{Bruno2023_A2142}, although we have undertaken an independent post-processing and analysis. A single LoTSS pointing (P239+27) covers Abell~2142, with the cluster reasonably close to the field phase centre at an angular distance of $0.27\degree$. Due to a combination of factors, both the relatively low elevation of Abell~2142 in the ongoing LoTSS coverage and the fact that the cluster was co-observed with the Corona Borealis supercluster, the pointing was covered by four separate observing runs, performed on 2018~Sept.~15, 2020~Oct.~25, 2020~Oct.~31, and 2020~Nov.~13. The total on-source time is 32\,hours. In accordance with standard LoTSS observing strategy, data were taken using the full International LOFAR Telescope \citep[ILT;][]{vanHaarlem2013} in \texttt{HBA\_DUAL\_INNER} mode, covering the frequency range 120$-$168~MHz. In this work we make use of the data from the Dutch LOFAR array only: the Core and Remote stations, encompassing baselines out to around $80$~km.

All observations were processed using the standard LoTSS pipeline\footnote{\url{github.com/mhardcastle/ddf-pipeline/}}, which is described in detail by \cite{Shimwell2019_LOTSS_PaperII,Shimwell2022_LOTSS_PaperIII} and \cite{Tasse2021}. Following full wide-field processing, the data were subjected to the extraction and self-calibration process described by \cite{vanWeeren2021} to both enable efficient post-processing and improve the quality of the data products in the region around the target. See \cite{Bruno2023_A2142} for details of the data processing and the extraction/self-calibration procedure.

\subsection{Flux Scaling}
Our MeerKAT observations of Abell~2142 had the flux density scale set by PKS~J1939$-$6342 and are thus tied to the \cite{Reynolds1994} scale, which is itself tied to the \cite{Baars1977} scale. LOFAR observations are tied to the \cite{ScaifeHeald2012} flux scale. As previously we used a polynomial fit to the conversion factors in Table~7 of \cite{Baars1977}, performed in log-linear space, to derive the scaling factor required to bring our MeerKAT observations into consistency with our LOFAR observations. The conversion factor at the reference frequency of our MeerKAT observations ($\nu_{\rm{ref}} = 1283$~MHz) yielded by this polynomial fit was 0.968.

To verify the flux scale of our LOFAR dataset, we employed the routine discussed in detail by \cite{Hardcastle2016} and \cite{Shimwell2019_LOTSS_PaperII}. To summarise the process, we generated an image at a resolution of 6\,arcsec using \texttt{WSclean} before extracting a source catalogue using the Python Blob Detection and Source Finder software \citep[\texttt{PyBDSF};][]{MohanRafferty2015}. The catalogue is compared with the full-field LoTSS image catalogue derived from P239+27, filtered to select only compact sources with a signal-to-noise ratio ${\rm SNR} \geq 7$ common to both catalogues, and then a linear regression best-fit is performed in the flux:flux plane to derive the bootstrap factor. For this extracted LOFAR dataset, the bootstrap factor yielded by this routine is $0.998$. This correction factor is marginally higher than that reported by \cite{Bruno2023_A2142}, who obtained a value of $0.8793$. However, the representative 10\% uncertainty in the LOFAR flux density scale \citep[see][]{Shimwell2022_LOTSS_PaperIII} --- which we adopt for this work --- can account for such differences without significantly affecting measurements and results; we also adopt a typical 5 per cent uncertainty in the flux scale of our MeerKAT images.

\subsubsection{Source Subtraction and Final Imaging}
To isolate the diffuse emission from the halo for further analysis, it was necessary to subtract the many compact or extended radio galaxies in the field. In our previous papers from this census, we subtracted specific sources which were in the vicinity of the cluster. However, in the case of Abell~2142 we took a slightly different approach. Given the recent detection of significant diffuse emission filling much of the cluster volume, reported by \cite{Bruno2023_A2142}, we wished to probe a region encompassing at least a 1\,Mpc radius from the cluster centre. As such, we subtracted the emission associated with \textit{all} radio galaxies in the field, including T1, T2, and 7C~1557+2712, although the complex morphologies meant we would likely end up with non-negligible residuals.

We adopted a two-stage process for subtracting these sources. Firstly, to isolate emission associated with compact sources or small-scale extended sources (up to around 1\,arcmin in extent) we generated a clean component model by re-imaging our data with \texttt{WSclean}, employing an inner \textit{uv}-cut of $1500\lambda$ ($\sim 2.3$\,arcmin), designed to filter the halo while retaining some sensitivity to extended emission.

This clean-component model was then filtered to exclude T1, T2, and 7C~1557+2712, as well as other moderately-extended radio sources, before predicting and subtracting from our data. We then re-imaged with our standard $80\lambda$ inner \textit{uv}-cut, and filtered the clean-component model to select only those components associated with T1, T2, 7C~1557+2712, and other moderately-resolved radio galaxies not subtracted previously; this model was then predicted and subtracted from our data. At this point, we were confident that any residuals not associated with cluster-scale diffuse emission would be either negligible and/or well spatially-separated from the main halo, and could be manually excluded from our later analysis. We then imaged our subtracted data with \texttt{WSclean} using a common inner \textit{uv}-cut of $80\lambda$, a combination of manual- and auto-masking, \texttt{robust} $-0.5$ weighting and two different \textit{uv}-tapers to make our science images. These \textit{uv}-tapers were tuned and combined with image-plane smoothing to achieve a final resolution of 25\,arcsec and 60\,arcsec.

\subsection{X-ray: XMM-Newton and Chandra}
In this work we make use of public \textit{XMM-Newton} and \textit{Chandra} observations previously considered by several other studies \citep[e.g.][and references therein]{Bruno2023_A2142}. The \textit{XMM-Newton} observations include a targeted pointing on the cluster centre (ObsID~0674560201; 2011~July) as well as pointings on the cluster outskirts (ObsIDs 0694440101, 0694440501, 0694440601, 0694440201; 2012~July) from the \textit{XMM} Cluster Outskirts Project \citep[X-COP;][]{Eckert2017_XCOP}. The total on-source time across the five pointings is 195\,ks. We used the publicly released final data products available from the X-COP webpage\footnote{\url{https://dominiqueeckert.wixsite.com/xcop}}, specifically the adaptively-smoothed, vignetting-corrected and background-subtracted X-ray surface brightness map in the [0.7$-$1.2]~keV range. We refer the reader to \cite{Rossetti2013_Abell2142}, \cite{Tchernin2016_XCOP}, and \cite{Ghirardini2018_XCOP} for technical details on the processing of these observations. We also made use of the temperature map obtained from the central pointing in \cite{Rossetti2013_Abell2142}, and we refer the reader to their paper for details on the production of this map.

For the \textit{Chandra} observations used in this work, we use the data products generated by \cite{Bruno2023_A2142}. These use ObsIDs 5005, 15186, 16564, 16565, for a total clean on-source time of 186.9\,ks, which overwhelmingly covers the cluster centre.

\section{Results}\label{sec:results}

\subsection{Radio continuum properties}
\begin{figure*}
\begin{center}
\includegraphics[width=0.99\textwidth]{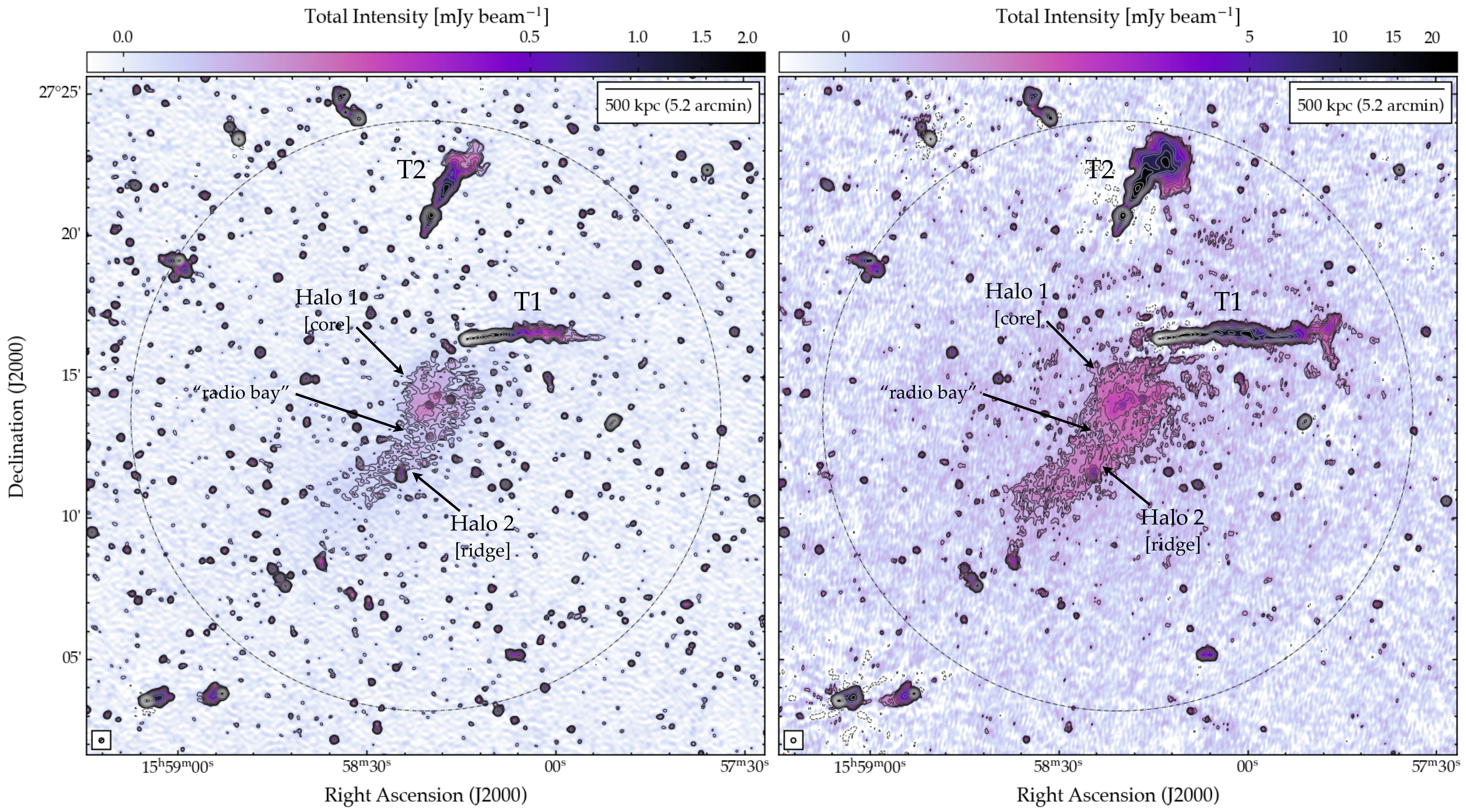}
\caption{Radio continuum images of Abell~2142 with MeerKAT (\textit{left}, 1283\,MHz) and LOFAR (\textit{right}, 143\,MHz) at 10\,arcesc resolution. Colourscale ranges from $-2\sigma$ to $300\sigma$ on an \texttt{arcsinh} stretch to emphasise diffuse emission. Contours start at $4\sigma$ and scale by a factor of $\sqrt{2}$, where $\sigma = 7.5 \, (80.3) \, \upmu{\rm Jy} \, {\rm beam}^{-1}$ at 1283\,MHz (143\,MHz). Dashed contours denote the $-3\sigma$ level. The dot-dashed circle traces a radius of 1\,Mpc around the centre of Abell~2142. Labels identify features and sources that will be discussed in this paper.}
\label{fig:A2142_radio_fullres}
\end{center}
\end{figure*}

We present our full-resolution (10\,arcsec) radio maps of Abell~2142 in Figure~\ref{fig:A2142_radio_fullres}, showing our MeerKAT map at 1283\,MHz in the left panel and our LOFAR map at 143\,MHz in the right panel. We show contours starting at $4\sigma$ and scaling by a factor of $\sqrt{2}$, where $\sigma$ is the off-source rms noise in the corresponding map: $\sigma = 7.5 ~ \upmu$Jy beam$^{-1}$ at 1283\,MHz and $80 ~ \upmu$Jy beam$^{-1}$ at 143\,MHz. Table~\ref{tab:img_summary} summarises the properties of the images presented in this paper.

\begin{table}
\footnotesize
\centering
\caption{Summary of image properties for images of Abell~2142. Images marked with a $^{\dag}$ were produced after source-subtraction, with a combination of \emph{uv}-tapering and image-plane convolution used to achieve the desired resolution. The quoted RMS noise values were derived as the average of several off-source regions in the vicinity of the phase centre. \label{tab:img_summary}}
\begin{tabular}{lccccc}
\hline
Telescope    & Freq. & \texttt{Robust} & RMS noise & Resolution & PA \\
  	     & $[$MHz$]$ & &  $[\upmu$Jy beam$^{-1}]$  & $[$arcsec$]$ & $[\degree]$ \\
\hline\hline
\multirow{3}{*}{MeerKAT} & \multirow{3}{*}{1283} & $-0.5$ & 7.5  & 10  & 0 \\
                                                & & $-0.5^{\dag}$ & 15  & 25 & 0 \\
                                                & & $-0.5^{\dag}$ & 55  & 60 & 0 \\
\hline
\multirow{3}{*}{LOFAR} & \multirow{3}{*}{143} & $-0.5$  & 80.3  & 10 & 0 \\
                                               & & $-0.5^{\dag}$  & 172  & 25 & 0 \\
                                               & & $-0.5^{\dag}$  & 285  & 60 & 0 \\
\hline
\end{tabular}
\end{table}

Figure~\ref{fig:A2142_radio_fullres} demonstrates the excellent performance and sensitivity of MeerKAT even at above Declination $+27 \degree$, as we recover many compact radio sources in the vicinity of Abell~2142, as well as numerous resolved radio galaxies in the field, plus the diffuse emission from the multi-component halo. After our direction-independent and direction-dependent calibration routine, residual contamination from calibration errors and deconvolution artefacts is minimal. The LOFAR image of Abell~2142 also achieves good sensitivity, resolution and dynamic range, although residual artefacts are more prominent. Fewer compact sources are visible compared to our MeerKAT map, although the extended emission from the radio galaxies associated with the field, as well as the diffuse emission from the halo, is universally more extended, reflecting the steep spectral index of the emission.

The two tailed radio galaxies, T1 and T2, are well-recovered by both MeerKAT and LOFAR, although differences are visible in the extent of the emission recovered. This likely implies an ultra-steep spectrum in these regions, reflecting aged electrons far from the host AGN. These sources are discussed in more detail later in Section~\ref{sec:radio_environment}. 

The main two halo components -- `Halo 1' and `Halo 2', following the nomenclature of \cite{Venturi2017_A2142}, hereafter `H1' and `H2' -- are both clearly detected by both MeerKAT and LOFAR even at 10\,arcsec resolution in Figure~\ref{fig:A2142_radio_fullres}, although the halo is visibly less extended at 1283\,MHz compared to 143\,MHz. The emission is elongated along a north-west/south-east axis following the thermal emission of the ICM; measuring above the $4\sigma$ contour we measure a largest angular scale of 400\,arcsec at 1283\,MHz and 470\,arcsec at 143\,MHz. At the redshift of Abell~2142, these angular scales correspond to a physical extent of 626\,kpc and 753\,kpc respectively.

From Figure~\ref{fig:A2142_radio_fullres}, the two primary components of the halo are both well-resolved. The `core' (H1) has considerably higher surface brightness than the `ridge' (H2), and shows a reasonably well-defined `peak' region, whereas H2 is more uniform in surface brightness and shows no clear peak. Additionally, the decrease in surface brightness that roughly divides H1 and H2, which \cite{Bruno2023_A2142} refer to as the `radio bay', is seen clearly in our MeerKAT and LOFAR maps.

\subsection{Source-subtracted images}
\begin{figure*}
\begin{center}
\includegraphics[width=0.9\textwidth]{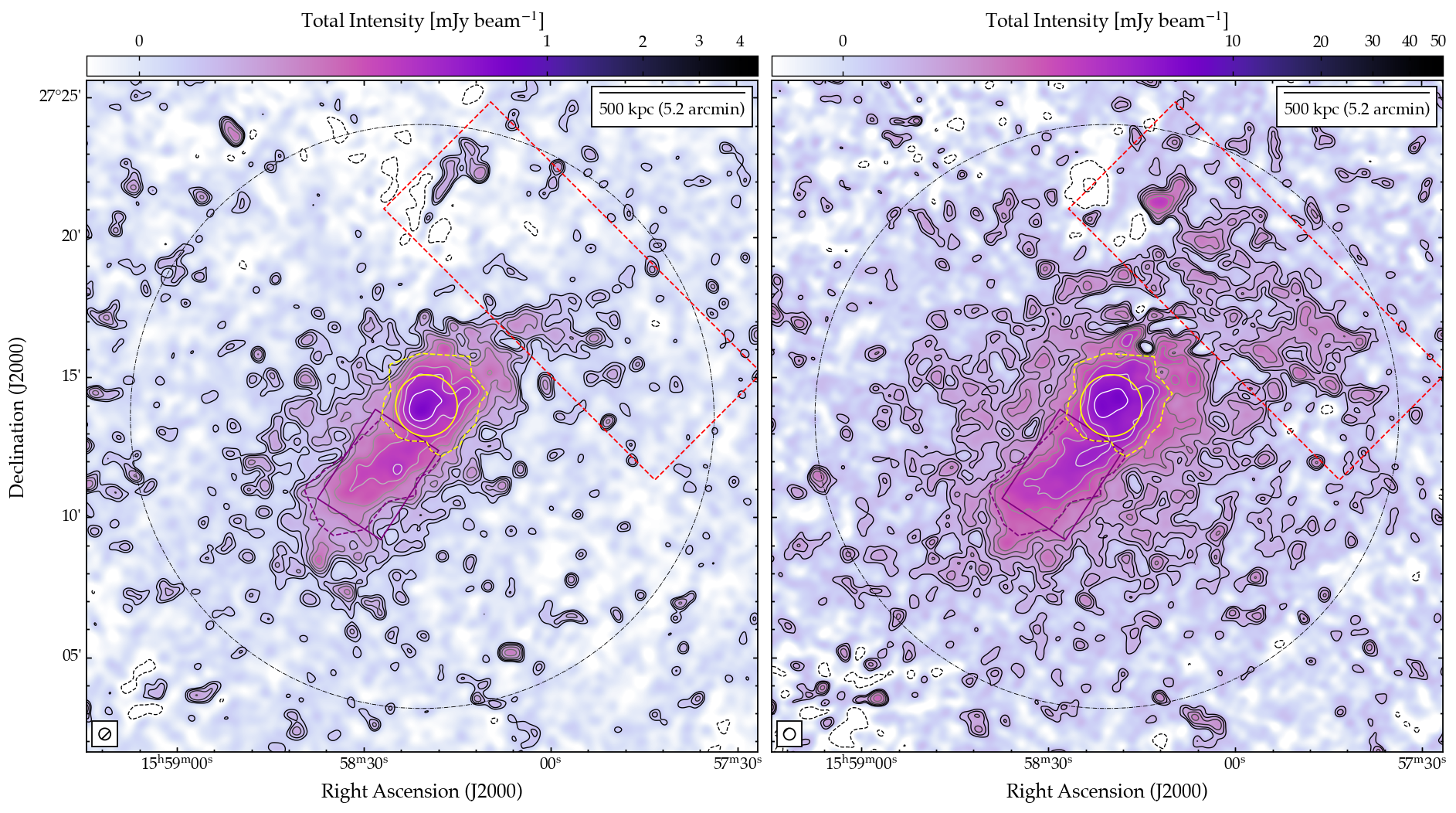}
\includegraphics[width=0.9\textwidth]{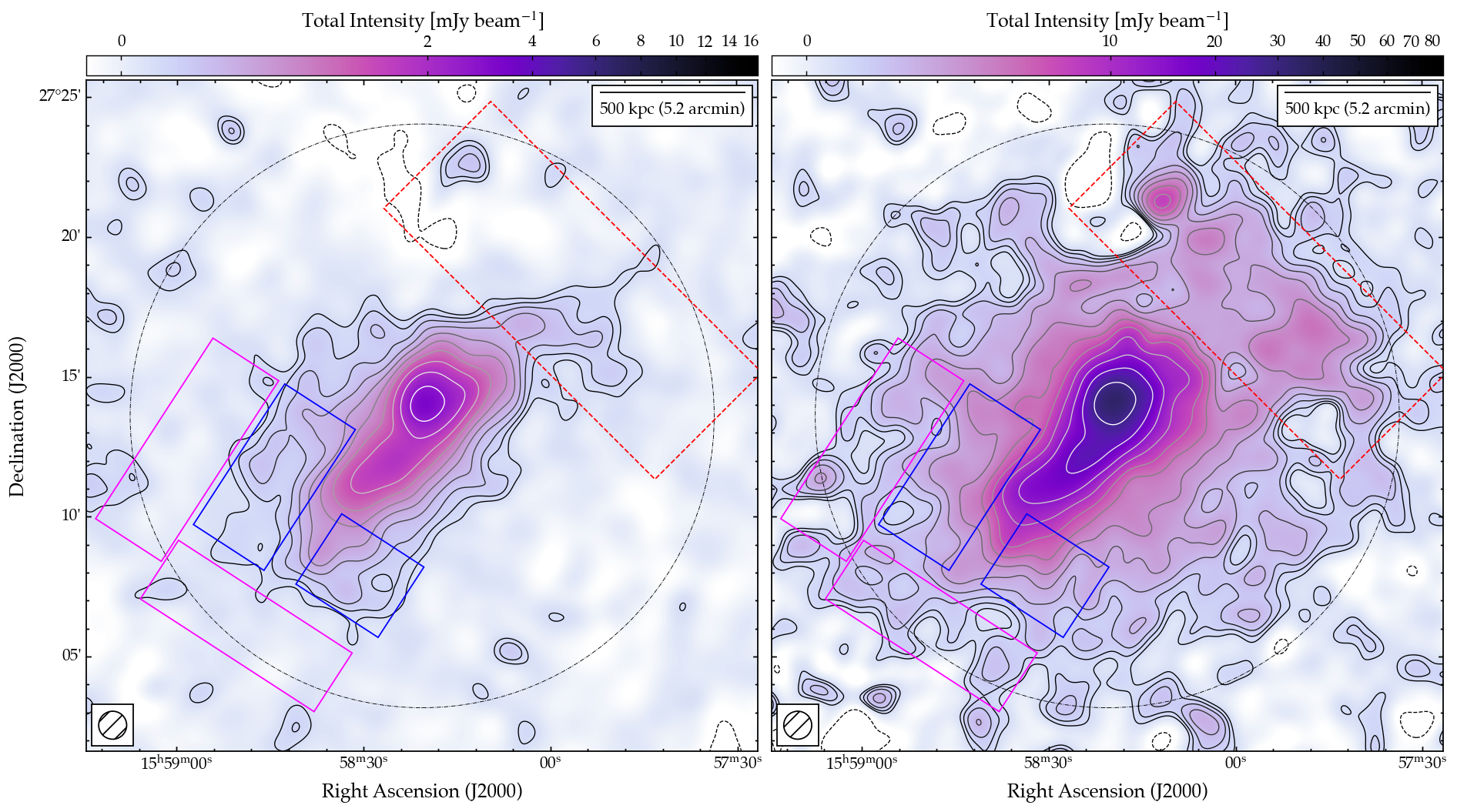}
\caption{Source-subtracted radio continuum images of Abell~2142 with MeerKAT (\textit{left}, 1283\,MHz) and LOFAR (\textit{right}, 143\,MHz) at 25\,arcesc resolution (\textit{top}) and 60\,arcsec resolution (\textit{bottom}). Colourscale ranges from $-2\sigma$ to $300\sigma$ on an \texttt{arcsinh} stretch to emphasise diffuse emission. The first contour is shown at the $2\sigma$ level, then contours scale by a factor of $\sqrt{2}$ from $3\sigma$, where $\sigma = 15 \, (172) \, \upmu{\rm Jy} \, {\rm beam}^{-1}$ at 1283\,MHz (143\,MHz). Dashed contour denotes the $-3\sigma$ level. The dot-dashed circle traces a radius of 1\,Mpc around the centre of Abell~2142. Note that all discrete sources have been subtracted from the visibilities before imaging, as described in the text. Solid-line boxes and dashed polygons denote the areas used to measure the flux of the H1 (yellow), H2 (purple), and H3 (blue and magenta) as described later in the text. Dashed red box indicates regions where residuals from T1 and T2 are significant, and as such are excluded from later analysis.}
\label{fig:A2142_radio_sub}
\end{center}
\end{figure*}

Our source-subtracted radio maps of Abell~2142 are shown in Figure~\ref{fig:A2142_radio_sub}, at 25\,arcsec resolution (\textit{top row}) and 60\,arcsesc resolution (\textit{bottom row}).

From our 25\,arcsec maps, both H1 and H2 are recovered at far greater signal-to-noise ratio. The diffuse emission is more filled out, with H1 extending to overlap with T1. The `radio bay' is filled at this resolution, although it remains somewhat visible in Figure~\ref{fig:A2142_radio_sub}. Further, patchy diffuse emission starts to appear between T1 and T2, although it is unclear whether this represents diffuse emission from the halo or fossil plasma associated with T2. At this resolution, we also start to recover diffuse emission extending toward the north-east and south-west of both H1 and H2: the emerging third component reported by \cite{Bruno2023_A2142}, `Halo 3' (hereafter `H3'). 

Finally, at 60\,arcsec resolution, components H1 and H2 become blended and the surface brightness distribution becomes far smoother; thus the radio bay is no longer distinct. At 1283\,MHz we see a significant component of H3 emerging to the north-east of H1 and H2, extending roughly 320\,arcsec from the primary north-west/south-east axis of the halo, corresponding to a distance of 509\,kpc. Conversely, to the south-west, the H3 component extends out to around 234\,arcsec from the primary axis, corresponding to a physical scale of 373\,kpc. This asymmetry is clearly visible in the outer surface brightness contours, which are spaced further apart in the north-easterly direction but are tightly packed in the south-westerly direction.

In contrast, from our LOFAR map at 143\,MHz, almost the entire 1\,Mpc radius is filled with diffuse synchrotron emission, including the region between T1 and T2 which is filled with diffuse radio emission. There remains a slight asymmetry, as the emission is more extended along the primary north-west/south-east axis compared to a north-east/south-west direction. Regardless, the maximum extent of the diffuse emission recovered by LOFAR at 143\,MHz is of the order of 2\,Mpc, consistent with the results reported by \cite{Bruno2023_A2142}. However, it is important to note that the contrast in the extent of the emission recovered by MeerKAT and LOFAR as seen in Figure~\ref{fig:A2142_radio_sub} does not necessarily imply an ultra-steep spectrum: taking a $2\sigma$ limit from our MeerKAT map and our typical measured LOFAR surface brightness of $\langle S_{\rm 143\,MHz} \rangle = 1.4$\,mJy~beam$^{-1}$, we derive an upper-limit spectral index of $\alpha \leq -1.17$. Deeper MeerKAT observations would be required to fully probe the extent of the halo in Abell~2142.


\subsection{The Radio Environment of Abell 2142}\label{sec:radio_environment}
The radio environment of the Abell~2142 field is very rich. In Figure~\ref{fig:fullfield_images} we show a full-field image with zooms on several sources of potential scientific interest that are unassociated with the cluster itself, such as Odd Radio Circle 4 \citep[ORC-4;][]{Norris2021_ORCs_Galaxies,Norris2021_ORCs_PASA} and the giant radio galaxy associated with LEDA~1783783. While we do not discuss these objects in this paper, they are included as they may be of broader interest to the scientific community, as well as demonstrating the excellent performance of MeerKAT even in the wide field beyond the PB FWHM.

\subsubsection{The brightest cluster galaxies}
\begin{figure}
\begin{center}
 \includegraphics[width=0.95\linewidth]{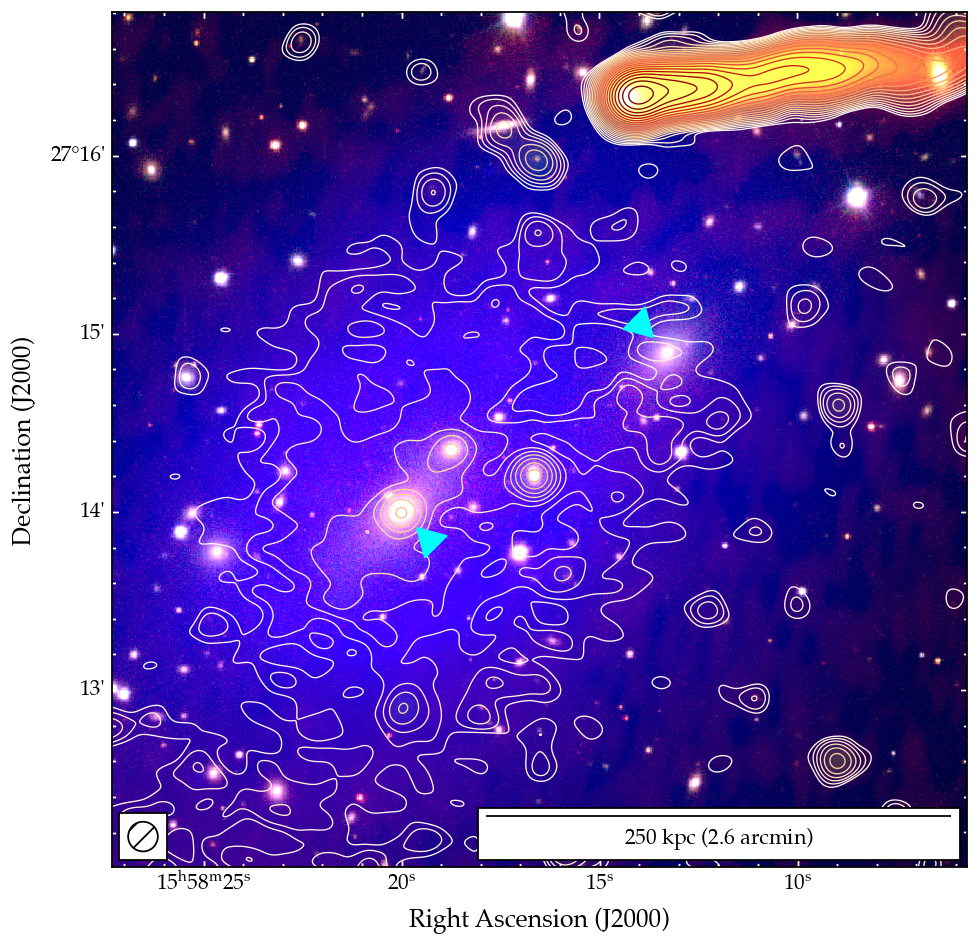}
\caption{Colour-composite image of the brightest cluster galaxies of Abell~2142, as per Figure~\ref{fig:A2142_Composite}. Contours denote the radio surface brightness measured by MeerKAT at 1283\,MHz, as per Figure~\ref{fig:A2142_radio_fullres}. Cyan markers point to the brightest cluster galaxies, BCG1 to the south-east and BCG2 to the north-west. BCG2 hosts no significant radio counterpart, whereas BCG1 hosts a compact radio counterpart.}
\label{fig:RGB_BCGs}
\end{center}
\end{figure}

Abell~2142 is known to host two brightest cluster galaxies (BCG) within the main cluster core; we present a zoom on this region of our colour-composite map in Figure~\ref{fig:RGB_BCGs}. The brightest of which, `BCG1' is a reasonably faint ($r$-band magnitude $r_{\rm mag} = 14.52$) galaxy also catalogued in the SDSS as SDSS12~J155820.00+271400.2, with a spectroscopic redshift $z_{\rm spec} = 0.0908$ \citep{Alam2015_SDSS-DR12}. BCG1 is hosted by the primary, richest sub-cluster within the larger Abell~2142 structure and lies at the centre of the gravitational potential well close to the innermost cold front \citep[e.g.][]{Okabe2008_weaklensing,Wang_Markevitch_2018}.

BCG2 lies further to the north-west of the cluster centre, toward the outer edge of the diffuse emission from H1 (as traced at 10\,arcsec resolution). This galaxy is likewise faint, catalogued as SDSS12~J155813.29+271453.2 with $r_{\rm mag} = 14.93$ and lies at $z_{\rm spec} = 0.0965$ \citep{Alam2015_SDSS-DR12}. While the relative velocity of BCG2 is reasonably high, around 1819\,km\,s$^{-1}$ above the cluster systematic velocity of 27,117\,km\,s$^{-1}$, the cluster velocity dispersion of $\sigma_{v} \sim 1200$\,km\,s$^{-1}$ \citep[e.g.][]{Oegerle1995,Munari2014} means the relative velocity of BCG2 is only around $1.4 \times \sigma_{v}$.

Neither MeerKAT nor LOFAR detect any significant radio emission associated with BCG2 -- the emission coincident with BCG2 in Figure~\ref{fig:RGB_BCGs} is diffuse emission from the main halo component H1. However, both MeerKAT and LOFAR detect significant emission associated with BCG1, which is unresolved at 10\,arcsec resolution. This compact counterpart remains unresolved even at the native resolution of our maps, around 3\,arcsec to 6\,arcsec. Exploration of the existing LoTSS data using the full ILT would be required to resolve this radio source. We measure a flux density of $S_{\rm 1283 \, MHz} = 0.258 \pm 0.014 \, \upmu$Jy and $S_{\rm 143 \, MHz} = 1.34 \pm 0.16 \,$mJy respectively with MeerKAT at 1283\,MHz and LOFAR at 143\,MHz, for a spectral index $\alpha = -0.76 \pm 0.06$.

The $k$-corrected radio power $P_{\nu}$ at frequency $\nu$ is expressed as:
\begin{equation}\label{eq:radio_lum}
    P_{\nu} = 4 \pi \,  D_{\rm L}^2 \, S_{\nu} \, (1 + z)^{-(1 + \alpha)}
\end{equation}
where $D_{\rm L}$ is the luminosity distance to the object and $S_{\nu}$ is the flux density at frequency $\nu$. For BCG1, the redshift of $z_{\rm spec} = 0.0908$ corresponds to a $D_{\rm L} = 399.1$~Mpc given our cosmology. Hence, from Equation~\ref{eq:radio_lum} yields a 1.4~GHz radio luminosity of $P_{\rm 1.4~GHz} = (5.11 \pm 0.24) \times 10^{21}$~W~Hz$^{-1}$ for BCG1.

\subsubsection{The embedded radio galaxies T1 and T2}
\begin{figure}
\begin{center}
\includegraphics[width=0.95\linewidth]{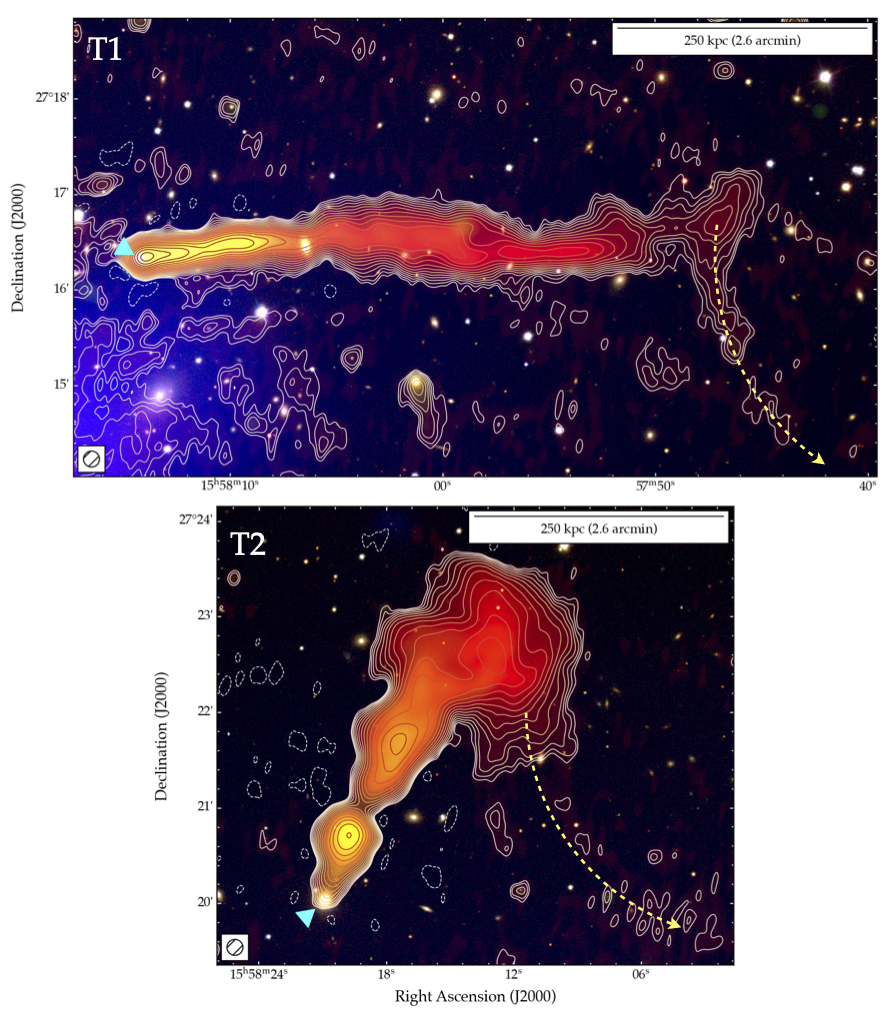}
\caption{Colour-composite images of the tailed radio galaxies T1 (\textit{top}) and T2 (\textit{bottom}) embedded in the ICM of Abell~2142, as per Figure~\ref{fig:A2142_Composite}. Contours denote the radio surface brightness measured by LOFAR at 143\,MHz as per Figure~\ref{fig:A2142_radio_fullres}. Cyan markers point to the host galaxies. Yellow curves indicate the direction of further faint ultra-steep spectrum emission that is only partially recovered at this resolution, which likely traces ancient fossil plasma from these radio galaxies.}
\label{fig:RGB_T1-T2}
\end{center}
\end{figure}

The two most spectacular resolved radio galaxies in Abell~2142 are the tailed radio galaxies T1 and T2 \citep[following the nomenclature of][]{Venturi2017_A2142}. Both show clear evidence of interaction with the ICM due to their complex morphology. We show colour-composite images of these radio galaxies in Figure~\ref{fig:RGB_T1-T2}. While a detailed study of these radio galaxies is left for future work (Bruno et al., in preparation), it is pertinent to briefly comment on these sources.

T1 is the radio galaxy B2~1556+27 \citep{Colla1972_B2cat,Owen1993_AbellClusters} hosted by SDSS16~J155814.31+271619.5 \citep[$z_{\rm spec} = 0.09540$;][]{Ahumada2020_SDSS-DR16}. In the radio regime, this source presents as a head-tail radio galaxy around 310\,arcsec (527\,kpc) in length at 1283\,MHz; at 143\,MHz the tail is somewhat longer, stretching out to 393\,arcsec (668\,kpc) as traced by the $4\sigma$ contour in Figure~\ref{fig:RGB_T1-T2}. 

The tail is not uniform in surface brightness but exhibits small scale undulations along its length, as commented on by \cite{Venturi2017_A2142}. Such features are exhibited by many tailed radio galaxies in clusters \citep[e.g.][]{Wilber2018_Abell1132,Rudnick2021_MysTail,Riseley2022_A3266} and provide clear indications that the plasma is being subjected to cluster weather. At 1283\,MHz MeerKAT recovers the brightest two-thirds of the emission, which exhibits a generally steep spectrum with a gradient from $\alpha = -0.61 \mathbf{\pm 0.05}$ near the core to a steepest value of $\alpha = -2.49 \mathbf{\pm 0.09}$ in the central part of the tail. This change in spectrum is visible in the colour change from yellow, orange through red in Figure~\ref{fig:RGB_T1-T2}.

The latter third of the tail is not detected by MeerKAT at the $3\sigma$ level, suggesting a spectrum $\alpha \lesssim -2.5$; this region appears slightly offset from the primary axis of emission, and at the very end the diffuse radio tail bends sharply from a roughly east-west direction to a north-south axis. Such features are becoming increasingly common in clusters thanks to the new generation of highly-sensitive low-frequency interferometers, and can be used to trace the cluster weather \citep[e.g.][]{Wilber2020_LOFARclusters,Giacintucci2022_A3562,Riseley2022_A3266,Lee2023_A514} Further patchy diffuse emission is visible in this direction, as indicated by the yellow arrow in Figure~\ref{fig:RGB_T1-T2}, but is poorly recovered even by LOFAR at 143\,MHz in our full-resolution maps.

T2 lies to the north of the cluster centre, beyond the densest region of the ICM. Despite this however, its morphology is complex, with an initial `bulb'-like structure extending north-west away from the host galaxy, SDSS16~J155820.88+272001.4 \citep[$z_{\rm spec} = 0.08953$;][]{Ahumada2020_SDSS-DR16}. The bulb narrows before the tail widens and then fans out into the larger diffuse tail which extends to the north-west.

While there is a slightly flatter-spectrum region ($\alpha = -0.71 \mathbf{\pm 0.06}$) associated with the host galaxy, the bulb exhibits a steep spectrum with a typical spectral index of $\alpha = -0.99 \mathbf{\pm 0.05}$. Where the tail widens again, the spectrum is steeper with a typical value of $\alpha = -1.41 \mathbf{\pm 0.06}$, and finally where the tail fans out into the more extended diffuse tail, the spectrum is again far steeper with a typical value of $\alpha = -2.08 \mathbf{\pm 0.06}$. MeerKAT does not recover the full extent of the diffuse fan region, although in some places the spectrum reaches a value of ${\mathbf{\alpha} =} -2.5 \mathbf{\pm 0.1}$. This spectral steepening is visible in the colour change from yellow, orange through red in Figure~\ref{fig:RGB_T1-T2}. 

To the south-west of the diffuse fan, there is patchy diffuse emission that is poorly-recovered by LOFAR at 143\,MHz in our 10\,arcsec resolution maps (the yellow arc in Figure~\ref{fig:RGB_T1-T2}), although this emission is recovered at greater signal-to-noise when tapering to lower resolution. At 60\,arcsec resolution, this emission extends into the larger-scale halo and fills much of the region between the fan and T1. It is difficult to conclusively say therefore that the diffuse emission recovered between T1 and T2 at low resolution is purely diffuse halo emission, purely ancient diffuse plasma from T2, or some combination thereof.

\begin{figure}
\begin{center}
 \includegraphics[width=0.95\linewidth]{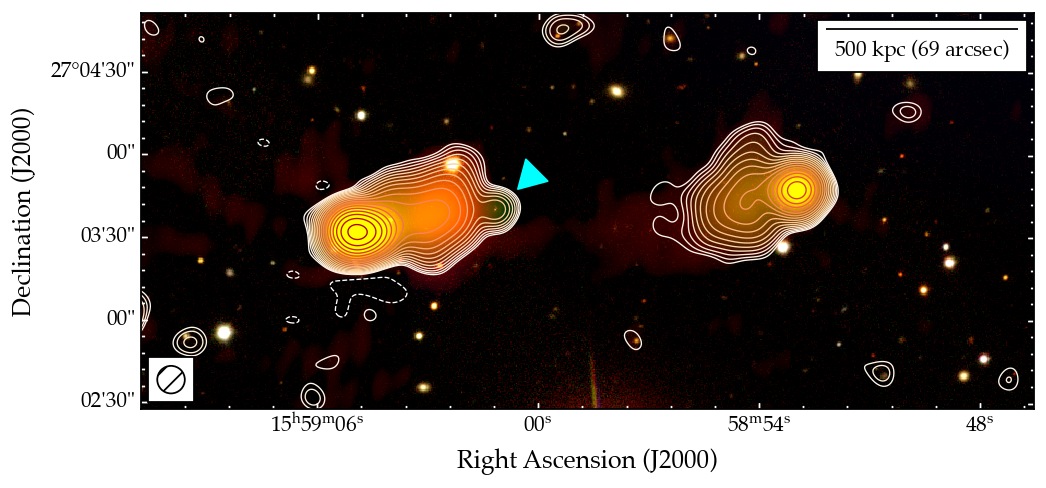}
\caption{Colour-composite image of the FR-II type radio galaxy 7C~1557+2712. Colourscale is as Figure~\ref{fig:A2142_Composite}, with yellow-through-red colours tracing an increasingly steep spectrum. The contours denote the radio surface brightness measured by MeerKAT at 1283\,MHz as per Figure~\ref{fig:A2142_radio_fullres}. Cyan marker points to the candidate host galaxy ($z_{\rm phot} = 0.760 \pm 0.047$) which has a compact radio counterpart detected by MeerKAT at 1283\,MHz but not by LOFAR at 143\,MHz, suggesting an inverted spectrum.}
\label{fig:RGB_FR2}
\end{center}
\end{figure}

\subsubsection{7C~1557+2712: a background giant radio galaxy}
This radio galaxy lies to the south-west of the cluster, at a projected distance beyond 1\,Mpc. In the radio, it exhibits a clear Fanaroff-Riley Class II \citep[FR-II;][]{FanaroffRiley1974} morphology with a pair of diametrically-opposed bright hotspots and fainter diffuse lobes. We present a zoom on our colour-composite map in Figure~\ref{fig:RGB_FR2}.

In general the radio emission from this source is steep-spectrum with the west and east hotspots exhibiting a spectral index of $\alpha = -0.79$ and $-0.81$, respectively. The more extended lobe emission is steeper spectrum, with a gradient toward the centre, showing typical values from $-0.93$ to $-1.39$ in the west and $-0.88$ to $-1.37$ in the east. The uncertainty in spectral index is 0.05 for each of these values. This steep spectrum is visible in the yellow-through-orange colours in Figure~\ref{fig:RGB_FR2}.

We also note the presence of a compact radio component just on the inner edge of the eastern lobe, which is detected by MeerKAT at 1283\,MHz (flux density $S_{\rm 1283 \, MHz} = 0.34 \pm 0.02$\,mJy) but not by LOFAR at 143\,MHz. Taking a $2\sigma$ limit from LOFAR, this indicates an inverted spectrum as we find $\alpha > 0.34$. This is also demonstrated in the apparent green colour in Figure~\ref{fig:RGB_FR2}.

Finally, we note that the compact radio counterpart has a faint optical counterpart, SDSS12~J155901.14$+$270340.4, with an $r$-band magnitude $23.1 \pm 0.2$ in SDSS DR12 \citep{Alam2015_SDSS-DR12}. No spectroscopic redshift is available, but the photometric redshift of $z_{\rm phot} = 0.760 \pm 0.047$ indicates that this is a background radio galaxy unassociated with Abell~2142. This galaxy is also catalogued in DR8 from the Dark Energy Spectroscopic Instrument (DESI) Legacy Imaging Surveys \citep{Duncan2022_DESI-DR8} with a consistent (albeit independently-derived) photometric redshift of $z_{\rm phot} = 0.856 \pm 0.120$. Taking the more conservative $z_{\rm phot}$ from SDSS DR12, the angular separation between hotspots (169\,arcsec) corresponds to a projected linear size of 1209\,kpc, placing 7C~1557+2712 in the category of a giant radio galaxy.

\section{Analysis: The Multi-Component Halo in Abell 2142}\label{sec:analysis}

\subsection{Spectral properties}
One of the key goals of our census is to perform a systematic precision study of the spectral properties of mini-haloes, as this provides a key window into the underlying particle acceleration mechanism. In the case of Abell~2142, this is more complex than for our previous mini-halo clusters MS~1455.0$+$2232 \citep{Riseley2022_MS1455} and Abell~1413 \citep{Riseley2023_A1413} due to the complex multi-component nature of the halo. As such, a multi-faceted approach is warranted.

\subsubsection{Integrated spectrum}\label{sec:integrated_spectrum}
We investigate the integrated spectral properties of the multi-component (mini-)halo using two different sets of regions. In the initial study by \cite{Venturi2017_A2142}, those authors determined the integrated fluxes of H1 and H2 using two boxes which roughly encompassed the emission (see their Fig.~6). The recent study by \cite{Bruno2023_A2142} used different boundaries for H1, considering this component as a sphere centred on BCG1 with a radius equivalent to the distance to the inner cold fronts, $R_{\rm H1} ~ 110$\,kpc. When measuring the flux of H2, both \citeauthor{Bruno2023_A2142} and \citeauthor{Venturi2017_A2142} used a similar region size, encompassing the $\sim3\sigma$ level of the radio `ridge' when viewed at 25\,arcsec resolution. \citeauthor{Bruno2023_A2142} then use two sets of regions to study the integrated spectrum of H3, an `inner' and an `outer' region set (see their Fig.~6).

\begin{figure*}
\begin{center}
 \includegraphics[width=0.99\textwidth]{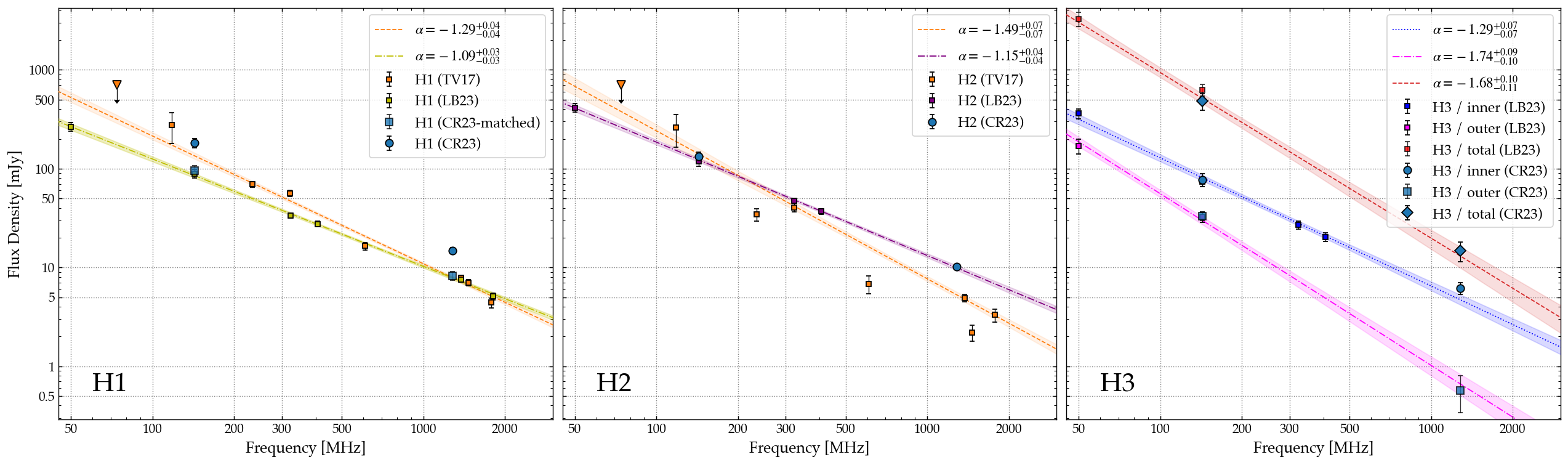}
\caption{Integrated spectral index measurements and fits for the three components: H1, H2 and H3, as presented in Table~\ref{tab:integrated_flux} and discussed in the text. Dashed/dotted/dot-dashed lines show the best-fit spectral index, with the shaded region denoting the $1\sigma$ uncertainty, colourised according to the extraction region in Figures~\ref{fig:A2142_radio_sub}. References `TV17', `LB23' and `CR23' refer to \cite{Venturi2017_A2142}, \cite{Bruno2023_A2142} and this work, respectively. Orange fits were performed solely on the measurements from \cite{Venturi2017_A2142}, whereas the yellow and purple fits to H1 and H2, as well as all fits to H3, were performed using the combination of our new measurements plus those from \cite{Bruno2023_A2142}. Note that the different sets of measurements for H1 were derived using different areas, as described in the text; the `CR23-matched' measurements denote those derived using the same region as \cite{Bruno2023_A2142}, whereas the `CR23' measurement was derived over the more full extent of H1. When fitting to the data from \cite{Venturi2017_A2142}, the 74\,MHz measurement for H1 and H2 is an upper limit and was thus excluded from the initial fit. Likewise, the measurements at 234\,MHz, 608\,MHz, and 1377\,MHz for H2 were derived from data with sparser \textit{uv}-coverage, and were excluded from the fit. See \cite{Venturi2017_A2142} for further details.}
\label{fig:A2142_SED_H1H2H3}
\end{center}
\end{figure*}

To provide a fair comparison with both previous studies, we adopt both approaches to measuring the integrated flux density of the halo components. For H1 and H2, we report integrated flux densities measured on our 25\,arcsec source-subtracted map (Figure~\ref{fig:A2142_radio_sub}, \textit{top}) and derived using regions shown in the same Figure, two polygonal shapes (displayed with dashed lines) that roughly encompass the full extent of H1 and H2. We also use the same regions as \citeauthor{Bruno2023_A2142} to measure the integrated flux density of H1 and H2 within the `core' and ridge regions (displayed in Figure~\ref{fig:A2142_radio_sub}, \textit{top} with solid lines). Finally, for H3 we report measurements made using the `inner' and `outer' regions of \citeauthor{Bruno2023_A2142}, measured from our images at 60\,arcsec resolution; these regions are overlaid in thw bottom panel of Figure~\ref{fig:A2142_radio_sub}. All measurements were made using the \texttt{RadioFluxTools} python library\footnote{Available at \url{https://gitlab.com/Sunmish/radiofluxtools}.}. These measurements are presented in Table~\ref{tab:integrated_flux}.

\begin{table}
\footnotesize
\renewcommand{\arraystretch}{1.08}
\centering
\caption{Integrated flux density measurements for H1, H2 and the inner and outer regions of H3. These were performed using the regions presented in Figure~\ref{fig:A2142_radio_sub}, as described in the text. We also include total integrated measurements for H3, derived using radial profiles (Section \ref{sec:radial_profiles}) here and in the literature. References `TV17' and `LB23' respectively refer to \cite{Venturi2017_A2142} and \cite{Bruno2023_A2142}. Measurements marked with a $^{\dag}$ were excluded from fitting due to sparser \textit{uv}-coverage, as described in the text. \label{tab:integrated_flux}}
\begin{tabular}{c | cc | c}
\hline
Component   &   Frequency    &   Flux Density     &   Reference   \\
            &  $[$MHz$]$     &   $[$mJy$]$        &               \\
\hline\hline
\multicolumn{4}{c}{Regions comparable to TV17}\\
\hline\hline
\multirow{10}{*}{H1}          &  1778          &   $4.4 \pm 0.5$    & TV17         \\  
            &  1465          &   $7.0 \pm 0.5$    &  TV17         \\  
            &  1377          &   $7.9 \pm 0.4$    &  TV17         \\  
            &  1283          &   $14.9 \pm 1.0$   &  This work    \\ 
            &   608          &   $16.5 \pm 1.4$   &  TV17         \\  
            &   322          &   $56.2 \pm 4.1$   &  TV17         \\  
            &   234          &   $69.3 \pm 4.9$   &  TV17         \\  
            &   143          &   $182.6 \pm 19.7$ &  This work    \\ 
            &   118          &   $275.0 \pm 94.1$ &  TV17         \\  
            &    74          &   $< 700$          &  TV17         \\ 
\hline
\multirow{10}{*}{H2}          &  1778          &   $3.3 \pm 0.5$    & TV17         \\ 
            &  1465          &   $2.2 \pm 0.4$    &  TV17         \\ 
            &  1377$^{\dag}$ &   $4.9 \pm 0.4$    &  TV17         \\  
            &  1283          &   $10.2 \pm 0.6$   &  This work    \\ 
            &   608$^{\dag}$ &   $6.8 \pm 1.4$    &  TV17         \\  
            &   322          &   $40.5 \pm 4.1$   &  TV17         \\ 
            &   234$^{\dag}$ &   $34.5 \pm 4.9$   &  TV17         \\ 
            &   143          &   $132.3 \pm 13.9$ &  This work    \\ 
            &   118          &   $260.0 \pm 94.1$ &  TV17         \\  
            &    74          &   $< 700$          &  TV17         \\ 
\hline\hline
\multicolumn{4}{c}{Regions matched to LB23}\\
\hline\hline
\multirow{8}{*}{H1}          &  1810          &   $5.1 \pm 0.4$    &  LB23          \\
            &  1380          &   $7.5 \pm 0.5$    &  LB23          \\  
            &  1283          &   $8.2 \pm 0.8$    &  This work    \\ 
            &   407          &   $27.6 \pm 1.8$   &  LB23          \\ 
            &   323          &   $33.6 \pm 2.1$   &  LB23          \\  
            &   143          &   $95.8 \pm 12.0$  &  This work    \\ 
            &   143          &   $89.1 \pm 8.9$   &  LB23          \\ 
            &    50          &   $266.4 \pm 27.6$ &  LB23          \\ 
\hline
\multirow{6}{*}{H2}          &  1283          &   $10.1 \pm 0.6$    &  This work    \\ 
            &   407          &   $37.0 \pm 2.4$   &  LB23          \\ 
            &   323          &   $47.5 \pm 2.9$   &  LB23          \\ 
            &   143          &   $131.4 \pm 13.9$ &  This work    \\ 
            &   143          &   $117.8 \pm 11.8$ &  LB23          \\  
            &    50          &   $414.4 \pm 43.0$ &  LB23          \\  
\hline
\multirow{6}{*}{H3 (inner)}  &  1283          &   $6.2 \pm 0.4$    &  This work    \\ 
            &   407          &   $20.3 \pm 2.0$   &  LB23          \\  
            &   323          &   $27.0 \pm 2.5$   &  LB23          \\ 
            &   143          &   $76.9 \pm 11.5$  &  This work    \\ 
            &   143          &   $73.7 \pm 7.6$   &  LB23          \\ 
            &    50          &   $360.8 \pm 41.9$ &  LB23          \\ 
\hline
\multirow{4}{*}{H3 (outer)}  &  1283          &   $0.6 \pm 0.2$    &  This work    \\ 
            &   143          &   $33.0 \pm 3.11$  &  This work    \\ 
            &   143          &   $32.5 \pm 3.9$   &  LB23          \\ 
            &    50          &   $169.3 \pm 29.0$ &  LB23          \\
\hline\hline
\multicolumn{4}{c}{Integrated from resolved profiles}\\
\hline\hline
\multirow{5}{*}{H3 (total)}  &  1283          &   $14.8 \pm 3.4$    &  This work    \\ 
            &   143          &   $628.4 \pm 85.0$   &  LB23    \\
            &   143          &   $482.3 \pm 90.4$  &  This work    \\ 
            &   143          &   $391.6 \pm 7.9$   &  This work    \\
            &    50          &   $3271.2 \pm 528.3$ &  LB23          \\  

\hline
\end{tabular}
\end{table}

Figure~\ref{fig:A2142_SED_H1H2H3} presents these integrated spectrum measurements for the three components: H1, H2 and H3. For H1 and H2, it is immediately clear that there is a discrepancy in the measurements reported by \cite{Venturi2017_A2142} and \cite{Bruno2023_A2142}. For H1 this can be attributed largely to the differences in region size at low frequencies, as \cite{Bruno2023_A2142} considered a smaller region size; at higher frequencies the measurements reported by both previous works are consistent. For H2, however, both previous works considered a very similar region size, and so the discrepancies are likely attributable to differences in data quality and in particular \textit{uv}-filling factor on the shorter baselines.

This is particularly evident when comparing our new MeerKAT measurements for H2 (and to a lesser extent H1) with the VLA observations that provided the high-frequency measurements considered by \cite{Venturi2017_A2142} and \cite{Bruno2023_A2142}. Our deep MeerKAT observations have dense \textit{uv}-coverage on short baselines and thus recover more of the diffuse flux than the shallow VLA D-configuration observations. At 143\,MHz, our LOFAR observations recover integrated flux densities for H1 and H2 that are consistent with the overall trends reported by both \cite{Venturi2017_A2142} and \cite{Bruno2023_A2142} when considering equivalent regions. For H3, our 143\,MHz measurements are consistent with those reported by \cite{Bruno2023_A2142}; our 1283\,MHz MeerKAT measurements are similarly consistent with the overall trend, extending the lever-arm in frequency significantly.

For each component, the overall trend in the integrated flux density appears consistent with a single power-law behaviour up to the limit of the frequency coverage. As such we attempt to fit a power-law SED to these observations. For H1, H2 and both the inner and outer regions of H3, we combine our new measurements with those from \cite{Bruno2023_A2142} when performing the fit, eliminating the duplicate measurement at 143\,MHz as this was derived from the same dataset, albeit independently imaged and post-processed; for H1 and H2 we fit the measurements from \cite{Venturi2017_A2142} separately, without incorporating our new measurements, for reference.

When performing the fits, we used the `affine invariant' Markov-chain Monte Carlo (MCMC) ensemble sampler \citep{Goodman2010} from the \textsc{emcee} package \citep{ForemanMackey2013_EMCEE} to derive both the best-fit parameters and the $1\sigma$ uncertainty region. These fits are presented in Figure~\ref{fig:A2142_SED_H1H2H3}. When fitting to the measurements of \cite{Venturi2017_A2142}, we find $\alpha_{\rm H1} = -1.29 \pm 0.04$ and $\alpha_{\rm H2} = -1.47 \pm 0.08$, which are consistent with the values reported in their paper. We note that for H2, the measurements at 234\,MHz, 608\,MHz and 1377\,MHz were excluded from our fitting routine due to the relatively sparser \textit{uv}-coverage, as noted by \citeauthor{Venturi2017_A2142}. Likewise the 74\,MHz upper limit was excluded from the fit, although it provides useful constraints when exploring the allowed parameter space with \textsc{emcee}. 

With our new dataset we find a best-fit spectral index of $\alpha_{\rm H1} = -1.09 \pm 0.03$ and $\alpha_{\rm H2} = -1.15 \pm 0.04$ for H1 and H2, respectively. These are consistent with the values reported by \citeauthor{Bruno2023_A2142} (see their Fig.~6). Finally, for the inner and outer components of H3, we find $\alpha_{\rm H3, \, inner} = -1.30 \pm 0.07$ and $\alpha_{\rm H3, \, outer} = -1.74 \pm 0.10$. For the inner region of H3, our spectral index measurement is marginally flatter than the value of $\alpha_{\rm H3, \, inner} = -1.36 \pm 0.05$ reported by \citeauthor{Bruno2023_A2142}, although still consistent within the uncertainties. This marginal flattening is strongly influenced by our 1283\,MHz MeerKAT datapoint, which sits slightly above the trend, although within the $1\sigma$ uncertainty region. We attribute this to the excellent sensitivity and short-baseline coverage of our MeerKAT observations, which despite being at a higher frequency than the GMRT observations used by \citeauthor{Bruno2023_A2142}, provide improved recovery of diffuse radio emission.

Conversely, for the outer region of H3, our fitted spectral index is steeper than the value of $\alpha_{\rm H3, \, outer} = -1.57 \pm 0.20$ reported by \citeauthor{Bruno2023_A2142}, although again these values are consistent within the uncertainty. Similarly, this steepening is likely strongly influenced by our 1283\,MHz MeerKAT datapoint, which shows significant uncertainty. From Figure~\ref{fig:A2142_radio_sub}, we see that in the outer regions of H3 (magenta boxes), MeerKAT recovers some emission but mostly at a level $\lesssim3\sigma$. Significantly deeper MeerKAT L-band or UHF-band observations would be required to improve the confidence in this measurement; likewise deep observations with the uGMRT in Band~3 and/or Band~4 would prove insightful.

We also note that we report in Table~\ref{tab:integrated_flux} measurements of the total flux density of H3, derived from our radial profiles (see Section~\ref{sec:radial_profiles}). Uniting our new measurements with those reported by \cite{Bruno2023_A2142}, we are able to estimate the total integrated spectrum of H3. This is also presented in Figure~\ref{fig:A2142_SED_H1H2H3}; we derive a spectral index of $\alpha_{\rm H3, \, total} = -1.68 \pm 0.10$.

\subsubsection{Luminosity and Scaling Relations}
We can use these fits --- using H1, H2, and the `total' values for H3 --- to estimate the radio power using Equation~\ref{eq:radio_lum}. To place Abell~2142 appropriately in known scaling relation planes, we scale the fluxes of the three components to the chosen frequency according to the fitted power-law spectral index, and derive the radio power using the sum of the fluxes. For completeness, we place Abell~2142 in the scaling relation planes at both 1.4\,GHz and 150\,MHz \citep[e.g.][]{Duchesne2021_MWA2_ASKAP,Duchesne2021_A141_A3404,Cuciti2023_LoTSS-DR2_Planck}. Using our fitted SEDs in conjunction with Equation~\ref{eq:radio_lum}, we derive a radio luminosity of $P_{\rm 1.4\,GHz} = (5.21 \pm 1.33) \times10^{23}$~W\,Hz$^{-1}$ at 1.4\,GHz and $P_{\rm 150\,MHz} = (1.29 \pm 0.33) \times10^{25}$~W\,Hz$^{-1}$ at 150\,MHz. 


\begin{figure}
\begin{center}
 \includegraphics[width=0.95\linewidth]{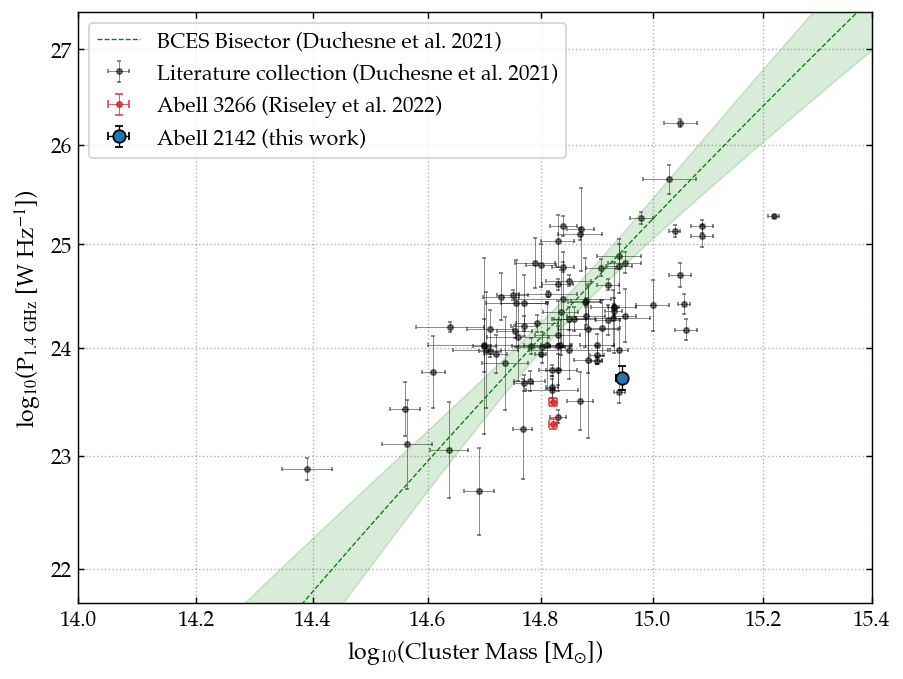}
 \includegraphics[width=0.95\linewidth]{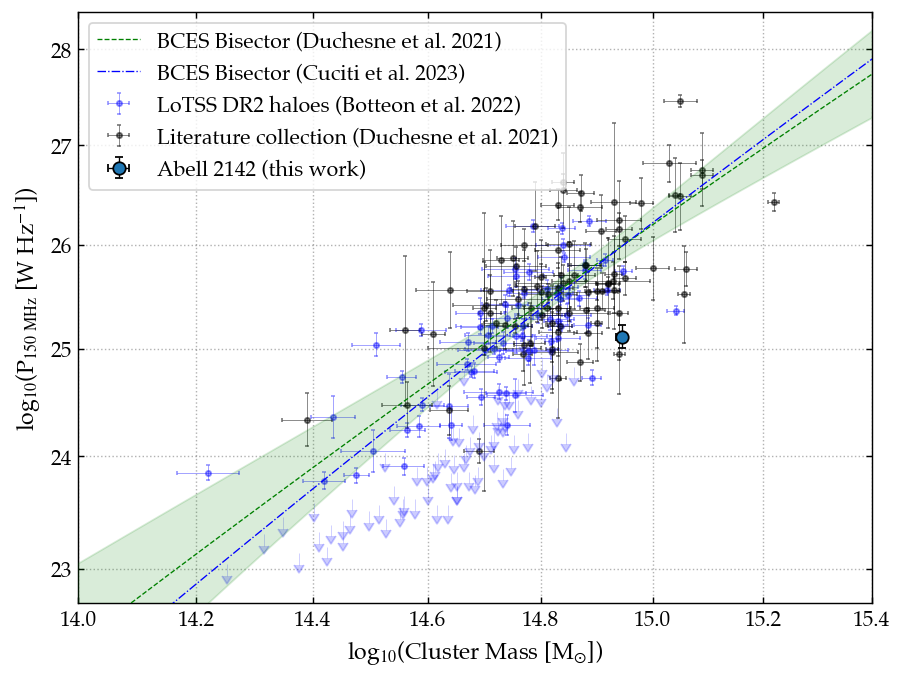}
\caption{Power scaling planes between cluster mass $M_{500}$ and radio halo power at 1.4\,GHz (\textit{top}) and 150\,MHz (\textit{bottom}). Green shaded regions mark the 95\% confidence limits.}
\label{fig:power_scaling}
\end{center}
\end{figure}

Figure~\ref{fig:power_scaling} presents the power-scaling planes between cluster mass $M_{500}$ and radio power at both 1.4\,GHz and 150\,MHz. We include other clusters with known haloes from the literature, using collections of previous works \citep[compiled by][and scaled to both 1.4\,GHz and 150\,MHz]{Duchesne2021_MWA2_ASKAP} as well as the recent catalogue of haloes from the LoTSS DR2 survey region \citep{Botteon2022_LOTSS_Planck,Cuciti2023_LoTSS-DR2_Planck}. We also overplot the BCES Bisector fits at 1.4\,GHz \citep{Duchesne2021_A141_A3404} and at 150\,MHz \citep{Duchesne2021_A141_A3404,Cuciti2023_LoTSS-DR2_Planck}. From Figure~\ref{fig:power_scaling} we see that while Abell~2142 is broadly consistent with the trend, albeit lying low in the plane: compared to the fitted scaling relations Abell~2142 is under-luminous by a factor 16 at 1.4\,GHz and around a factor 7.5 at 150\,MHz.

\subsubsection{Resolved spectral properties}\label{sec:alfa_maps}
\begin{figure*}
\begin{center}
\includegraphics[width=0.9\textwidth]{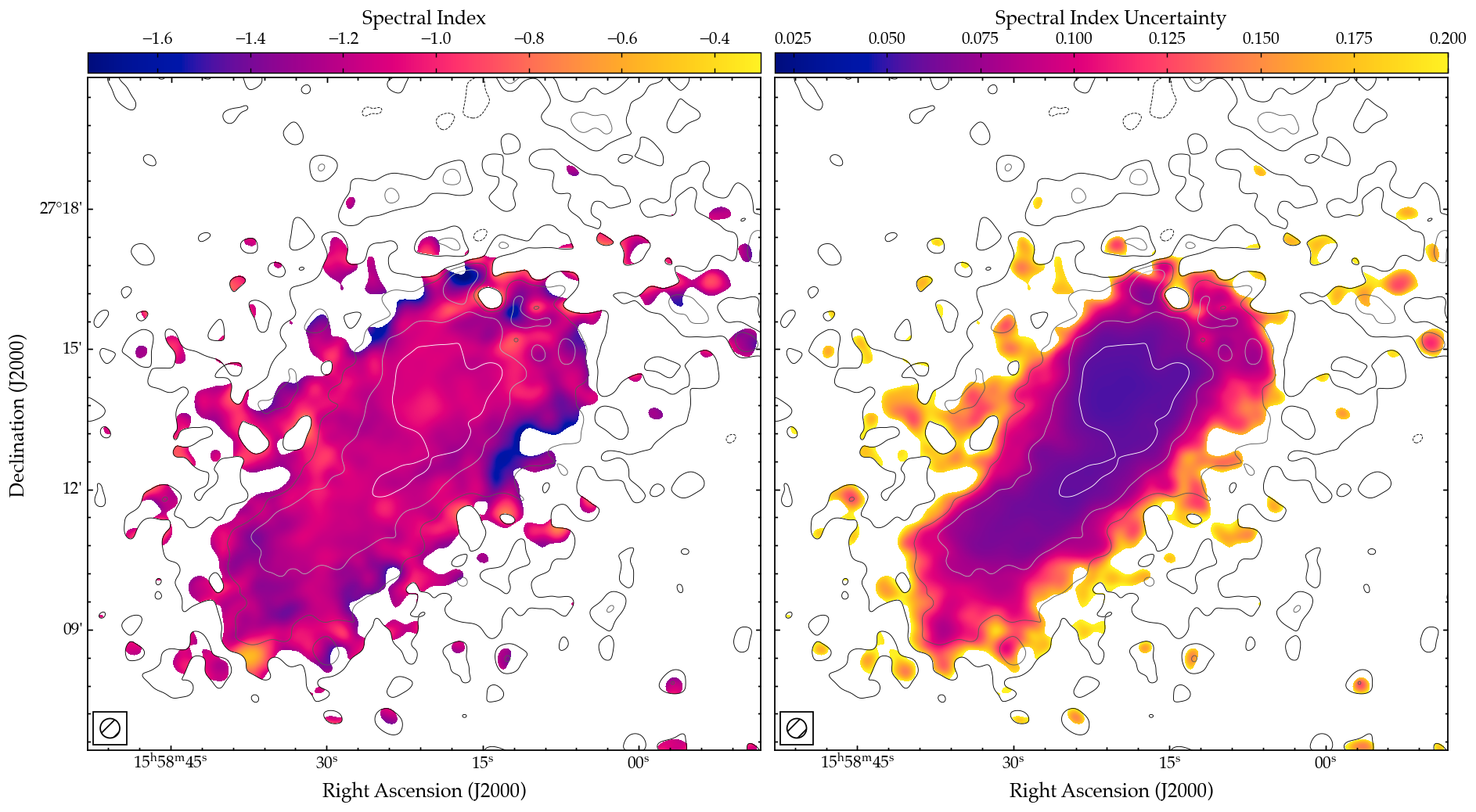}
\includegraphics[width=0.9\textwidth]{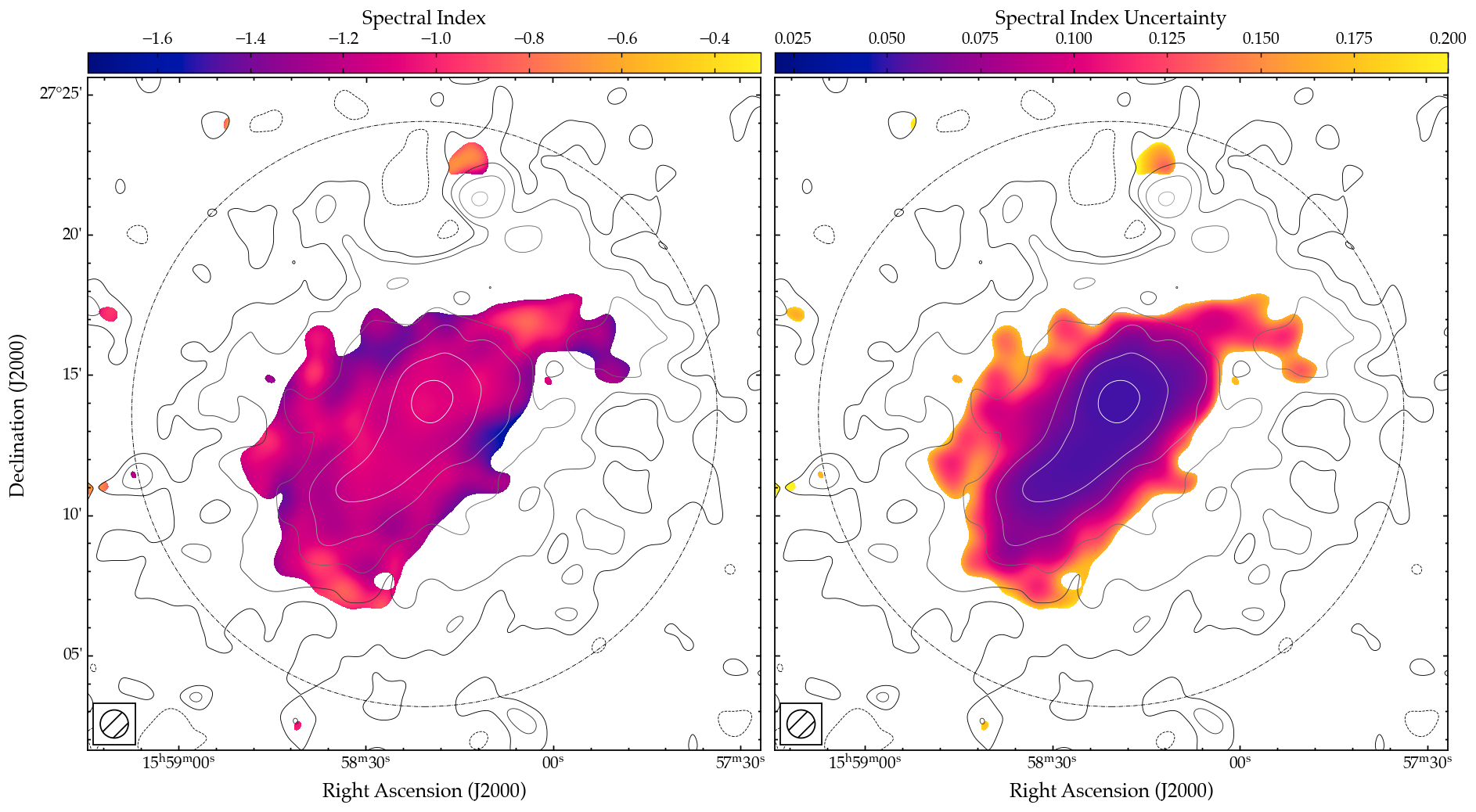}
\caption{Spectral index map of the radio halo in Abell~2142 at 25\,arcsec resolution (\textit{left}) along with the associated uncertainty (\textit{right}), derived using the maps presented in Figure~\ref{fig:A2142_radio_sub}. The lower panels show the same quantities (spectral index and associated uncertainty) at 60\,arcsec resolution.}
\label{fig:A2142_radio_alfa}
\end{center}
\end{figure*}

Figure~\ref{fig:A2142_radio_alfa} presents the two-point spectral index map between LOFAR at 143\,MHz and MeerKAT at 1283\,MHz, at a resolution of 25\,arcsec and 60\,arcsec, along with the associated uncertainty.

From Figure~\ref{fig:A2142_radio_alfa} we see that the spectral index is relatively uniform at 25\,arcsec resolution. Measuring from the inner region of the halo where we have the greater signal-to-noise, we find a median spectral index of $\langle \alpha \rangle = -1.16 \pm 0.06$. The standard deviation in the measured spectral index within this region is $\sigma_{\alpha} = 0.07$, comparable to the uncertainty.

We can use our maps at 25\,arcsec resolution, where we have the resolution to more clearly separate H1 and H2, to attempt to investigate whether there is any significant difference in the resolved spectrum between these two main components. For H1 we find an overall spectral index of $\alpha_{\rm H1} = -1.16 \pm 0.06$; for H2 we find $\alpha_{\rm H2} = -1.18 \pm 0.08$. While the standard deviation in spectral index is higher for H1, where $\sigma_{\alpha, {\rm H1}} = 0.09$, than for H2, where $\sigma_{\alpha, {\rm H1}} = 0.08$, these differences are not significant. As such, while Figure~\ref{fig:A2142_SED_H1H2H3} demonstrates that H2 has a steeper \textit{integrated} spectrum than H1, our resolved study indicates that the spectral properties are consistent when examined in detail.


At 60\,arcsec, shown in the lower panel of Figure~\ref{fig:A2142_radio_alfa} we see a similar situation. Overall, the halo shows a largely-uniform spectral index with a median value of $\langle \alpha \rangle = -1.17 \pm 0.06$ with a standard deviation of $\sigma = 0.11$, consistent with the profile at 25\,arcsec resolution. Overall, our spectral index maps appear to show similar trends to the lower-frequency spectral index maps between LOFAR LBA at 50\,MHz and HBA at 143\,MHz presented by \cite{Bruno2023_A2142}, when viewed at comparable resolution (their Fig.~8, top and middle rows). 

\cite{Bruno2023_A2142} report that the `radio bay' shows a significant flattening in the spectral index between 143\,MHz and 323\,MHz, based on their LOFAR HBA and GMRT observations. In this region, the spectral index transitions from a global value of $\sim -1.2$ to a local value of $-0.9$, although the uncertainty in their spectral index measurements is large in this area, typically $\sim 0.3$ \citep[see Fig.~A1 of][]{Bruno2023_A2142} so strictly-speaking the reported flattening is not statistically significant. In our spectral index maps, we do not see a spectral flattening in the `radio bay'. Given that we use the same low-frequency dataset, namely LOFAR HBA observations at 143\,MHz, but we use new broad-band MeerKAT L-band observations as our high-frequency reference, whereas \citeauthor{Bruno2023_A2142} use narrow-band GMRT 325\,MHz observations, we suggest that the apparent spectral flattening is driven by the comparatively limited \textit{uv}-coverage (and hence less optimal diffuse emission recovery) of the historic data\footnote{We note also that a similar situation was reported by \cite{Riseley2022_MS1455} for MS1455.}.

\subsection{Resolved Spatial Profiles}
As we are looking to explore the spectral properties of the different halo components (H1, H2, and H3) in order to study the underlying particle acceleration mechanism, we consider profiles measured along wedge-shaped regions toward the south-east, south-west, and  north-east of Abell~2142. The spacing of each annulus was chosen to be half the resolution element, i.e. 30\,arcsec. As the origin of these wedges, we take the peak of the diffuse radio emission (J2000 Right Ascension, Declination) $=$ (15$:$58$:$19.7, $+$27$:$14$:$07). We note that a similar analysis was performed by \cite{Bruno2023_A2142} on their LOFAR HBA and LBA data, although those authors considered only full-circular annuli; here we go further and consider these sectors to investigate asymmetries. These wedges are presented in the upper panels of Figure~\ref{fig:profiles_60arcsec}, where they are overlaid on X-ray surface brightness maps and radio contours for reference.

\subsubsection{Surface Brightness}\label{sec:radial_profiles}
\begin{figure*}
\begin{center}
\includegraphics[width=0.94\textwidth]{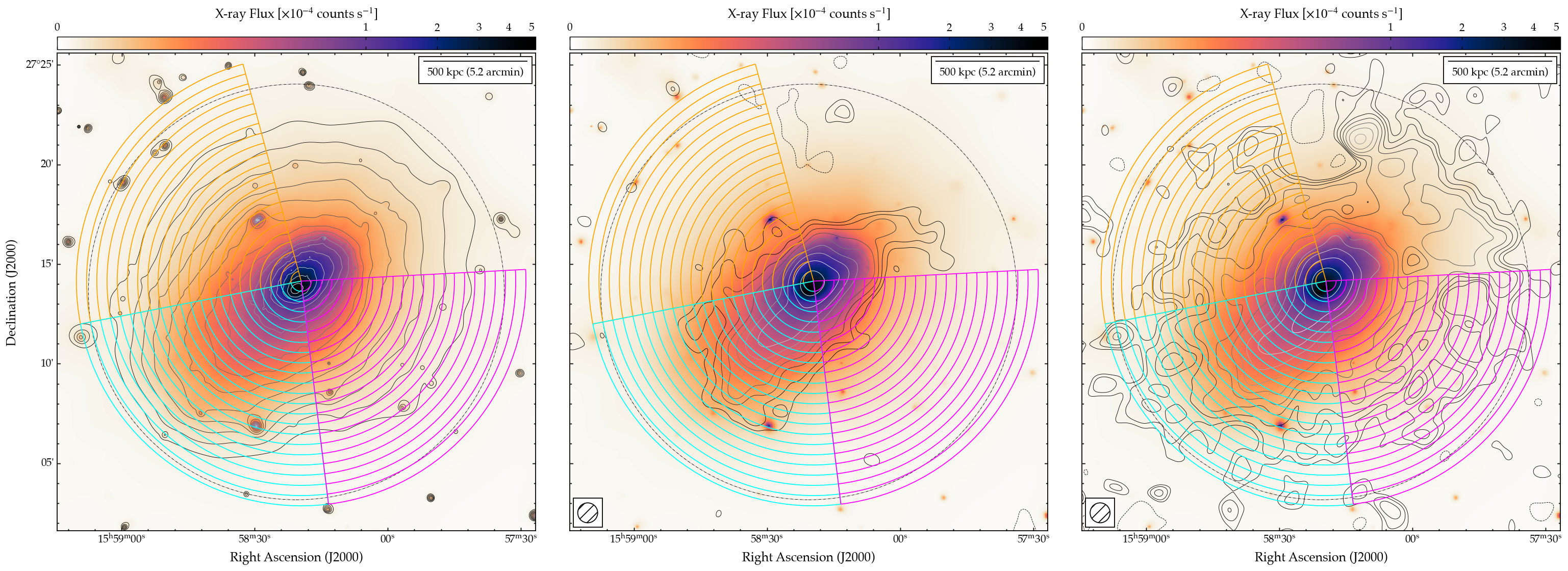}
 \includegraphics[width=0.95\textwidth]{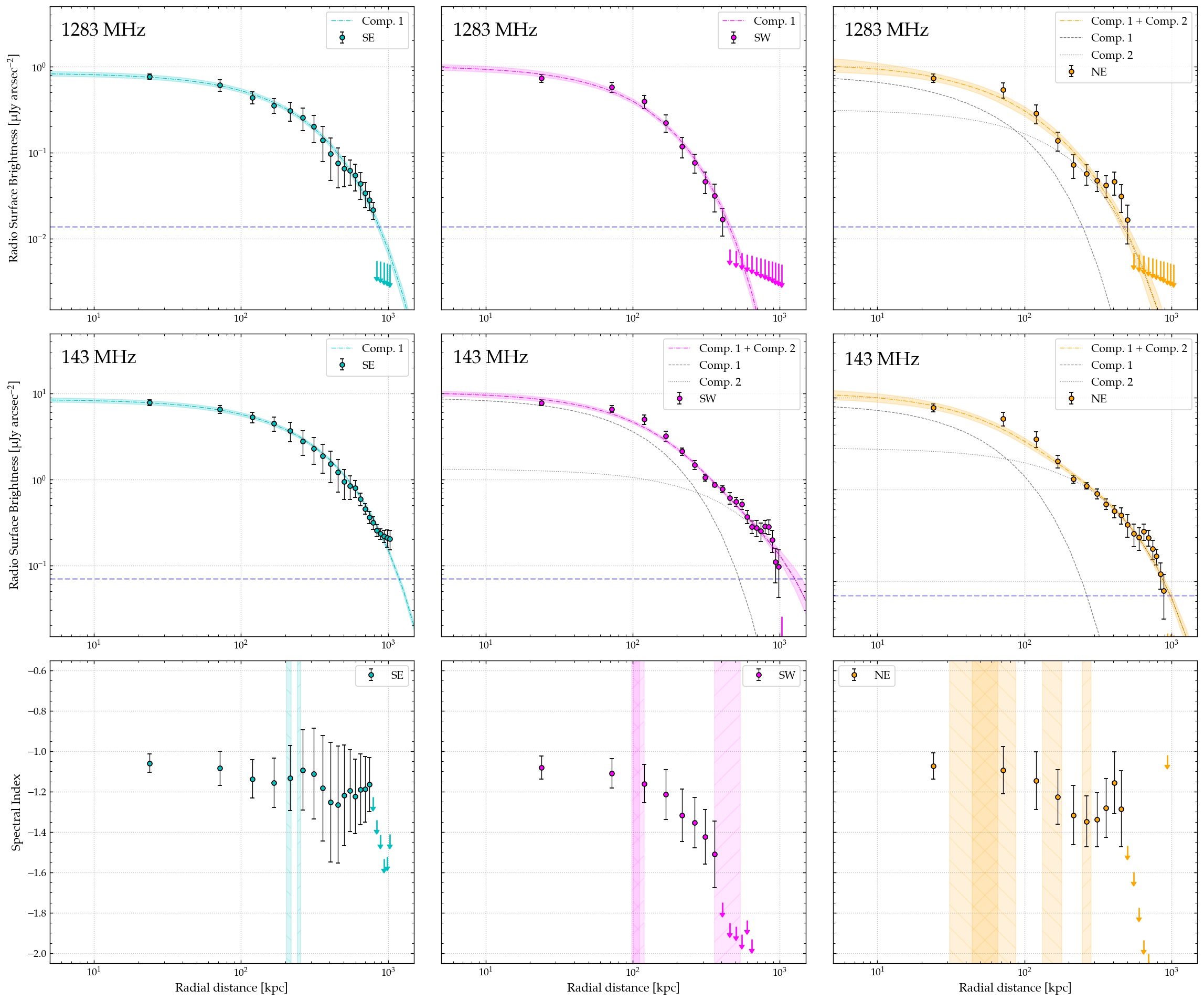}
\caption{Radial surface brightness profiles for the multi-component halo in Abell~2142, as measured by MeerKAT at 1283\,MHz and LOFAR at 143\,MHz, both at 60\,arcsec resolution. The wedge-shaped regions are displayed in the upper row for reference. Datapoints show the measured values, with arrows denoting upper limits of $1\sigma$; the uncertainties show only statistical errors. The horizontal dashed line indicates the $1\sigma$ level. Colourised dot-dashed curves show fits to the data, which describe either a single- or double-component circular exponential profile described in Equation~\ref{eq:efold}, with best-fit parameters presented in Table~\ref{tab:radial_profiles_60arcsec}. Where the two-component model provides a better description of our data, we show each component individually using dashed and dotted curves. The lower panels show the radial spectral index profile derived using the average surface brightness in each region at 1283\,MHz and 143\,MHz, with $3\sigma$ upper limits shown as arrows. Uncertainties include both statistical and systematic uncertainties. Shaded regions show the characteristic e-folding radius of each component, as plotted above and shown in Table~\ref{tab:radial_profiles_60arcsec}.}
\label{fig:profiles_60arcsec}
\end{center}
\end{figure*}

We present our radial surface brightness profiles along each of these directions in Figure~\ref{fig:profiles_60arcsec}, where we show MeerKAT 1283\,MHz profiles in the top row and LOFAR 143\,MHz profiles in the bottom row.

Typically, a standard circular exponential profile is used to model the surface brightness of haloes and mini-haloes. While exponential profiles are assumed by convention rather than being truly physically-motivated, their use is observationally-driven as they are straightforward to implement and rely on few free parameters, and have been shown to provide a good descriptor; as such, exponential profiles have been used in numerous studies \citep[e.g.][]{Orru2007_A2744_A2219,Murgia2009,Bonafede2017}. Similarly, the straighforward $\beta$-model continues to provide a good general descriptor of the X-ray profile of the ICM \citep[e.g.][]{Cavaliere_FuscoFemiano_1976,Henriksen_Mushotzky_1985,Ettori2000_BetaModel,Arnaud2009}. The circular exponential radio profile takes the form:
\begin{equation}\label{eq:efold}
I_{\rm R} (r) = \sum_i \, I_{{\rm R, c,} i} \, \exp\left( -r / r_{e, i} \right) \, , 
\end{equation} 
where $I_{\rm R, c}$ is the central radio surface brightness and $r_e$ is the $e$-folding radius, typically found to be $\sim R_{\rm H} / 2.6$ for a radio halo of radius $R_\text{H}$ \citep[e.g.][]{Bonafede2017}. Historically, most haloes have been fitted using a single component, $i = 1$; however with the advent of new instrumentation in the form of the SKA Pathfinders and Precursors, new observations are revealing the presence of numerous multi-component haloes \citep[e.g.][]{Bonafede2022_Coma-LOFAR-II,Riseley2022_A3266,Bruno2023_A2142} and several mini-haloes \citep[e.g.][]{Biava2021_RXCJ1720,Riseley2022_MS1455,Riseley2023_A1413,Lusetti2023}. Indeed, there is emerging evidence that these straightforward profiles provide an increasingly worse descriptor of the radial profile as signal-to-noise increases, although this is likely telling us that substructures and asymmetries only begin to emerge once a certain data quality is achieved \citep[see discussion by e.g.][]{Botteon2022_LOTSS_Planck,Campitiello2023_arXiv}.

As such, given the multi-component nature of the radio halo in Abell~2142, we attempted to fit both a single- and a double-component exponential profile to the measurements in Figure~\ref{fig:profiles_60arcsec}. In each sub-plot, we show the fit that provided the best recreation of our observational measurements; we favour a two-component fit only if it provides a better reproduction of our data compared to a one-component fit, and if the e-folding radii are statistically significantly different for each component. We again used \textsc{emcee} to marginalise over the parameter space allowed by our data and derive both the best-fit parameters and the $1\sigma$ uncertainty region. These were respectively defined using the $50^{\rm th}$ percentile and the $16^{\rm th}$/$84^{\rm th}$ percentiles. The best-fit parameters are reported in Table~\ref{tab:radial_profiles_60arcsec} and plotted in Figure~\ref{fig:profiles_60arcsec}; these profiles clearly support both the presence of multiple components in the Abell~2142 halo, as well as the interpretation that the halo is asymmetric. 

\begin{table*}
\small
\renewcommand{\arraystretch}{1.4}
\centering
\caption{Best-fit parameters for one- and two-component model fits to the radial surface brightness profiles (shown in Figure~\ref{fig:profiles_60arcsec}) along the south-east, south-west, and north-east wedges shown in the same Figure. We report the central surface brightness ($I_{R,c,i}$) and e-folding radius ($r_{e,i}$) for each component, according to Equation~\ref{eq:efold}, with the best-fit and $1\sigma$ uncertainty defined as the $50^{\rm th}$ percentile and the $16^{\rm th}$/$84^{\rm th}$ percentiles respectively. \label{tab:radial_profiles_60arcsec}}
\begin{tabular}{lll | cc | cc}
\hline
Direction   & Instrument  &  Frequency    &   \multicolumn{2}{c |}{Component~1}     &   \multicolumn{2}{c}{Component~2}     \\
            &   &               &   $I_{R,c,1}$     &   $r_{e,1}$   &   $I_{R,c,2}$     &   $r_{e,2}$   \\
            &   &   $[\rm MHz]$   &   $[\upmu$Jy~arcsec$^{-2}]$  & $[\rm{kpc}]$    &   $[\upmu$Jy~arcsec$^{-2}]$  & $[\rm{kpc}]$    \\
\hline\hline
\multirow{2}{*}{South-East (SE)}     &  MeerKAT &   1283    & $0.84^{+0.05}_{-0.05}$   &  $212^{+7}_{-8}$    &  $-$     &  $-$     \\
                                     &  LOFAR   &   143     & $8.58^{+0.48}_{-0.49}$   &  $248^{+7}_{-6}$    &   $-$    &  $-$     \\
\hline
\multirow{2}{*}{South-West (SW)}     &  MeerKAT &   1283    & $1.02^{+0.09}_{-0.07}$   &  $106^{+4}_{-4}$    &   $-$    & $-$      \\
                                     &  LOFAR   &   143     & $9.04^{+0.76}_{-0.71}$   &  $108^{+10}_{-9}$   & $1.33^{+0.50}_{-0.47}$      &  $429^{+134}_{-70}$     \\
\hline
\multirow{2}{*}{North-East (NE)}     &  MeerKAT &   1283    & $0.81^{+0.20}_{-0.14}$   & $63^{+24}_{-32}$    & $0.32^{+0.15}_{-0.18}$      &  $153^{+27}_{-18}$     \\
                                     &  LOFAR   &   143     & $8.66^{+1.37}_{-1.36}$   & $55^{+10}_{-10}$    & $2.83^{+0.39}_{-0.38}$      &  $265^{+19}_{-17}$     \\
\hline
\end{tabular}
\end{table*}

The SE profile indicates that at 60\,arcsec resolution, the two primary components --- the halo core (H1) and the ridge (H2) --- are not clearly distinguishable. The radial profile in the SE direction shows a smooth surface brightness profile that is well-described by a single component with an e-folding radius $r_e = 212^{+7}_{-8}$~kpc at 1283\,MHz and $r_e = 248^{+7}_{-6}$~kpc at 143\,MHz. The e-folding radius is similar at both frequencies, although the larger radius at 143\,MHz compared to 1283\,MHz is consistent with earlier evidence of spectral steepening toward the outskirts of the halo. While we note a slight surface brightness excess in the LOFAR surface brightness profile at 143\,MHz at very large radii ($r \gtrsim 800$\,kpc) which may suggest further ultra-diffuse components, residuals from the subtraction of the FR-II radio galaxy 7C~1557$+$2712 to the south-east prevent us from exploring this further.

The SW profile shows curious behaviour. At 1283\,MHz, the profile is well-described by a single exponential profile with an e-folding radius $r_{e} = 106^{+4}_{-4}$~kpc. Conversely at 143\,MHz LOFAR clearly recovers two components: the inner component shows a near-identical e-folding radius of $r_{e,1} = 108^{+10}_{-9}$~kpc, and the outer component shows a much larger e-folding radius of $r_{e,2} = 429^{+134}_{-70}$~kpc. While the uncertainty is large, given that $r_e \sim R_{\rm H} / 2.6$, this e-folding radius would imply a halo radius of order $1.12$\,Mpc, which is consistent with the halo extent recovered by LOFAR.

Finally, to the NE, both MeerKAT and LOFAR show clear evidence of two components in the radial profiles, as implied by the detection of extended emission in this direction shown in Figure~\ref{fig:A2142_radio_sub}. The inner component shows a consistent e-folding radius of $r_{e,1} = 63^{+24}_{-32}$~kpc at 1283\,MHz and $r_{e,1} = 55^{+10}_{-10}$~kpc at 143\,MHz. For the outer component, the MeerKAT profile suggests an e-folding radius of $r_{e,2} = 153^{+27}_{-18}$~kpc whereas the LOFAR profile at 143\,MHz yields a larger e-folding radius of $r_{e,2} = 265^{+19}_{-17}$~kpc. The significantly larger e-folding radius at 143\,MHz compared to 1283\,MHz further supports the interpretation of a steepening spectral index toward the outer regions of the radio halo. While we leave the detailed discussion of the interpretation to Section~\ref{sec:discussion}, this spectral steepening broadly supports the turbulent (re-)acceleration scenario, but the differences in e-folding radius in different regions may favour inhomogeneous (re-)acceleration.

Finally, we can use these fitted profiles to estimate the total flux of each component by integrating up to a radius of $3 r_{\rm e}$, which typically sets the outer boundary of radio haloes and contains some $\sim80\%$ of the total flux that would be obtained by integrating out to infinity. We use the simplified relation:
\begin{equation}\label{eq:flux_fit}
S_{\rm fit} = 0.8 \times 2 \, \pi \, I_{R,c} \, r_{\rm e}^2 \, 
\end{equation}
to derive the total flux of each component according to our fitted profiles, using the values of $I_{R,c,1}$, $r_{e,1}$, $I_{R,c,2}$, and $r_{e,2}$ from Table~\ref{tab:radial_profiles_60arcsec}. It is of particular interest to estimate the total flux of H3, which we could not do directly in Section~\ref{sec:integrated_spectrum} as H3 likely underlies the H1 and H2 regions.

However, by making a reasonable assumption that H2 does not provide significant contamination along the SW and NE wedges, we can use the parameters of the second fitted component to estimate the total integrated flux of H3. Thus, using Equation~\ref{eq:flux_fit} and the `Component 2' parameters from the SW profile, we find an estimated integrated flux density of $S_{\rm fit, \, 143\,MHz} ({\rm H3}) = 482.29 \pm 90.44$\,mJy at 143\,MHz. Alternatively, from the NE profile, we find values of $S_{\rm fit, \, 143\,MHz} ({\rm H3}) = 391.58 \pm 7.94$\,mJy at 143\,MHz and $S_{\rm fit, \, 1283\,MHz} ({\rm H3}) = 14.76 \pm 3.40$\,mJy at 1283\,MHz.

We note that the flux density estimates derived from our radial profiles are somewhat lower than the estimate derived by \cite{Bruno2023_A2142}, who found $S_{\rm fit, \, 143\,MHz} ({\rm H3}) = 628.4 \pm 85.0$\,mJy, although the uncertainties are large and overlapping. We also note that \citeauthor{Bruno2023_A2142} considered global radial profiles, and their flux density estimate derived from these profiles includes some contribution from H2 \citep[see discussion in Sec.~4.3 of][]{Bruno2023_A2142}. On the other hand, our estimates are derived from wedge-shaped profiles, somewhat mitigating any such contamination.

Our estimates of the total integrated flux density of H3 allow us to derive an estimate of the integrated spectral index. We find an integrated spectrum of $\alpha_{\rm H3} = -1.49 \pm 0.11$ or $-1.59 \pm 0.14$ depending whether we take the lesser or greater integrated flux density at 143\,MHz. These values are broadly consistent with the integrated spectrum presented in Figure~\ref{fig:A2142_SED_H1H2H3} \citep[as well as those reported by][]{Bruno2023_A2142}, further supporting the suggestion that H3 exhibits an ultra-steep radio spectrum. Deeper follow-up with MeerKAT and/or the uGMRT at complementary frequencies is strongly motivated.

\subsubsection{Spectral Index}
The lower panels of Figure~\ref{fig:profiles_60arcsec} show the radial spectral index profiles along the SE, SW and NE wedges, derived using the surface brightness profiles at 1283\,MHz and 143\,MHz in the above panels. The coloured bands represent the fitted e-folding radii for each component at each frequency, with forward-diagonal bands (i.e. `\slash\slash') denoting the fits at 143\,MHz and reverse-diagonal banding (i.e. `\textbackslash\textbackslash') denoting the fits at 1283\,MHz.

Along each arc, we see somewhat different behaviour. To the SE, along the direction of the radio ridge (H2), we see no clear evidence of a steepening spectrum out to the limit of our profiles (around 800\,kpc). Beyond the fitted e-folding radius, the uncertainty increases significantly but the average spectral index remains consistently around $-1.1$ to $-1.2$. Along the SW arc however, we see conclusive evidence of spectral steepening between the e-folding radii of the first and second components, from a spectral index of around $-1.1$ to $-1.4$. Beyond the e-folding radius of the second component, we can only place limits on the spectral index, but these suggest a dramatic steepening toward $\alpha \lesssim -1.7$. Along the NE arc, the behaviour is mixed. We see tentative evidence of spectral steepening between the limits of the inner and outer components, after which the spectrum appears to flatten again to a value of around $-1.1$. This is consistent with the minor surface brightness excess (compared to the fitted profiles) at large radii, but with the present data we cannot investigate this further. Deeper high-frequency observations would be required.

\subsection{Thermal/non-thermal comparison}
\subsubsection{Point-to-point correlation: the large-scale halo}
Being spatially co-located with the thermal plasma of the ICM, both radio haloes and mini-haloes show an intrinsic link between the non-thermal and thermal components. In practice this takes the form of a correlation between the radio surface brightness and X-ray surface brightness, which follows a functional form $I_{\rm R} \propto I_{\rm X}^b$ \citep[e.g.][]{Govoni2001}.

This correlation provides much insight into the thermal/non-thermal connection, and into the underlying particle acceleration mechanism at work. We are interested in studying the strength of the correlation, which indicates the strength of the connection; the spatial distribution of the correlation, i.e. whether one correlation is evident throughout, or whether there are coherent areas of different correlations, which could indicate different physical processes at work and/or different environmental conditions \citep[e.g.][]{Biava2021_RXCJ1720}; the spectral dependence of the slope, i.e. whether there is a change in correlation slope $b$ with frequency, which would indicate changes in the spectral index. Finally, we are \text{also} interested in quantifying the slope $b$, as this provides key insights into the particle acceleration mechanism at work.

For mini-haloes, previous studies overwhelmingly report a \textit{super}-linear point-to-point correlation slope, i.e. $b > 1$ \citep[][]{Ignesti2020_MHsample,Ignesti2021_2A0335,Biava2021_RXCJ1720,Riseley2022_MS1455,Riseley2023_A1413,Lusetti2023}. Conversely, haloes overwhelmingly show a \textit{sub}-linear slope, i.e. $b < 1$ \citep[e.g.][]{Hoang2019_A520,Duchesne2021_A141_A3404,Bonafede2022_Coma-LOFAR-II,Rajpurohit2021_MACSJ0717_Halo,Rajpurohit2021_A2744,Riseley2022_A3266}, implying that the non-thermal radio emission declines less rapidly than the thermal X-ray surface brightness. In other words, a sub-linear slope implies that the distribution of the non-thermal components is broader than that of the thermal components (and vice versa for a super-linear slope).

\begin{figure*}
\begin{center}
\includegraphics[width=0.9\linewidth]{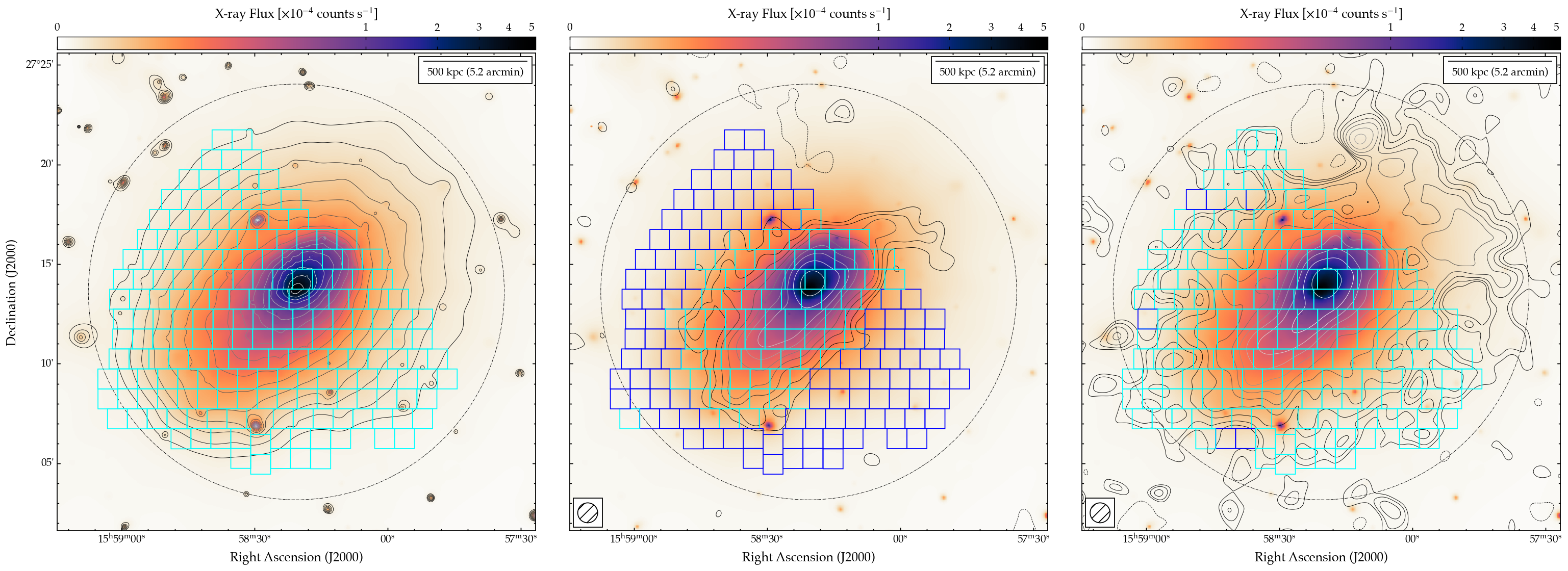}
\includegraphics[width=0.92\linewidth]{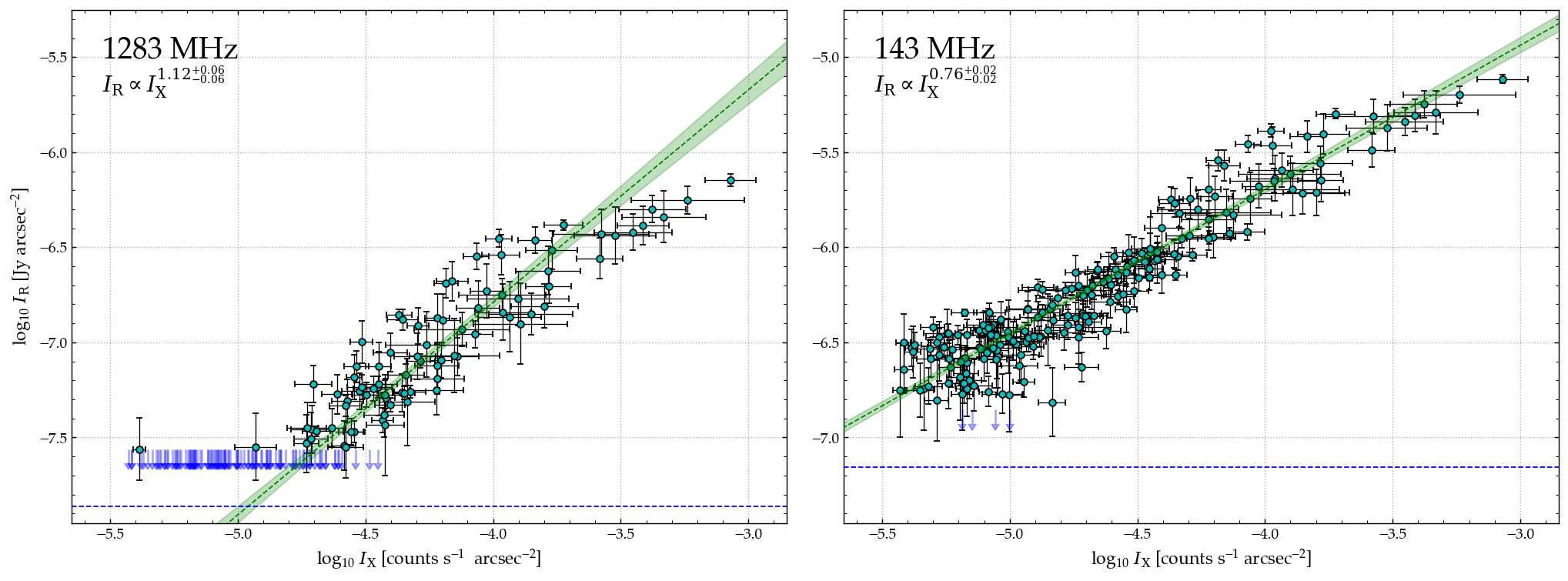}
\includegraphics[width=0.92\linewidth]{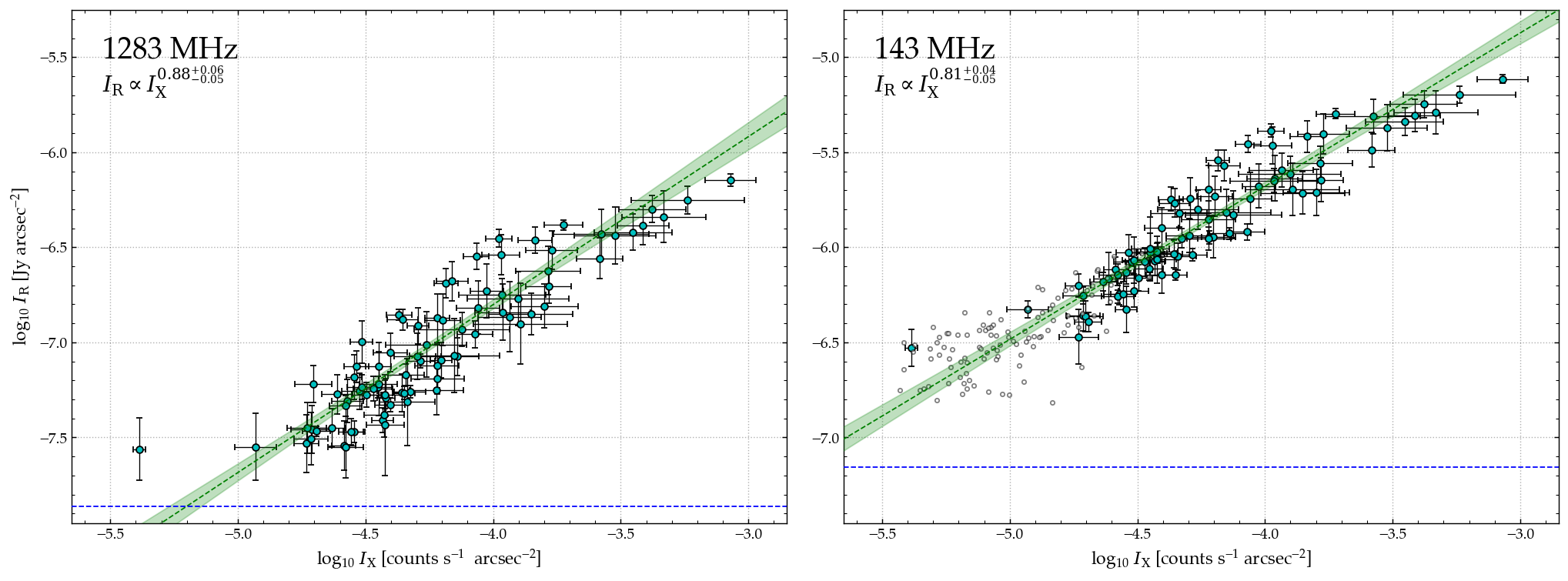}
\caption{\textit{Upper}: regions used to extract the surface brightness for the $I_{\rm{R}}/I_{\rm{X}}$ correlation. Colourscale is the \textit{XMM-Newton} surface brightness mosaic, with contours showing surface brightness measured by \textit{XMM-Newton}, MeerKAT and LOFAR from left to right. Boxes are colourised according to whether the surface brightness within is considered a measurement (cyan) or a $2\sigma$ limit (blue). \textit{Middle and lower}: Radio/X-ray surface brightness correlation ($I_{\rm{R}}/I_{\rm{X}}$) for Abell~2142 at 1283~MHz (\textit{left}) and 143~MHz (\textit{right}) at 60~arcsec resolution. The dashed blue line indicates the $1\sigma$ level. Dashed green line shows the best-fit power-law relation and $1\sigma$ uncertainty region shown shaded. The slope is indicated in the upper-left corner inset. The lower panels show the results using a common region set where both MeerKAT and LOFAR measure a radio surface brightness of at least $2\sigma$ (cyan markers); open datapoints indicate measurements excluded from the fit.}
\label{fig:ptp_radio_xray_60arcsec}
\end{center}
\end{figure*}

To probe this correlation, we cover the entire cluster with boxes of 60\,arcsec in all areas above an X-ray surface brightness of $2 \times 10^{-5}$\,counts\,s$^{-1}$. We then remove any regions which are contaminated by significant emission associated with X-ray point sources or radio galaxies; in the case of our maps, this is almost exclusively the regions contaminated by residuals from the embedded tailed radio galaxies T1 and T2, as all other sources have been well-subtracted. The region set used to probe the thermal/non-thermal connection is displayed in Figure~\ref{fig:ptp_radio_xray_60arcsec}, where the boxes are colourised according to whether the median surface brightness in the region is at least $2\sigma$, in which case it is considered a measurement (cyan boxes), or whether it is below $2\sigma$, in which case it is considered an upper-limit with a value of $2\sigma$ (blue boxes).

Figure~\ref{fig:ptp_radio_xray_60arcsec} presents the $I_{\rm R}/I_{\rm X}$ correlation plane for Abell~2142. The non-thermal/thermal components appear well-correlated at both 1283\,MHz and 143\,MHz; indeed, we find a strong correlation with Spearman and Pearson coefficients ($r_{\rm S}$ and $r_{\rm P}$, respectively) of $r_{\rm S} = r_{\rm P} = 0.91$ at 1283\,MHz and $r_{\rm S} = 0.92$ and $r_{\rm P} = 0.94$ at 143\,MHz.

We then proceed to quantify the slope of the correlation, fitting a power-law relation in log-log space using the functional form:
\begin{equation}\label{eq:ptp_correlation}
    {\rm{log}}(I_{\rm{R}}) = c + b \, {\rm{log}}(I_{\rm{X}}),
\end{equation}

We used the MCMC implementation of \texttt{Linmix}\footnote{Currently available at \url{https://linmix.readthedocs.io/en/latest/src/linmix.html}.} \citep{Kelly2007_LINMIX} to determine the optimal values of $b$ and $c$ at each frequency. While other programs exist \citep[e.g. the `point-to-point trend extractor', \texttt{PT-REX};][]{Ignesti2022_PTREX} we opt to use \texttt{Linmix} as it performs a Bayesian linear regression accounting for measurement uncertainties on both the independent and dependent variables, intrinsic scatter, and upper limits on the dependent variable. It is the ability to properly treat upper limits that makes \texttt{Linmix} our tool of choice for this census \citep{Riseley2022_MS1455,Riseley2023_A1413} and other point-to-point studies of haloes \citep[e.g.][]{Rajpurohit2021_MACSJ0717_Halo,Rajpurohit2021_A2744,Riseley2022_A3266} as frequently the available radio data is of inhomogeneous sensitivity and often the diffuse radio emission recovered from the halo does not cover the full extent of the thermal ICM. Thus, using \texttt{Linmix} also enables a direct comparison with other works in the literature.

\begin{table}
\small
\renewcommand{\arraystretch}{1.4}
\centering
\caption{Summary of our point-to-point correlation fit results performed using \texttt{Linmix}. Relations were fit between the X-ray surface brightness and either the radio surface brightness at the listed frequency or spectral index (each at 60\,arcsec resolution) as indicated in the first column. Here $b$ is the best-fit correlation slope, and $r_{\rm{S}}$ and $r_{\rm{P}}$ are the Spearman and Pearson correlation coefficients, respectively. The `full' fits were performed on the full region including upper limits, as presented in the upper row of Figure~\ref{fig:ptp_radio_xray_60arcsec}; the `matched' fits were performed using only regions where both MeerKAT and LOFAR measure a surface brightness of at least $2\sigma$, as described in the text. \label{tab:correlation_results}}
\begin{tabular}{>{\centering\arraybackslash}p{1.25cm} >{\centering\arraybackslash}p{1.25cm} >{\centering\arraybackslash}p{1.25cm} >{\centering\arraybackslash}p{1.25cm} >{\centering\arraybackslash}p{1.25cm} }
\hline
Image & Slope & Intrinsic scatter   & Spearman coeff. & Pearson coeff. \\
      & $b$   & $\sigma_{\rm int}$  & $r_{\rm{S}}$    & $r_{\rm{P}}$ \\
\hline\hline
1283~MHz (full)       & $1.12^{+0.06}_{-0.05}$ & $0.021^{+0.007}_{-0.005}$  &  0.91 &   0.91  \\
143~MHz (full)        & $0.76^{+0.02}_{-0.02}$ & $0.011^{+0.002}_{-0.002}$  &  0.92 &   0.94    \\
\hline
1283~MHz (matched)    & $0.88^{+0.06}_{-0.05}$ & $0.017^{+0.005}_{-0.004}$  &  0.91 &   0.91    \\ 
143~MHz (matched)     & $0.81^{+0.04}_{-0.05}$ & $0.011^{+0.003}_{-0.003}$  &  0.94 &   0.93    \\ 
\hline
$\alpha$              & $0.07^{+0.03}_{-0.03}$ & $-$  &  0.17 &   0.26    \\
\hline
\end{tabular}
\end{table}

Table~\ref{tab:correlation_results} reports the results of our fitting routine, where we use the $50^{\rm th}$ percentile and the $16^{\rm th}$/$84^{\rm th}$ percentiles to define the best-fit and uncertainties respectively. Our fitting routine yields a slope of $b_{\rm 1283 \, MHz} = 1.12^{+0.06}_{-0.05}$ for MeerKAT at 1283\,MHz and $b_{\rm 143 \, MHz} = 0.76^{+0.02}_{-0.02}$ for LOFAR at 143\,MHz. These fits are overlaid on the central panels of Figure~\ref{fig:ptp_radio_xray_60arcsec}. The slope of our point-to-point correlation derived at 143\,MHz is broadly consistent with the results reported by \cite{Bruno2023_A2142}, who derived a slope of $b_{\rm 143 \, MHz} = 0.82 \pm 0.03$ using the same dataset, albeit different images at a different resolution (128\,arcsec) and using a different gridding and brightness measurement method (\texttt{PT-REX}).

We also note that \texttt{Linmix} also allows us to estimate the intrinsic scatter ($\sigma_{\rm int}$) of our data. When considering this full region set, we find that the $\sigma_{\rm int}$ is almost a factor two greater at 1283\,MHz than 143\,MHz, as $\sigma_{\rm int, \, 1283 \, MHz} = 0.021^{+0.007}_{-0.005}$ and $\sigma_{\rm int, \, 143 \, MHz} = 0.011^{+0.002}_{-0.002}$.

However, given the difference in halo extent we re-ran our fitting routine at both frequencies, this time selecting only regions where both MeerKAT and LOFAR measure a surface brightness of at least $2\sigma$, and excluded upper limits from the fitting routine. These fits are shown in the lower panels of Figure~\ref{fig:ptp_radio_xray_60arcsec}. This time, both MeerKAT and LOFAR show a \textit{sub}-linear correlation slope, with LOFAR measuring a slope of $b_{\rm 143 \, MHz} = 0.88^{+0.06}_{-0.06}$ and MeerKAT measuring a slope of $b_{\rm 1283 \, MHz} = 0.81^{+0.05}_{-0.05}$. When considering this matched region set, the intrinsic scatter decreases marginally at 1283\,MHz --- although it is still consistent within the uncertainties --- and remains unchanged at 143\,MHz; we find $\sigma_{\rm int, \, 1283 \, MHz} = 0.017^{+0.005}_{-0.004}$ and $\sigma_{\rm int, \, 143 \, MHz} = 0.011^{+0.003}_{-0.003}$.

These values, along with the correlation coefficients at each frequency, are also shown in Table~\ref{tab:correlation_results}. Hence, when considering only this matched region set, we see a consistent sub-linear slope at both 1283\,MHz and 143\,MHz. This indicates that the super-linear slope seen at 1283\,MHz is strongly influenced by the sensitivity of the current MeerKAT data, and that deeper observations would be required to probe further into the cluster outskirts.

\begin{figure}
\begin{center}
\includegraphics[width=0.99\linewidth]{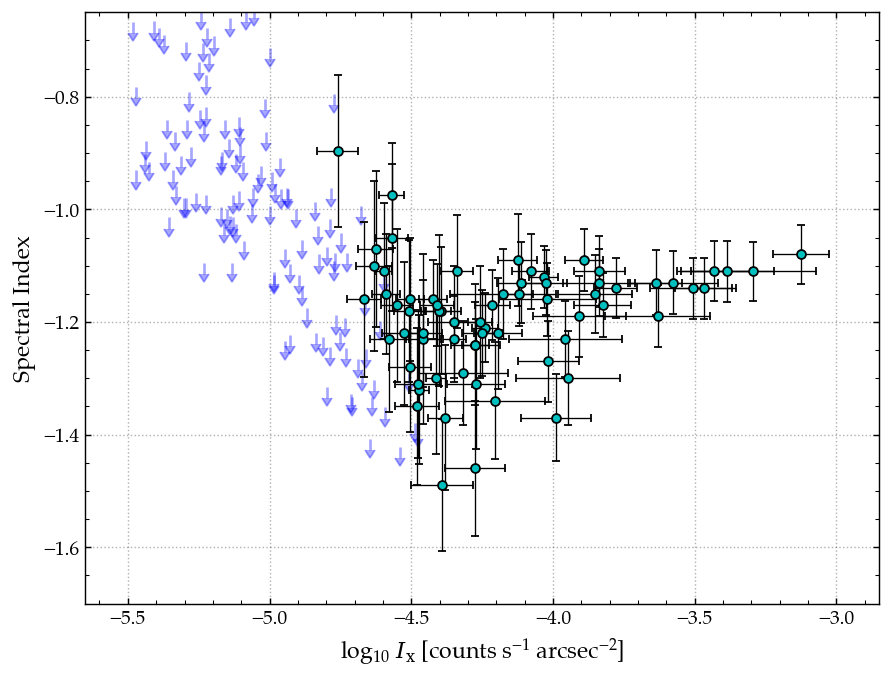}
\caption{Radio spectral index/X-ray surface brightness correlation $(\alpha^{1283~{\rm{MHz}}}_{143~{\rm{MHz}}} / I_{\rm X})$ for Abell~2142 at 60~arcsec resolution. Note the inverted $y$-axis, to facilitate comparison with previous similar studies. The strength of the correlation is negligible, with Spearman (Pearson) coefficients $r_{\rm S} = 0.17$ $(r_{\rm P} = 0.26)$. Blue markers indicate upper limits to the spectral index, which were excluded from the fitting and are shown here for illustrative purposes only. }
\label{fig:ptp_alfa}
\end{center}
\end{figure}

\subsubsection{Point-to-point correlation: halo spectral index}
Our high-quality multi-frequency data allow us to further investigate the correlation between the X-ray surface brightness and the radio spectral index, the $\alpha/I_{\rm X}$ correlation. Limits are difficult to account for in this correlation plane, as they can be either upper limits or lower limits depending on the relative sensitivity of radio datasets and the physical processes at work. Figure~\ref{fig:A2142_radio_alfa} demonstrates that for all regions where the MeerKAT surface brightness is above $3\sigma$, the surface brightness measured by LOFAR at 143\,MHz is likewise above $3\sigma$. In any case, we make use only of regions where we have spectral index measurements.

In Figure~\ref{fig:ptp_alfa} we show the $\alpha/I_{\rm X}$ correlation plane. We also plot upper limits throughout the rest of the radio halo using the measured 143\,MHz LOFAR surface brightness and a $2\sigma$ upper limit to the MeerKAT surface brightness at 1283\,MHz (blue arrows) for reference, although these limits were excluded from the fitting routine. The spectral index measurements appear to show no strong single correlation; indeed when taking all measurements we find Spearman (Pearson) coefficients of $r_{\rm S} = 0.17$ $(r_{\rm P} = 0.26)$, consistent with negligible correlation.

\begin{figure*}
\begin{center}
\includegraphics[width=0.95\linewidth]{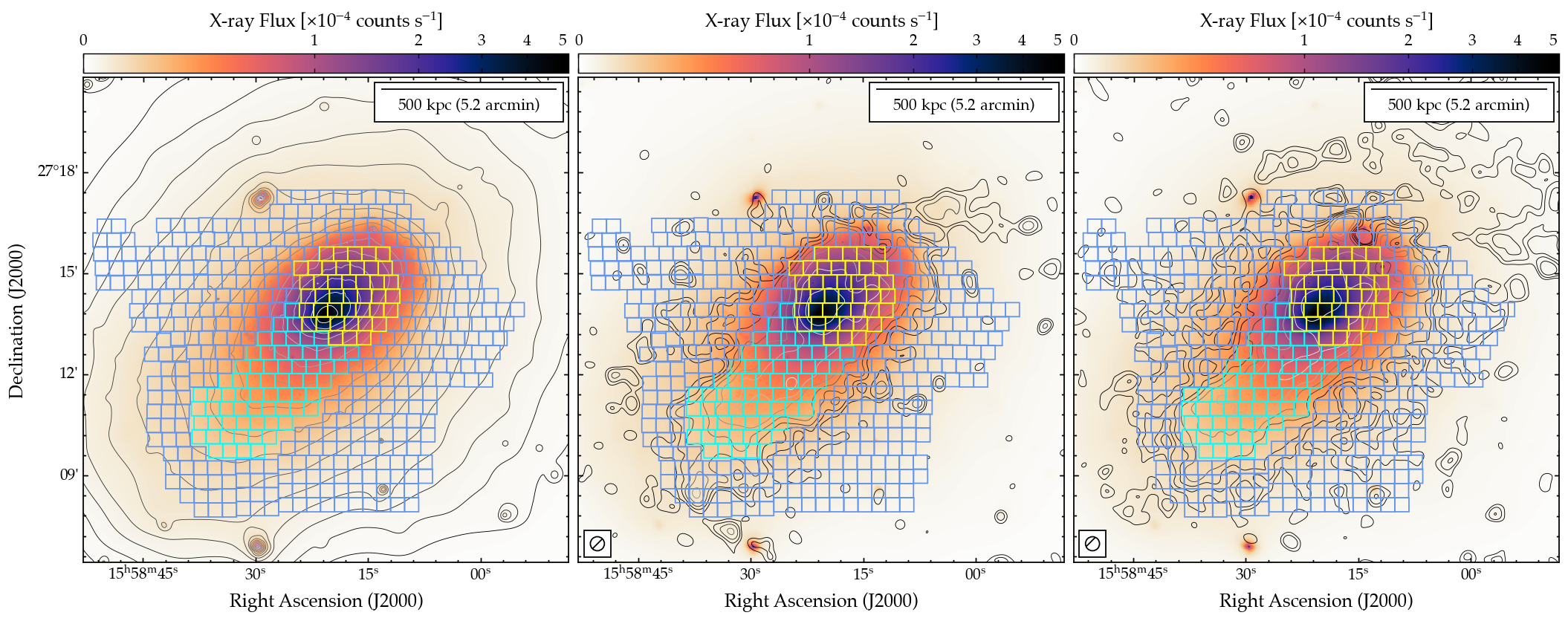}
\includegraphics[width=0.95\linewidth]{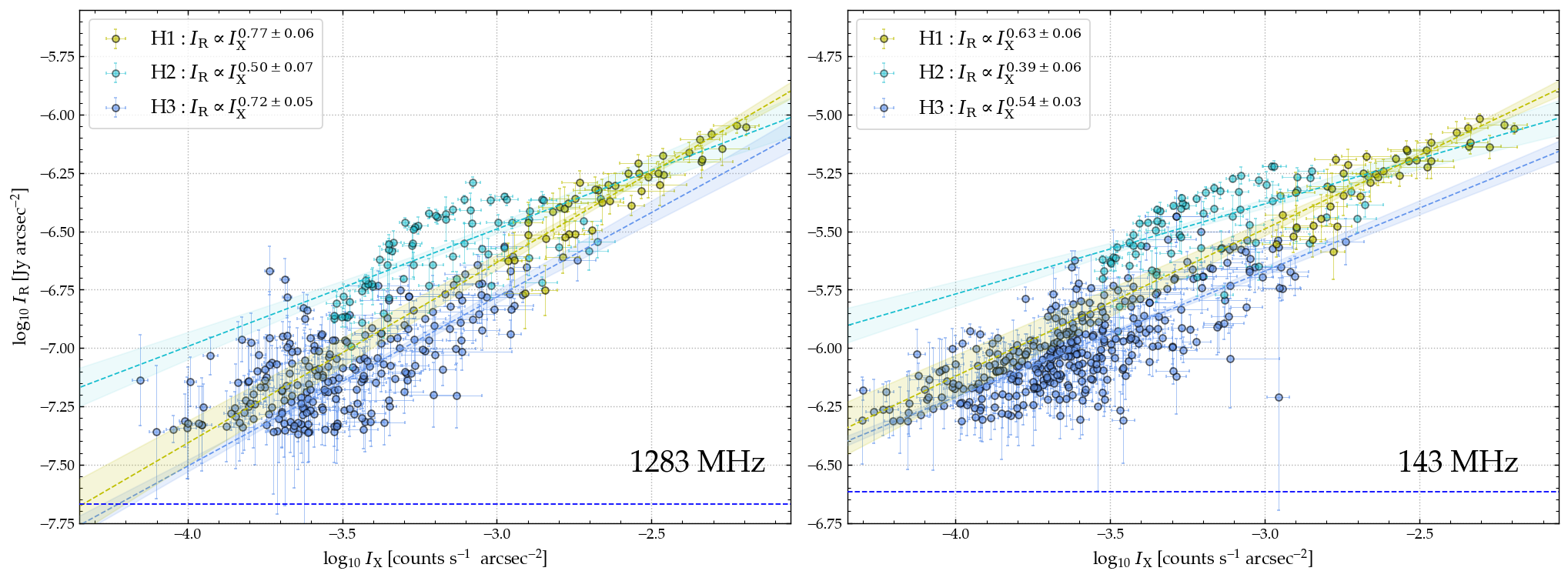}
\caption{The point-to-point thermal/non-thermal correlation for Abell~2142 at 25\,arcsec resolution. Top panels show the 25\,arcsec boxes used to probe the correlation, colourised according to the halo component (H1:yellow, H2:cyan, H3:powder blue). Colourscale shows the \textit{XMM-Newton} surface brightness, with contours denoting the surface brightness measured by \textit{XMM-Newton}, MeerKAT, and LOFAR (both at 25\,arcsec resolution). Radio contours start at $3\sigma$ and scale by a factor $\sqrt{2}$ as per Figure~\ref{fig:A2142_radio_sub}. Bottom panels show the $I_{\rm R}/I_{\rm X}$ plane for components H1, H2, and H3, as well as the best-fit slope derived by \texttt{Linmix}. The slopes are shown in the inset, as well as in Table~\ref{tab:correlation_results_subregions}.}
\label{fig:ptp_radio_xray_25arcsec_h1h2h3}
\end{center}
\end{figure*}

To quantify the correlation, we fit a power-law relation in log-linear space using the functional form:
\begin{equation}\label{eq:ptp_alfa}
    \alpha = c + b_{\alpha} \, {\rm{log}}(I_{\rm{X}})
\end{equation}

We again used \texttt{Linmix} to derive the best-fit values of $b_{\alpha}$ and $c$. These are reported in Table~\ref{tab:correlation_results}, with the $50^{\rm th}$ and $16^{\rm th}$/$84^{\rm th}$ percentiles used to define the best-fit and uncertainties. \texttt{Linmix} reports a best-fit slope of $b_{\alpha} = 0.07^{+0.03}_{-0.03}$, although we emphasise again that there is negligible overall correlation.

\subsubsection{Point-to-point correlation: halo subregions}
It has been demonstrated that some clusters hosting complex multi-component diffuse radio sources show different point-to-point correlations in different regions, which may highlight different particle acceleration mechanisms and/or different physical conditions. Such examples include the hybrid mini-halo in RXC~J1720.1+2638 \citep{Biava2021_RXCJ1720}, and the radio haloes in Coma \citep[e.g.][]{Bonafede2022_Coma-LOFAR-II} and Abell~2256 \citep{Rajpurohit2023_A2256-Halo}.

Inspired by these previous works, and considering that hints of a non-uniform $I_{\rm R}/I_{\rm X}$ slope were reported by \cite{Bruno2023_A2142} (see their Fig.~13), we investigate the point-to-point correlation in the different components of the radio halo in Abell~2142, namely `H1', `H2', and `H3'. We performed this investigation using our 25\,arcsec resolution maps as the higher resolution allows us to more clearly separate the different components. As previously, we covered the halo region with adjacent boxes of 25\,arcsec on a side, corresponding to a physical scale of 40\,kpc. This resulting region set, as well as the resulting point-to-point correlation plane, is shown in Figure~\ref{fig:ptp_radio_xray_25arcsec_h1h2h3}.

From Figure~\ref{fig:ptp_radio_xray_25arcsec_h1h2h3}, we see that the different components tend to occupy different regions of the $I_{\rm R}/I_{\rm X}$ correlation plane. The brighter halo component, H1, is tightly correlated and traces the brighter X-ray surface brightness regions, whereas the fainter halo component, H3, is more loosely correlated and traces the fainter X-ray surface brightness regions toward the cluster outskirts. Table~\ref{tab:correlation_results_subregions} reports the correlation coefficients and best-fit correlation slopes for these sub-regions.

\begin{table*}
\small
\renewcommand{\arraystretch}{1.4}
\centering
\caption{Summary of our point-to-point correlation fit results performed using \texttt{Linmix} on the sub-regions of the halo, namely H1, H2 and H3. Relations were fit between as indicated in the first column, with the radio data being at 25\,arcsec resolution. Here, $b$ is the best-fit correlation slope, and $r_{\rm{S}}$ and $r_{\rm{P}}$ are the Spearman and Pearson correlation coefficients, respectively. \label{tab:correlation_results_subregions}}
\begin{tabular}{l c c c c c }
\hline
Region & Relation & Slope & Intrinsic scatter & Spearman coeff. & Pearson coeff. \\
       &          & $b$   & $\sigma_{\rm int}$ &  $r_{\rm{S}}$ & $r_{\rm{P}}$ \\
\hline\hline
\multirow{4}{*}{H1} &   $I_{R,\,1283\,{\rm MHz}}/I_{\rm X}$ & $0.77^{+0.06}_{-0.06}$    & $0.001^{+0.001}_{-0.001}$  &   $0.93$ &  $0.90$   \\
                    &   $I_{R,\,143\,{\rm MHz}}/I_{\rm X}$  & $0.63^{+0.06}_{-0.06}$ & $0.002^{+0.001}_{-0.001}$  &   $0.89$ &  $0.87$   \\
                    &   $\alpha / T_{{\rm X},\, Chandra}$                 & $-0.049^{+0.015}_{-0.017}$   & $-$ &   $-0.39$ & $-0.42$   \\
                    &   $\alpha / T_{{\rm X},\, XMM}$                 & $-0.109^{+0.203}_{-0.240}$   & $-$ &   $-0.37$ & $-0.32$   \\
\hline
\multirow{4}{*}{H2} &   $I_{R,\,1283\,{\rm MHz}}/I_{\rm X}$     & $0.50^{+0.07}_{-0.07}$   & $0.014^{+0.003}_{-0.003}$  &    $0.72$ & $0.66$  \\ 
                    &   $I_{R,\,143\,{\rm MHz}}/I_{\rm X}$      & $0.39^{+0.06}_{-0.06}$   & $0.011^{+0.003}_{-0.002}$  &    $0.61$ & $0.61$  \\ 
                    &   $\alpha / T_{{\rm X},\, Chandra}$                     & $-0.004^{+0.018}_{-0.019}$  & $-$  &  $-0.02$ & $0.00$  \\
                    &   $\alpha / T_{{\rm X},\, XMM}$                 & $+0.039^{+0.101}_{-0.108}$  & $-$  &  $-0.39$ & $-0.42$   \\
\hline
\multirow{2}{*}{H3} &   $I_{R,\,1283\,{\rm MHz}}/I_{\rm X}$     & $0.72^{+0.05}_{-0.04}$    &  $0.026^{+0.004}_{-0.004}$  &  $0.58$ &   $0.64$ \\ 
                    &   $I_{R,\,143\,{\rm MHz}}/I_{\rm X}$      & $0.54^{+0.03}_{-0.03}$    & $0.013^{+0.002}_{-0.002}$   &  $0.71$ &   $0.73$ \\
\hline
\end{tabular}
\end{table*}

Indeed, we find a very strong correlation for H1, as we find correlation coefficients of $r_{\rm S} = 0.93$ and $r_{\rm P} = 0.90$ at 1283\,MHz and $r_{\rm S} = 0.89$ and $r_{\rm P} = 0.87$ at 143\,MHz. For H2 and H3 we find a moderate-to-strong correlation; for H2 we find $r_{\rm S} = 0.72$ and $r_{\rm P} = 0.66$ at 1283\,MHz and $r_{\rm S} = 0.61 = r_{\rm P} = 0.61$ at 143\,MHz; for H3, $r_{\rm S} = 0.58$ and $r_{\rm P} = 0.64$ at 1283\,MHz and $r_{\rm S} = 0.71$ and $r_{\rm P} = 0.73$ at 143\,MHz. 

We used \texttt{Linmix} to fit for the correlation slope, the results of which are presented in Table~\ref{tab:correlation_results_subregions} and overlaid in Figure~\ref{fig:ptp_radio_xray_25arcsec_h1h2h3}. The correlation is well-characterised by a sub-linear slope at all frequencies in all regions H1, H2 and H3. However, the slope is not constant, but rather changes between frequencies and between regions. 

For H1 we find a slope of $b_{\rm 1283 \, MHz} = 0.77^{+0.06}_{-0.06}$ and $b_{\rm 143 \, MHz} = 0.623^{+0.06}_{-0.06}$. The intrinsic scatter is also extremely low, with $\sigma_{\rm int} = 0.001 \, (0.002)$ at 1283\,MHz (143\,MHz). Proceeding to the south-east of Abell~2142, H2 shows a significantly flatter slope of $b_{\rm 1283 \, MHz} = 0.50^{+0.07}_{-0.07}$ and $b_{\rm 143 \, MHz} = 0.39^{+0.06}_{-0.06}$. In this region, the intrinsic scatter is around an order of magnitude larger, with $\sigma_{\rm int} = 0.014 \, (0.011)$ at 1283\,MHz (143\,MHz).

Finally, for the larger-scale component H3, we find the slope steepens again to $b_{\rm 1283 \, MHz} = 0.72^{+0.05}_{-0.04}$ and $b_{\rm 143 \, MHz} = 0.54^{+0.03}_{-0.03}$. For H3, the intrinsic scatter increases further, with $\sigma_{\rm int} = 0.026 \, (0.013)$ at 1283\,MHz (143\,MHz). The uncertainties on the intrinsic scatter are sufficiently small that each of these increases is statistically significant.

Overall there is no strong systematic trend in the slope proceeding from H1 through to H3, although the steepening of the correlation slope in each region toward higher frequency suggests spectral steepening. The significant increase in the intrinsic scatter $\sigma_{\rm int}$ and weakening correlation coefficient moving from H1 to H2 and H3 may be interpreted in a variety of ways, although it provides further support to the interpretation for spectral steepening; similarly, the difference of the correlation slope for different regions can be caused by the interaction of several different processes, which we will discuss further in Section~\ref{sec:discussion}. 

\subsubsection{Pseudo-point-to-point correlation: spectral index and X-ray temperature}
Continuing on from our investigation of the $I_{R}/I_{\rm X}$ point-to-point correlation for the three different halo components, which reveals that H1 and H2 show significantly different correlation slopes, we also examined the correlation between radio spectral index and ICM temperature for the three halo components.

We used temperature maps derived from both \textit{Chandra} and \textit{XMM-Newton} for this analysis, respectively presented by \cite{Bruno2023_A2142} and \cite{Rossetti2013_Abell2142}. The available \textit{Chandra} data do not cover the full extent of the halo, meaning temperature measurements are not available for the vast majority of H3 and around 40\% of H2; conversely, the \textit{XMM-Newton} data cover a wider area, allowing us to probe the full extent of H2. We omit H3 from this investigation as neither \textit{Chandra} nor \textit{XMM-Newton} cover the full extent, nor do we recover sufficient extent with MeerKAT at 1283\,MHz to meaningfully probe the spectral index.

\begin{figure*}
\begin{center}
 \includegraphics[width=0.8\linewidth]{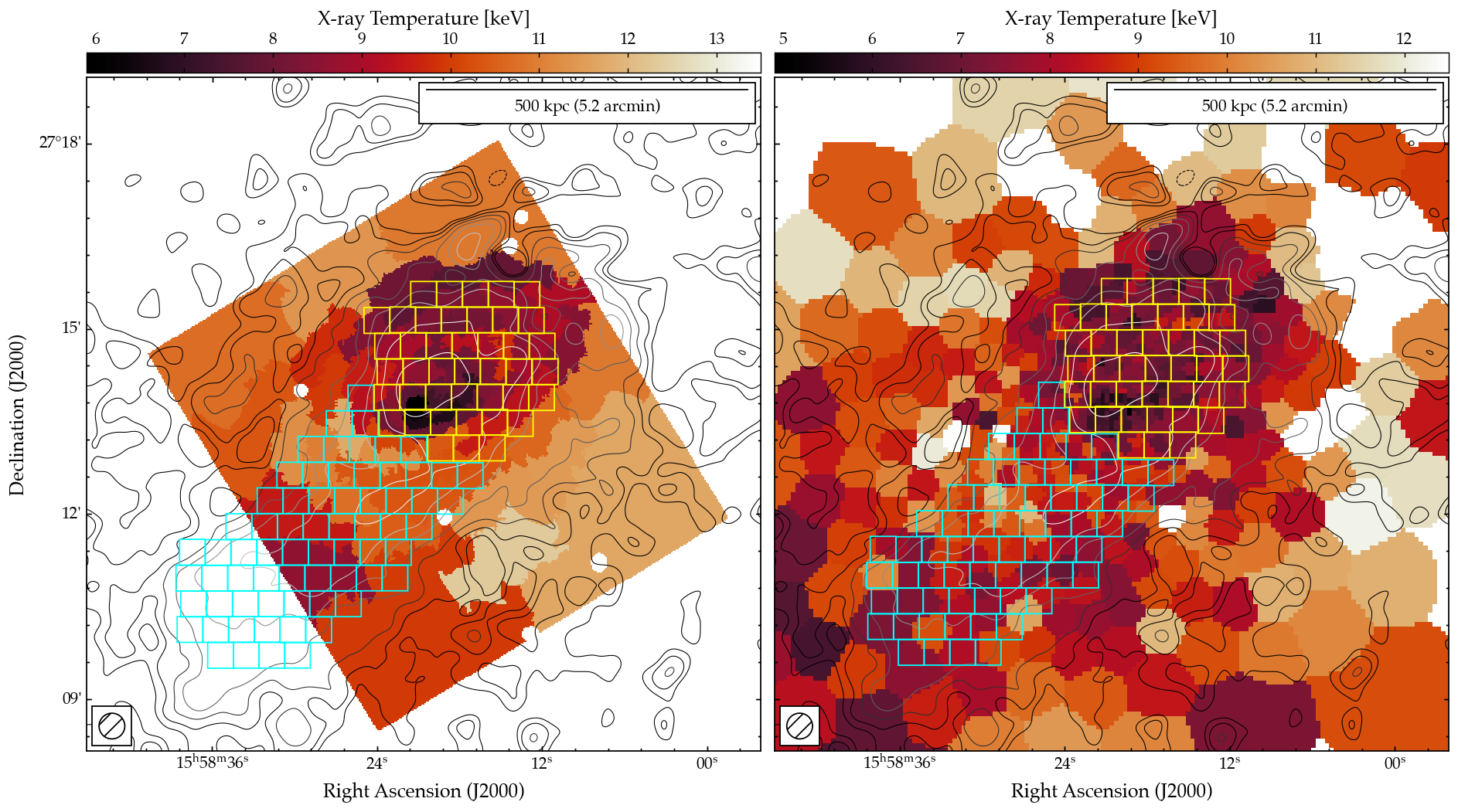}
 \includegraphics[width=0.45\linewidth]{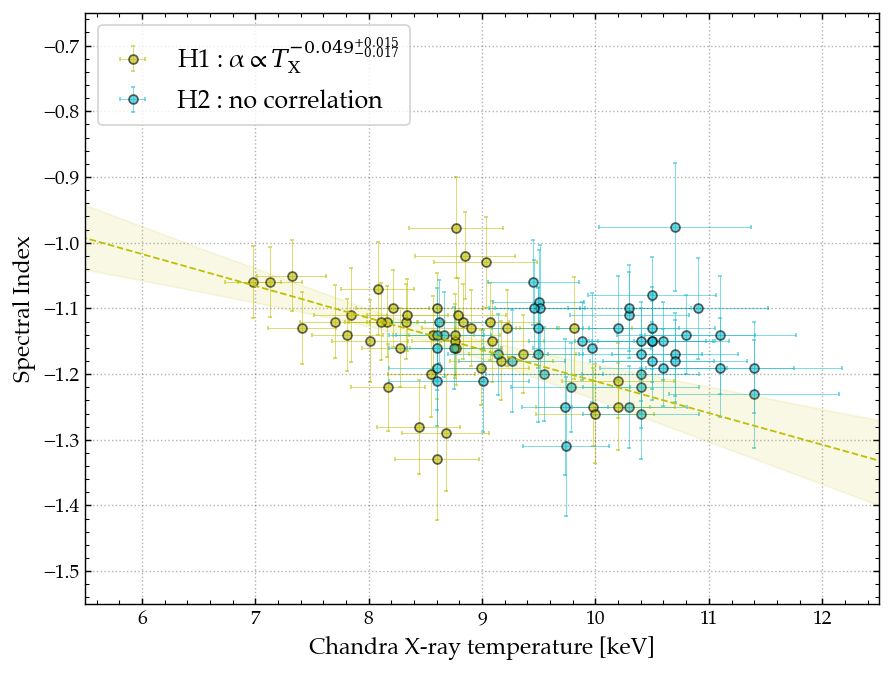}
 \includegraphics[width=0.45\linewidth]{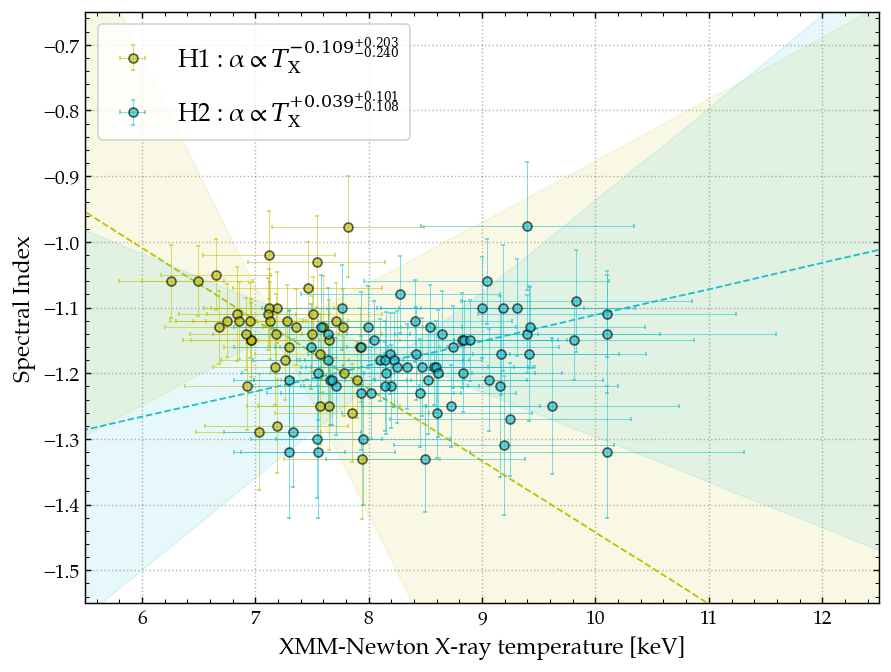}
\caption{\textit{Upper:} X-ray temperature maps derived from \textit{Chandra} (\textit{left}) and \textit{XMM-Newton} (\textit{right}) observations. Contours show 143\,MHz LOFAR data at 25\,arcsec resolution. Coloured boxes show the regions covering H1 (yellow) and H2 (cyan) used to probe the $\alpha / \mathbf{T_{\rm X}}$ correlation. \textit{Lower:} point-to-point correlations between radio spectral index and X-ray temperature measured by \textit{Chandra} (left) and \textit{XMM-Newton} (right). The dashed lines indicate the best-fit relation, with the shaded region denoting the $1\sigma$ uncertainty; the slopes are shown in the inset in each subplot.}
\label{fig:ptp_alfa_kt_h1h2}
\end{center}
\end{figure*}

Figure~\ref{fig:ptp_alfa_kt_h1h2} shows the ICM temperature maps from \textit{Chandra} and \textit{XMM-Newton}, overlaid with LOFAR contours at 143\,MHz for reference, and the regions used to quantify the $\alpha / T_{\rm X}$ correlation. The lower panels of Figure~\ref{fig:ptp_alfa_kt_h1h2} show the results of our point-to-point correlation analysis for both \textit{Chandra} and \textit{XMM-Newton}, as well as the correlation between X-ray temperature measurements derived from both instruments. This third plot is used to verify our results. We again used \texttt{Linmix} to derive the best-fit correlation and quantify the uncertainty.

From the point-to-point analysis, we see that for H1 (yellow boxes and datapoints) there is an anti-correlation between the X-ray temperature and the radio spectral index; that is to say that hotter regions show a steeper radio spectrum. We find a moderate anticorrelation, with Spearman (Pearson) coefficients of $r_{\rm S} = -0.39$ $(r_{\rm P} = -0.42)$ when using the \textit{Chandra} temperature map and $r_{\rm S} = -0.37$ $(r_{\rm P} = -0.32)$ when using \textit{XMM-Newton} temperature map. The slope is $b = -0.049^{+0.015}_{-0.017}$ for \textit{Chandra} and $b = -0.109^{+0.203}_{-0.240}$ when using \textit{XMM-Newton}. While the uncertainty in the fitted slope is large for the \textit{XMM-Newton} temperature map, the correlation is significant and shows the same behaviour as the relation using \textit{Chandra}.

For H2 (cyan boxes and datapoints) we find contrasting behaviour: the $\alpha / T_{\rm X}$ plane shows no correlation when using \textit{Chandra} temperature map $(r_{\rm S} = -0.02$, $r_{\rm P} = 0.00)$, but a positive correlation is found when using \textit{XMM-Newton} temperature map. In the case of \textit{XMM-Newton}, the correlation is weak $(r_{\rm S} = 0.30$, $r_{\rm P} = 0.28)$ and the fitted slope again shows large uncertainty $b = 0.039^{+0.101}_{-0.108}$, but the data do show a correlation. We attribute this difference primarily to the difference in coverage: as is visible in the left-hand panel of Figure~\ref{fig:ptp_alfa_kt_h1h2}, the \textit{Chandra} observations do not cover the full extent of H2. As such, we place more confidence in the \textit{XMM-Newton} point-to-point analysis of H2, although we emphasise that these results in the $\alpha/T_{\rm X}$ correlation are tentative and due to the large uncertainties no firm conclusions can be drawn.

\section{Discussion: on the nature of the halo in Abell 2142}\label{sec:discussion}
\subsection{On the radial profiles}
Our radial profiles (Figure~\ref{fig:profiles_60arcsec}, Table~\ref{tab:radial_profiles_60arcsec}) allow us to clearly confirm the presence of two components oriented along a north-east/south-west axis, perpendicular to the primary merger axis along which the original two components were detected by \cite{Venturi2017_A2142}. For the inner component we find consistent e-folding radius at both 143\,MHz and 1283\,MHz, although the inner component appears to be a factor $\sim2$ larger to the SW than the NE. Given that we took the peak of the halo brightness as the origin point for our radial profiles, this perhaps suggests asymmetry rather than a position error on our part. 

We also find that the outer component is significantly more extended at 143\,MHz compared to 1283\,MHz, having a $\sim70\%$ larger e-folding radius to the NE, although we do not find significant evidence of a second component to the SW at 1283\,MHz. This is consistent with the asymmetry in the halo evident at 1283\,MHz but not at 143\,MHz in Figure~\ref{fig:A2142_radio_sub}, and may reflect asymmetry in the spectral index. Such an interpretation is reflected in the radial spectral index profiles of Figure~\ref{fig:profiles_60arcsec}, where the spectral index steepens significantly with increasing radius toward the SW but not appreciably toward the NE. Overall, deeper high-frequency follow-up is strongly motivated.

Conversely, the picture is different along the SE wedge. While two components are clearly visible in our radio continuum maps (Figure~\ref{fig:A2142_radio_sub}) and have been previously reported in the literature \citep{Venturi2017_A2142,Bruno2023_A2142}, the radial profiles at both 143\,MHz and 1283\,MHz favour a single-component exponential profile at both 60\,arcsec resolution and 25\,arcsec resolution (though 25\,arcsec profiles are not shown in this paper). Moreover, the profiles do not show even tentative evidence of departure from a single-exponential profile.

However, it is worth noting that the best-fit e-folding radius is large, a factor $2-4$ greater than that recovered for the inner component along the SW and NE wedges; it is consistent with the e-folding radius recovered for the outer component along the NE arc at 143\,MHz. This may suggest a more complex distribution than two clearly distinct components distributed in the plane of the sky. This would appear to be supported by the spectral index profile, which shows no evidence of steepening with increasing radial distance from the halo surface brightness peak.

\subsection{On the point-to-point correlations}
Our results demonstrate that the slope of the $I_{\rm R}/I_{\rm X}$ correlation steepens significantly between 143\,MHz and 1283\,MHz, although these quantities remain very strongly correlated at both frequencies. The steepening of the correlation slope implies spectral steepening further away from the cluster centre, and the sub-linear slope recovered at both frequencies --- when considering the matched region set for the most rigorous interpretation of our data --- implies that the non-thermal surface brightness declines less rapidly than the thermal surface brightness.

Relatively few radio haloes have the high-quality data required to investigate the $I_{\rm R}/I_{\rm X}$ correlation in detail across a broad frequency range, and of those that do, the picture is mixed. Some clusters show a change in the correlation slope with frequency, although \cite{Rajpurohit2021_MACSJ0717_Halo} report a correlation slope that \textit{steepens} toward higher frequencies for MACS~J0717.5+3745, whereas the radio haloes in CLG~0217+70 and Abell~2256 show correlation slopes that \textit{flatten} toward higher frequencies \citep[respectively][]{Hoang2021_CLG0217,Rajpurohit2023_A2256-Halo}. Conversely, some cluster radio haloes show no significant frequency dependence in the $I_{\rm R}/I_{\rm X}$ correlation slope \citep[e.g.][]{Hoang2019_A520,Rajpurohit2021_A2744}.

For the different regions in the halo, namely H1, H2 and H3, we observe that H1 and H3 show a broadly consistent slope in the $I_{\rm R}/I_{\rm X}$ correlation, whereas H2 shows a significantly flatter $I_{\rm R}/I_{\rm X}$ correlation. All regions show a markedly steeper slope at higher frequencies, suggesting that the spectrum of all components steepens toward higher frequencies. The intrinsic scatter increases dramatically, and the correlation strength decreases significantly, moving from H1 to H2 to H3. This difference in sub-regions of the halo is similar to the behaviour in Abell~2744, where \cite{Rajpurohit2021_A2744} observed differing correlations in the northern and southern components of the radio halo; in the case of Abell~2142 it may indicate a transition to an increasingly turbulent and/or inhomogeneous ambient environment.

As for the $\alpha/I_{\rm X}$ correlation plane, the interpretation is non-trivial. Overall, there is no significant correlation, although our data tentatively suggest that the spectral index is steeper and shows more significant fluctuations in regions of fainter X-ray flux, i.e. toward the cluster outskirts. However, as can be seen from the many upper limits, significantly deeper high-frequency data is required to probe this correlation in more detail and test this hypothesis. Such a picture would be consistent with many other haloes \citep[e.g.][]{Rajpurohit2021_MACSJ0717_Halo,Rajpurohit2023_A2256-Halo,Bonafede2022_Coma-LOFAR-II,Santra2023_A521}.

In the literature, very few haloes possess sufficiently high-quality X-ray and multi-frequency radio observations to compare Abell~2142 with, and such studies have been performed on only a handful of clusters. From these studies, there is no general consensus: Abell~2255 shows a weak anti-correlation between X-ray temperature and radio spectral index \citep{Botteon2020_Abell2255}, much like H1. Conversely, the halo component associated with the primary sub-cluster in the complex system Abell~2744 also shows a weak positive correlation \citep{Rajpurohit2021_A2744}, and similarly a weak positive correlation between X-ray temperature and radio spectral index was found for Abell~521 by \cite{Santra2023_A521}. Regardless, we emphasise that in all of these aforementioned clusters, the interpretation is the same: \textit{hotter} regions of the ICM tend to show \textit{flatter} radio spectral index; this is the same behaviour that we observe for H2 in Abell~2142.

For H1 however, we see the converse behaviour --- an anti-correlation --- implying that hotter regions of the ICM tend to show a \textit{steeper} radio spectral index. This result is confirmed by both \textit{Chandra} and \textit{XMM-Newton}, lending support to its veracity, although to the best of our knowledge this is the first case in which such an anti-correlation has been reported in the literature.

\subsection{The overall picture}
Overall, our results demonstrate that the radio halo in Abell~2142 is highly complex. We confirm the presence of multiple components through our analysis of maps, radio profiles, and point-to-point correlations. Any scenario that attempts to explain our results must replicate the following observables:
\begin{itemize}
    \item The presence of multiple components within the radio halo, which show different spatial distributions.
    \item An integrated spectrum that follows no departures from a single steep-spectrum power-law behaviour up to the highest frequency of our data, but shows an increasingly steep spectrum proceeding from H1 to H2 and H3.
    \item A spectral index profile that is broadly homogeneous but shows evidence of (i) steepening and (ii) increased variance towards larger radii, at least towards the SW and potentially the NE of the cluster centre.
    \item A sub-linear point-to-point correlation between $I_{\rm R}/I_{\rm X}$ which shows (i) a steeper slope at higher frequencies and (ii) different correlations in different regions of the halo.
    \item Both an anti-correlation and a positive correlation between the X-ray temperature and radio spectral index in different regions of the halo.
\end{itemize}

These results are most naturally explained in a scenario where turbulence plays a significant role in generating the multi-component radio halo \citep[e.g.][]{Brunetti2001,Brunetti2008_A521}. 

Turbulence in clusters is an intermittent process both temporally and spatially, linked to ongoing dynamical activity --- which could arise from mergers or sloshing, for example, although AGN feedback into the ICM via jets and outflows also plays a role. These motions are sub-sonic but super-Alfv{\'e}nic and so turbulent energy is typically injected on scales comparable to the cluster core, of the order of a few hundred kpc, and cascades down onto smaller scales generating turbulent structures on small, resonant scales \citep[see e.g.][and references therein]{Brunetti_Jones_2014}.

As such, ICM turbulence is highly inhomogeneous. The injection and evolution of turbulent energy will have a dependence on the local dynamics and physical conditions, which will then dictate the thermal/non-thermal connection. Thus, depending on the size of the region considered --- whether the whole cluster, or sub-regions, for example --- one may expect to see different spatial components with different relations to the ICM.

It is also important to consider that as we are dealing with column-integrated quantities, it is highly likely that we are mixing different turbulent structures along the line of sight. More stringent constraints and measurements of the turbulent motions along the line of sight would be required to break this degeneracy, but this requires high spectral resolution X-ray information \citep[e.g.][]{Inogamov_Sunyaev_2003,Clerc2019_ICM_Xray_mapping,Zhang2023_arXiv}. Currently only the \textit{X-ray Imaging and Spectroscopy Mission} \citep[XRISM;][]{Ishisaki2018_XRISM} provides this capability.

While the more homogeneous turbulent (re-)acceleration models generally reproduce the statistical halo properties \citep[e.g.][and references therein]{Cassano2023_LoTSS_Planck}, resolved studies like ours have shown that the realities are far more complex. Reconciling the power-law integrated spectrum with the evidence of spectral steepening provided by our radial profiles and point-to-point correlations is relatively straightforward if we invoke an inhomogeneous turbulent (re-)acceleration framework. Inhomogeneities in the magnetic field topography and/or acceleration timescale within the emitting volume may act to smooth and blur spectral discontinuities when averaging over the full volume of the region.

Invoking this scenario also provides a natural explanation for the spectral steepening, presence of multiple components, and the sub-linear $I_{\rm R}/I_{\rm X}$, as well as the positive $\alpha / T_{\rm X}$ correlation for H2. The presence of a cold front 1\,Mpc from the cluster centre toward the SE \citep{Rossetti2013_Abell2142} supports this interpretation, suggesting the presence of turbulence --- or at least previous turbulence during the formation of the cold front --- along the region of H2.

However, on the face of it the anti-correlation between $\alpha / T_{\rm X}$ for H1 is more challenging to explain within a turbulent acceleration scenario. It is also worth noting that the average temperature within the region of H1 is significantly lower than that measured for H2 (see Figure~\ref{fig:ptp_alfa_kt_h1h2}); this, combined with the different correlations between $I_{\rm R}/I_{\rm X}$ and $\alpha / T_{\rm X}$ suggest that we are not simply seeing an extension of the same phenomenon, but we are perhaps seeing either different evolutionary stages of the same phenomenon and/or we are tracing different physical conditions present in different regions of the ICM. 

The presence of three distinct cold fronts within H1 may provide the basis for an explanation of this anti-correlation. Cold fronts represent bulk motions of cold gas within the ICM, which would provide a source of energy input into the ICM \citep[e.g.][]{Ghizzardi2010} and are linked to core sloshing and particle acceleration in several clusters \citep[e.g.][Biava et al., submitted]{Savini2018,Riseley2022_MS1455}. As such, it follows that in sloshing cool core --- or \textit{remnant} cool core, such as that suggested to exist in Abell~2142 --- we might expect a flatter spectrum where the ICM shows a lower temperature due to the presence of cold fronts. This is the correlation we observe for H1. Further theoretical work involving bespoke simulations, as well as more detailed substructure analysis would be required to understand this further.

\section{Conclusions}\label{sec:conclusions}
This paper is the third in a series presenting results from the `MeerKAT-meets-LOFAR' mini-halo census, a statistically-significant and observationally-homogeneous census of 13 clusters observed with MeerKAT and LOFAR HBA, combined with archival X-ray observations from \textit{XMM-Newton} and/or \textit{Chandra}. In this paper we report on the galaxy cluster Abell~2142, which hosts a complex three-component radio halo previously reported as a potential `mini-halo-plus-halo' type system, and shows evidence of extreme core sloshing at X-ray wavelengths, with three cold fronts in the cluster core and a fourth at a distance of around 1\,Mpc to the south-east of the cluster centre. 

We have presented new observations with MeerKAT L-band (1283\,MHz) combined with recently-published LOFAR HBA observations at 143\,MHz and archival deep \textit{XMM-Newton} and \textit{Chandra} data, which provide a window into the thermal properties of the ICM. Despite its position at Declination around $+27\degree$, our MeerKAT observations achieve high resolution, dynamic range and sensitivity, with a representative off-source rms noise of $7.5 \, \upmu$Jy~beam$^{-1}$ at 10\,arcsec resolution. This excellent imaging performance allows us to detect the many point sources in the region of the cluster, as well as the spectacular embedded tailed radio galaxies which show clear evidence of interaction with the ICM.


After subtracting all radio sources from the field and enhancing our sensitivity to diffuse emission through the use of a $uv$-taper, we map the multi-component halo at 25\,arcsec and 60\,arcsec resolution. This allows us to recover diffuse emission on scales up to $\sim 2$\,Mpc at 143\,MHz; at 1283\,MHz the emission is less extended, up to $\sim 1.5$\,Mpc, reflecting the steep-spectrum nature of the diffuse emission. Our data also confirm the multi-component nature of the halo in Abell~2142, as we clearly recover the full extent of H1 and H2, as well as recovering the brighter component of H3.

We use our new MeerKAT data to extend the spectral modelling from \cite{Bruno2023_A2142}. We confirm that the integrated spectra show single power-law behaviour up to 1283\,MHz, with a spectral index of $\alpha_{\rm H1} = -1.09 \pm 0.03$ for H1, $\alpha_{\rm H2} = -1.15 \pm 0.04$ for H2, and $\alpha_{\rm H3, inner} = -1.30 \pm 0.07$ and $\alpha_{\rm H3, outer} = -1.74 \pm 0.10$ for the inner and outer regions of H3. These results confirm the ultra-steep spectrum nature of H3, as well as demonstrate that H3 shows spectral steepening. We also estimate the total integrated spectrum of H3 to be $\alpha_{\rm H3, \, total} = -1.68 \pm 0.10$. From these integrated spectra, we derive integrated luminosities of $P_{\rm 1.4\,GHz} = (5.21 \pm 1.33) \times10^{23}$~W\,Hz$^{-1}$ at 1.4\,GHz and $P_{\rm 150\,MHz} = (1.29 \pm 0.33) \times10^{25}$~W\,Hz$^{-1}$ at 150\,MHz. Thus, we confirm that Abell~2142 lies low in established power scaling planes between cluster mass and halo radio power, lying around a factor 16 (7.5) below the best-fit correlation relations at 1.4\,GHz (150\,MHz).

Conversely, insofar as we have the 1283\,MHz sensitivity to map them, the resolved spectra are relatively smooth. While they show some fluctuations, there is no clear evidence of gradients which would constitute steepening. However, with the available MeerKAT observations we do not recover the full extent of H3.

We study radial profiles of the radio surface brightness at 1283\,MHz and 143\,MHz as well as the radio spectral index along three different directions: to the south-east, south-west and north-east of the cluster centre, as measured from the surface brightness peak of the halo. Along the SE direction, while we clearly see two components (H1 and H2) in our continuum maps, we only see evidence of a single exponential profile in the surface brightness, and no evidence of spectral steepening. Towards the SW and NE directions --- perpendicular to the main axis of the cluster --- however, we see clear evidence of multiple components in our radial profiles. To the SW we see two components at 143\,MHz and only one component at 1283\,MHz, whereas to the NE we see two components at both 1283\,MHz and 143\,MHz; it is along this direction that we also visibly detect additional emission in our continuum maps. We take this to signify the emergence of H3, which would appear to be highly asymmetric based on these profiles. To the SW the spectral index profile shows clear steepening, whereas only a marginal steepening-then-flattening is seen to the NE.

We perform point-to-point correlation investigations between the radio surface brightness and X-ray surface brightness $(I_{\rm R}/I_{\rm X})$ and radio spectral index and X-ray surface brightness $(\alpha/I_{\rm X})$ for the halo as a whole. For the $I_{\rm R}/I_{\rm X}$ plane, when considering only those regions where both MeerKAT and LOFAR recover flux above the $2\sigma$ level, we recover a sub-linear correlation with a slope $b_{\rm 1283\,MHz} = 0.88^{+0.05}_{-0.05}$ at 1283\,MHz and $b_{\rm 143\,MHz} = 0.81^{+0.04}_{-0.05}$ at 143\,MHz. We see a tentative, but very weak, correlation in the $\alpha/I_{\rm X}$ plane, with lower $I_{\rm X}$ tracing regions with (i) steeper radio spectrum and (ii) increasing spectral variance.

We also investigate the point-to-point correlations for H1, H2 and H3 separately, studying the radio surface brightness/X-ray surface brightness $(I_{\rm R}/I_{\rm X})$ plane and the correlation between radio spectral index and X-ray temperature $(\alpha/T_{\rm X})$. For each region, we find a sub-linear $I_{\rm R}/I_{\rm X}$ correlation that steepens at higher frequencies, signifying spectral steepening; we note that H1 and H3 show a very similar $I_{\rm R}/I_{\rm X}$ correlation slope, but H2 shows a markedly flatter $I_{\rm R}/I_{\rm X}$ correlation. We are able to study the $\alpha/T_{\rm X}$ correlation only for H1 and H2 due to the available X-ray data. However, we find that H1 shows a moderate anti-correlation --- that is, hotter regions show steeper radio spectrum --- while H2 shows the inverse, a moderate positive correlation: hotter regions show flatter radio spectrum.

Finally, we attempt to interpret what our results tell us about the nature of the multi-component radio halo in Abell~2142 in the context of particle acceleration mechanisms. Many of our results are naturally replicated by an interpretation of the turbulent (re-)acceleration framework, although we require some combination of inhomogeneities in magnetic field topography, acceleration timescale and/or ambient environment to provide a better explanation. However, the \textit{anti}-correlation between $\alpha/T_{\rm X}$ for H1 (the `core') appears challenging to explain within a typical turbulent acceleration scenario. The typically lower temperature within this region, as well as the presence of three cold fronts within this region, may provide the basis for investigating this further, and we consider it likely that we are seeing either different evolutionary stages of the same physical phenomenon and/or tracing different microclimates in different regions. However, bespoke simulations and theoretical work would be required to investigate this further, and deeper study of this rich and intriguing cluster --- including deeper high-frequency observations with MeerKAT and/or the uGMRT --- is strongly motivated.

\section*{Acknowledgements}
CJR, A.~Bonafede and NB acknowledge financial support from the ERC Starting Grant `DRANOEL', number 714245. A.~Botteon acknowledges financial support from the European Union - Next Generation EU. EB acknowledges support from DFG FOR5195. FL acknowledges financial support from the Italian Minister for Research and Education (MIUR), project FARE, project code R16PR59747, project name FORNAX-B. FL acknowledges financial support from the Italian Ministry of University and Research $-$ Project Proposal CIR01$\_$00010.

The MeerKAT telescope is operated by the South African Radio Astronomy Observatory, which is a facility of the National Research Foundation, an agency of the Department of Science and Innovation. We wish to acknowledge the assistance of the MeerKAT science operations team in both preparing for and executing the observations that have made our census possible.

LOFAR is the Low Frequency Array designed and constructed by ASTRON. It has observing, data processing, and data storage facilities in several countries, which are owned by various parties (each with their own funding sources), and which are collectively operated by the ILT foundation under a joint scientific policy. The ILT resources have benefited from the following recent major funding sources: CNRS-INSU, Observatoire de Paris and Universit\'e d'Orl\'eans, France; BMBF, MIWF-NRW, MPG, Germany; Science Foundation Ireland (SFI), Department of Business, Enterprise and Innovation (DBEI), Ireland; NWO, The Netherlands; The Science and Technology Facilities Council, UK; Ministry of Science and Higher Education, Poland; The Istituto Nazionale di Astrofisica (INAF), Italy.

This research made use of the Dutch national e-infrastructure with support of the SURF Cooperative (e-infra 180169) and the LOFAR e-infra group. This work is co-funded by the EGI-ACE project (Horizon 2020) under Grant number 101017567. The J\"{u}lich LOFAR Long Term Archive and the German LOFAR network are both coordinated and operated by the J\"{u}lich Supercomputing Centre (JSC), and computing resources on the supercomputer JUWELS at JSC were provided by the Gauss Centre for Supercomputing e.V. (grant CHTB00) through the John von Neumann Institute for Computing (NIC). 

This research made use of the University of Hertfordshire high-performance computing facility and the LOFAR-UK computing facility located at the University of Hertfordshire and supported by STFC [ST/P000096/1], and of the Italian LOFAR IT computing infrastructure supported and operated by INAF, and by the Physics Department of Turin university (under an agreement with Consorzio Interuniversitario per la Fisica Spaziale) at the C3S Supercomputing Centre, Italy.

Funding for the Sloan Digital Sky Survey IV has been provided by the Alfred P. Sloan Foundation, the U.S. Department of Energy Office of Science, and the Participating Institutions. SDSS-IV acknowledges support and resources from the Center for High Performance Computing at the University of Utah. The SDSS website is \url{www.sdss.org}.

SDSS-IV is managed by the Astrophysical Research Consortium for the Participating Institutions of the SDSS Collaboration including the Brazilian Participation Group, the Carnegie Institution for Science, Carnegie Mellon University, Center for Astrophysics | Harvard \& Smithsonian, the Chilean Participation Group, the French Participation Group, Instituto de Astrof\'isica de Canarias, The Johns Hopkins University, Kavli Institute for the 
Physics and Mathematics of the Universe (IPMU) / University of Tokyo, the Korean Participation Group, Lawrence Berkeley National Laboratory, Leibniz Institut f\"ur Astrophysik Potsdam (AIP), Max-Planck-Institut f\"ur Astronomie (MPIA Heidelberg), Max-Planck-Institut f\"ur Astrophysik (MPA Garching), Max-Planck-Institut f\"ur Extraterrestrische Physik (MPE), National Astronomical Observatories of China, New Mexico State University, New York University, University of Notre Dame, Observat\'ario Nacional / MCTI, The Ohio State University, Pennsylvania State University, Shanghai Astronomical Observatory, United Kingdom Participation Group, Universidad Nacional Aut\'onoma de M\'exico, University of Arizona, University of Colorado Boulder, University of Oxford, University of Portsmouth, University of Utah, University of Virginia, University of Washington, University of Wisconsin, Vanderbilt University, and Yale University.

Finally, we acknowledge the developers of the following python packages (not mentioned explicitly in the text), which were used extensively during this project: \texttt{aplpy} \citep{Robitaille2012}, \texttt{astropy} \citep{Astropy2013}, \texttt{cmasher} \citep{vanderVelden2020}, \texttt{colorcet} \citep{Kovesi2015}, \texttt{matplotlib} \citep{Hunter2007}, \texttt{numpy} \citep{Numpy2011} and \texttt{scipy} \citep{Jones2001}.

\section*{Data Availability}
The images underlying this article will be shared on reasonable request to the corresponding author. Raw MeerKAT visibilities for SCI-20210212-CR-01 are in the public domain and can be accessed via the SARAO archive (\url{https://apps.sarao.ac.za/katpaws/archive-search}). Raw LOFAR visibilities can be accessed via the LOFAR Long-Term Archive (LTA; \url{https://lta.lofar.eu}). The \textit{XMM-Newton} data used in this paper are available through the Science Data Archive (\url{https://www.cosmos.esa.int/web/xmm-newton/xsa}), alternatively processed data products can be downloaded via the X-COP webpage (\url{https://dominiqueeckert.wixsite.com/xcop}). \textit{Chandra} data are available via the Chandra Data Archive (\url{https://cxc.harvard.edu/cda/}).

\bibliographystyle{aa}
\bibliography{A2142_MeerKAT-meets-LOFAR}

\begin{appendix}
\section{Wide-Field MeerKAT Images}
In this appendix, we show the full-field direction-dependent calibrated MeerKAT at 1283\,MHz. Selected sources are highlighted that are unassociated with Abell~2142, but may be of broader interest to the community. Specifically, these are the giant radio galaxy associated with LEDA~1783783 at redshift $z = 0.086$ \citep[e.g.][]{Albareti2017_SDSS-DR13} and ORC-4 at redshift $z = 0.385$ \citep[][]{Norris2021_ORCs_Galaxies,Norris2021_ORCs_PASA}.

In the case of LEDA~1783783, the angular span of the lobes is 9.33~arcmin corresponding to a physical span of around 870~kpc at $z = 0.086$, placing this firmly in the category of giant radio galaxies. We note however that this should be considered a lower limit to the extent of the radio emission as this does not account for the curvature of the extended lobe emission, which is particularly evident in the southern lobe.

For ORC-4, our MeerKAT maps show a double-ring structure as well as the compact radio source associated with the host galaxy. The second, fainter ring is slightly offset to the east of the primary, brighter ring; this second ring is barely visible in the GMRT maps presented by \cite{Norris2021_ORCs_PASA} but with the improved signal-to-noise afforded by our MeerKAT observations (despite the source position well beyond the PB FWHM), it is detected at sufficient signal-to-noise to be confirmed. In total the span of the double-ring structure is 1.81~arcmin, corresponding to a physical scale of 551~kpc at the galaxy redshift. For the primary ring alone, the span is 1.44~arcmin, equivalent to a physical span of 438~kpc.

The compact central AGN has a flux density $S_{\rm 1283 \, MHz} = 0.59 \pm 0.04$~mJy; taking the 325~MHz flux density measurement for this compact source from \cite{Norris2021_ORCs_PASA}, we find a two-point spectral index of $\alpha = -0.48 \pm 0.09$. Thus, we find a $k$-corrected radio power of $P_{\rm 1283 \, MHz} = (2.41 \pm 0.16) \times 10^{23}$~W~Hz$^{-1}$ for this compact radio source.

\begin{figure*}
\begin{center}
 \includegraphics[width=0.75\linewidth]{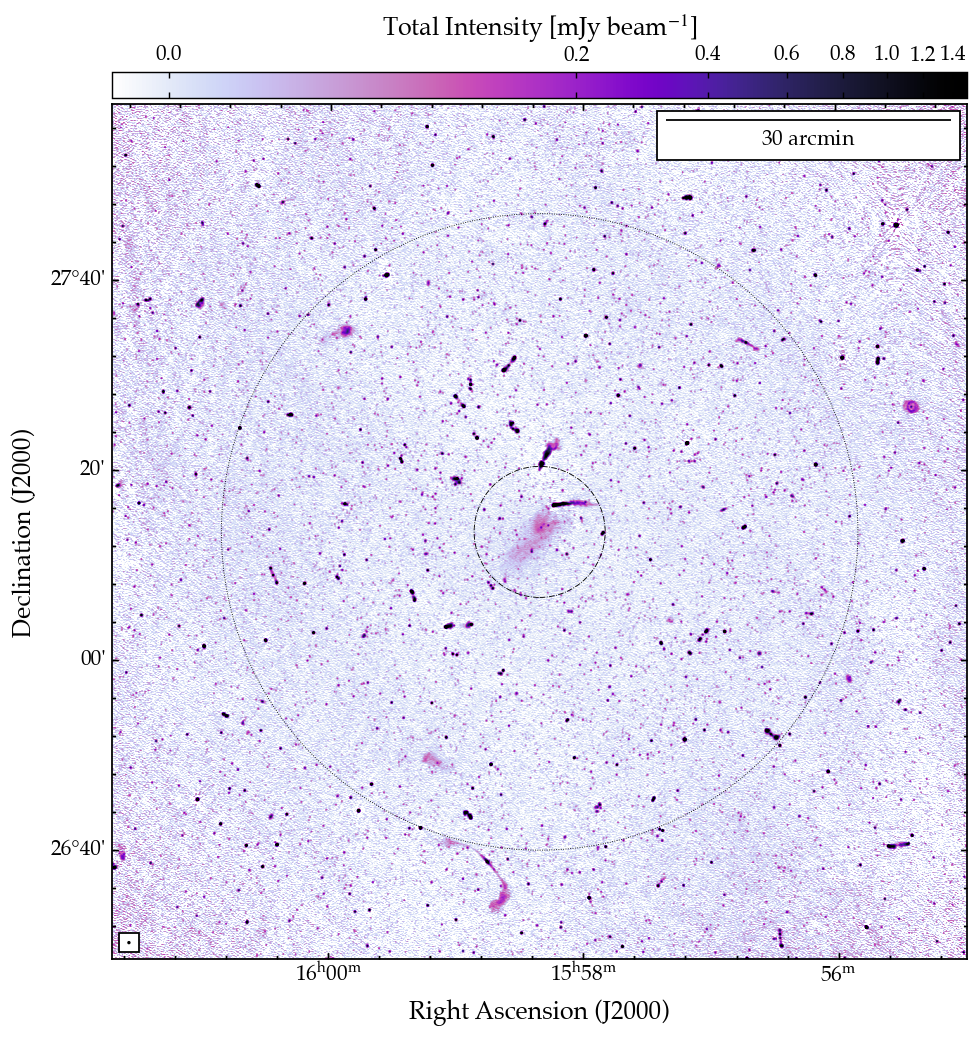}
 \includegraphics[width=0.4\linewidth]{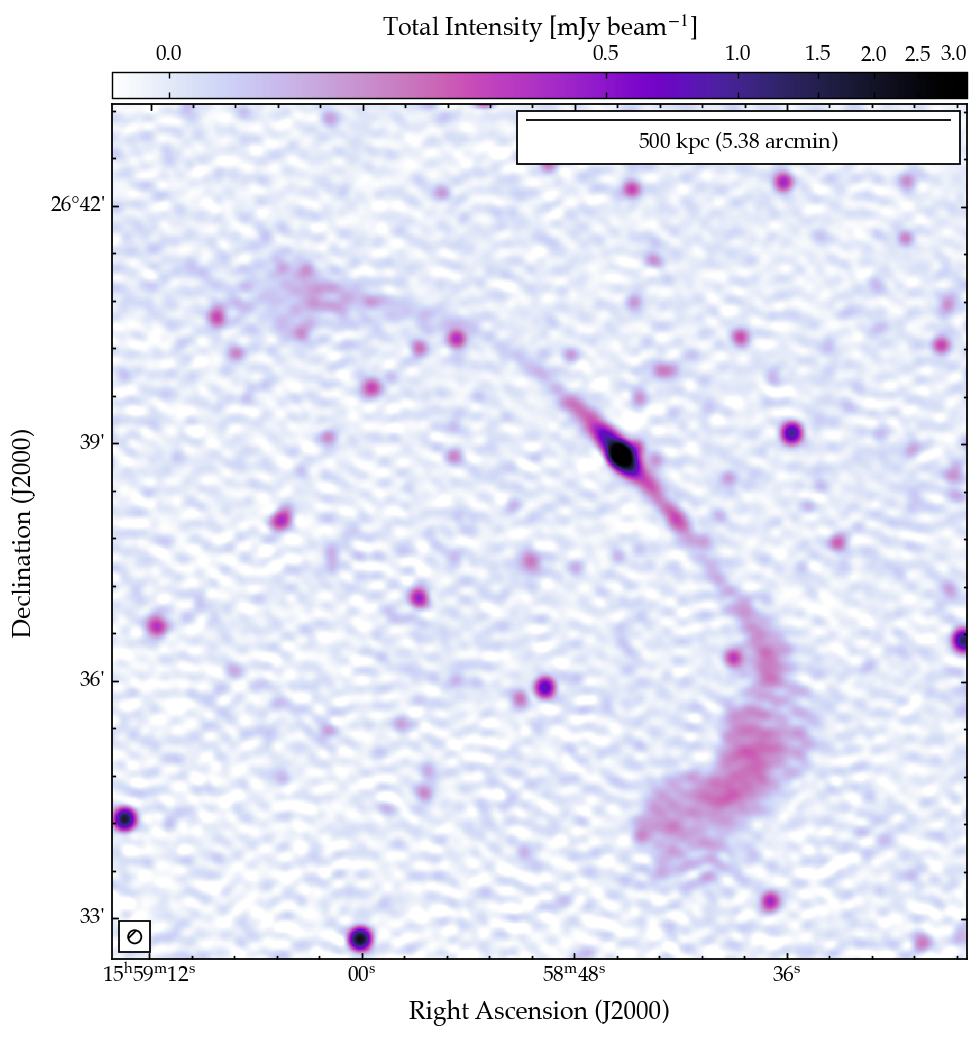}
 \includegraphics[width=0.4\linewidth]{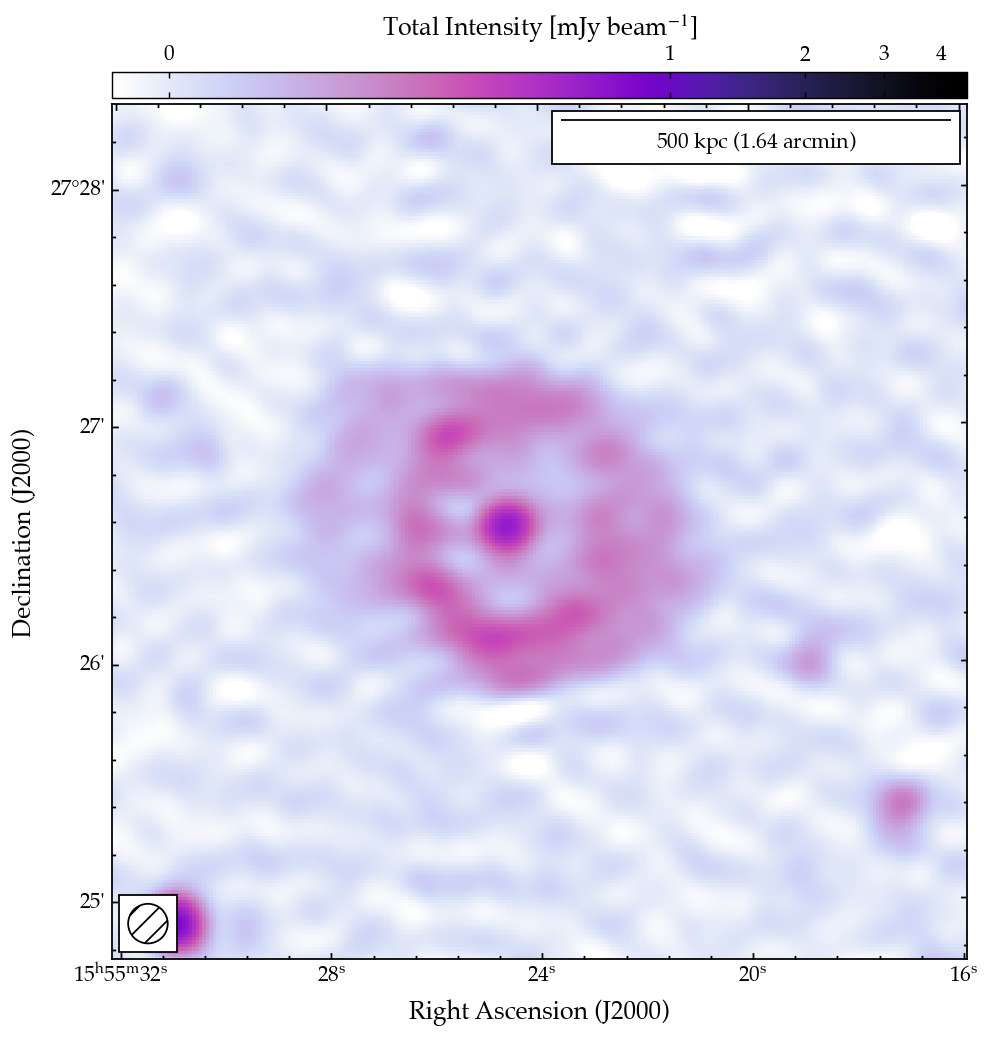}
\caption{Direction-dependent calibrated MeerKAT images produced from our observations of Abell~2142. \textit{Upper:} full-field image centred on the cluster, with the inner circle denoting a 1~Mpc radius around the cluster centre, and the outer circle showing the MeerKAT PB FWHM at 1283~MHz (67~arcmin). \textit{Lower:} zooms on two selected sources of interest, the giant radio galaxy associated with LEDA~1783783 at redshift $z = 0.086$ \citep[\textit{left}, e.g.][]{Albareti2017_SDSS-DR13} and ORC-4 at redshift $z = 0.385$ \citep[\textit{right}, e.g.][]{Norris2021_ORCs_Galaxies,Norris2021_ORCs_PASA}. In each of the lower panels, the colourscale ranges from $-2\sigma_{\rm local}$ to $250\sigma_{\rm local}$ where $\sigma_{\rm local} = 12.8 \, (18.2) \, \upmu$Jy~beam$^{-1}$ in the left (right) panel; the scalebar corresponds to a physical scale of 500~kpc at the indicated host galaxy redshift.}
\label{fig:fullfield_images}
\end{center}
\end{figure*}

\end{appendix}

\end{document}